\documentclass[12pt]{article}
\pdfoutput=1

\usepackage{draft,hyperref,tikz,cancel,subfig,float}

\usepackage{graphicx,etoolbox}
\robustify{\includegraphics}

\usepackage[nosort]{cite}

\usetikzlibrary{snakes}
\usetikzlibrary{shapes.misc}

\newtheorem{conj}{Conjecture}

\begin{document}

\begin{titlepage}

\rightline{ CALT-TH 2017-015 }

\begin{center}

\hfill \\
\hfill \\
\vskip 1cm

\title{
Carving Out the End of the World
\\
\large or (Superconformal Bootstrap in Six Dimensions)
}

\author{Chi-Ming Chang${\includegraphics[scale=0.02]{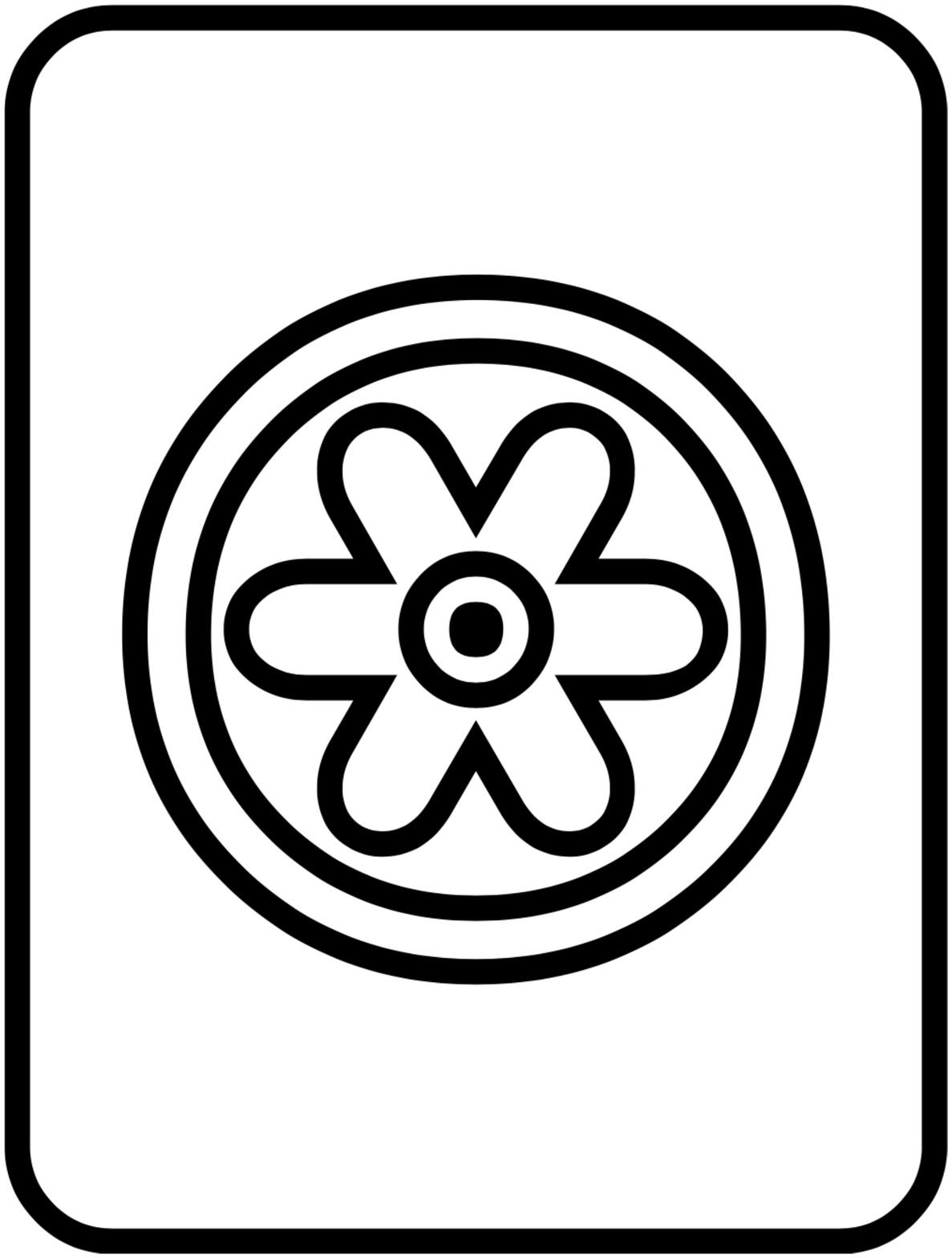}}$, Ying-Hsuan Lin${\includegraphics[scale=0.02]{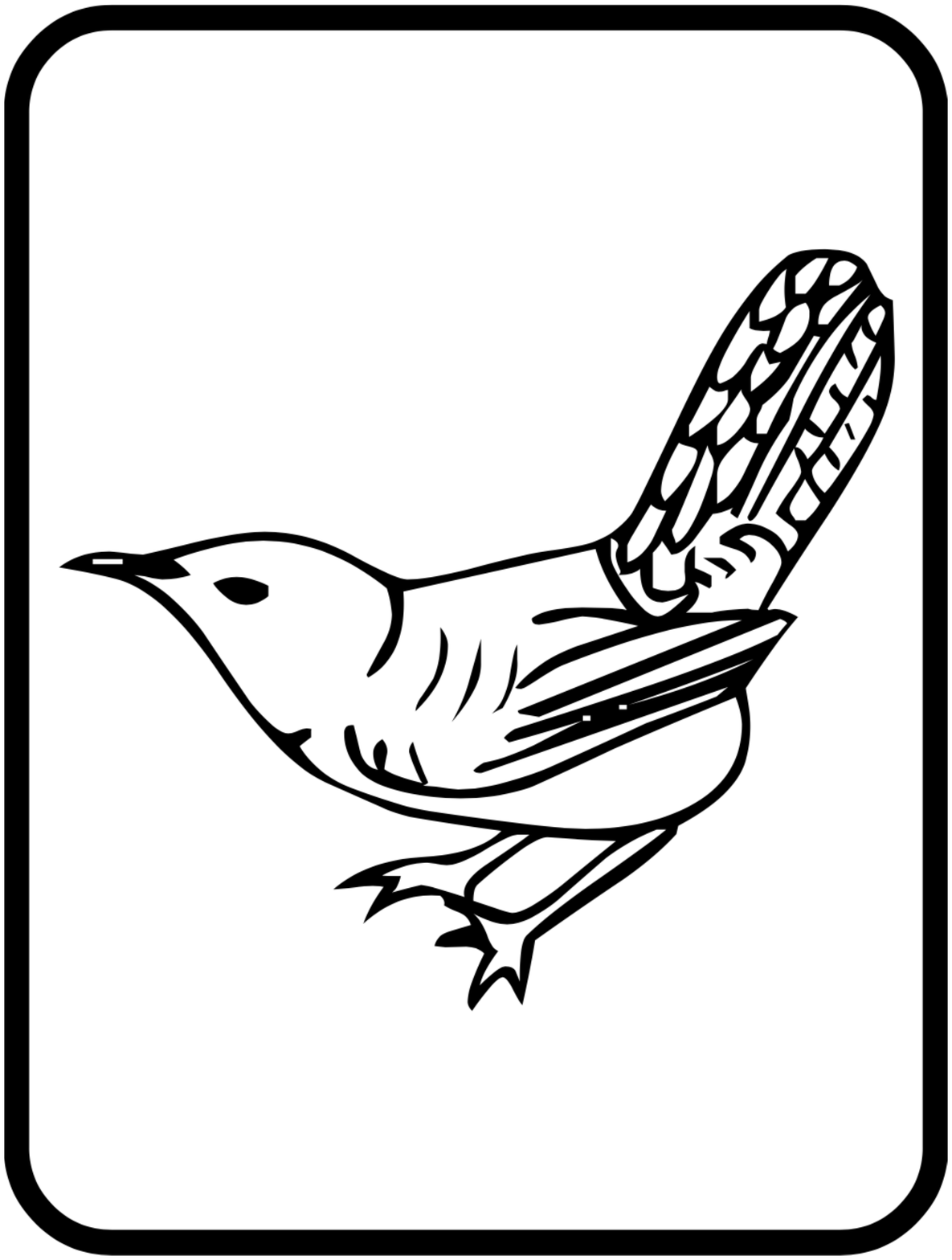}}$
}

\address{${\includegraphics[scale=0.02]{bing.pdf}}$Center for Quantum Mathematics and Physics (QMAP) \\
University of California, Davis, CA 95616, USA}
\address{${\includegraphics[scale=0.02]{bird.pdf}}$Walter Burke Institute for Theoretical Physics \\ California Institute of Technology,
Pasadena, CA 91125, USA}

\email{wychang@ucdavis.edu, yhlin@caltech.edu}

\end{center}

\abstract{We bootstrap ${\cal N}=(1,0)$ superconformal field theories in six dimensions, by analyzing the four-point function of flavor current multiplets.  Assuming $E_8$ flavor group, we present universal bounds on the central charge $C_T$ and the flavor central charge $C_J$.  Based on the numerical data, we conjecture that the rank-one E-string theory saturates the universal lower bound on $C_J$, and numerically determine the spectrum of long multiplets in the rank-one E-string theory.  We comment on the possibility of solving the higher-rank E-string theories by bootstrap and thereby probing M-theory on AdS${}_7\times{\rm S}^4$/$\bZ_2$.
} 

\vspace{.5in}

\begin{center}
{\it Dedicated to John Schwarz on the occasion of his 75th birthday.}
\end{center}

\vfill

\end{titlepage}

\eject

\tableofcontents

\section{Introduction and summary}

Conformal field theories in six dimensions parent a plethora of conformal field theories in lower dimensions through compactification.  A primal example is the compactification of ${\cal N} = (2,0)$ theories on Riemann surfaces to class $\cal S$ theories in four dimensions \cite{Gaiotto:2009hg,Gaiotto:2009we}.  While no argument exists for the necessity of supersymmetry, all known interacting conformal field theories in six dimensions are in fact superconformal.\footnote{A class of non-supersymmetric AdS$_7$ vacua in the massive type IIA supergravity was recently proposed as potential duals to non-supersymmetric 6d CFTs\cite{Apruzzi:2016rny} (we thank Xi Yin for pointing this out to us).  However, it is unclear whether those vacua are stable.  The proposal also violates the strong version of the weak gravity conjecture\cite{Ooguri:2016pdq}.}  It follows from representation theory that these interacting theories have neither marginal nor relevant deformations \cite{Minwalla:1997ka,Bhattacharya:2008zy,Louis:2015mka,Buican:2016hpb,Cordova:2016emh,Cordova:2016xhm}.  Moreover, no known interacting theory admits a classical limit (hence essentially strongly coupled), or arises in the infrared limit of renormalization group flows from a Lagrangian theory.  For these reasons, only a scarcity of tools exists for extracting physical quantities in these theories.

The conformal bootstrap aims to extract physical observables in strongly coupled conformal field theories, using only the basic assumptions: unitarity, (super)conformal symmetry, and the associativity of operator product expansions (OPEs)\cite{Polyakov:1974gs,Ferrara:1973yt,Mack:1975jr,Belavin:1984vu}.  The past decade has seen substantial developments of numerical bootstrap techniques -- most notably the linear functional method -- in constraining conformal field theories \cite{Rattazzi:2008pe,Rychkov:2009ij,Poland:2010wg,Poland:2011ey,ElShowk:2012ht,ElShowk:2012hu,Beem:2013qxa,Kos:2013tga,El-Showk:2014dwa,Chester:2014fya,Kos:2014bka,Caracciolo:2014cxa,Chester:2014mea,Bae:2014hia,Beem:2014zpa,Chester:2014gqa,Simmons-Duffin:2015qma,Kos:2015mba,Chester:2015qca,Beem:2015aoa,Iliesiu:2015qra,Lemos:2015awa,Lin:2015wcg,Kos:2016ysd,Li:2016wdp,Collier:2016cls,Lin:2016gcl,Lemos:2016xke,Li:2017ddj,Collier:2017shs}.
In particular, the bootstrap has been applied to ${\cal N}=(2,0)$ superconformal symmetry in six dimensions, and substantial evidence was found to support the conjecture that the bootstrap bound on the central charge is saturated by the $A_1$ theory, which arises in the infrared limit of the worldvolume theory of two coinciding M5 branes \cite{Beem:2015aoa}.  For theories that saturate the bootstrap bounds,  the linear functional method determines the scaling dimensions and OPE coefficients of all the operators that contribute to the correlators under analysis\cite{ElShowk:2012hu}.  By incorporating more and more correlators, the conformal bootstrap potentially solves these theories completely.\footnote{The mixed correlator bootstrap refines the constraints on the space of unitary conformal field theories \cite{Kos:2014bka,Kos:2016ysd,Li:2016wdp,Li:2017ddj}.
}

In this paper, we apply the conformal bootstrap to study yet another interesting class of six-dimensional conformal field theories -- the E-string theories -- which arise in the infrared limit of the worldvolume theory of M5 branes lying inside an ``end-of-the-world'' M9 brane\cite{Ganor:1996mu,Seiberg:1996vs}.  These ${\cal N} = (1,0)$ theories have tensionless string excitations charged under an $E_8$ flavor symmetry, and are related to various lower-dimensional conformal field theories.  For instance, upon compactification on a circle with the presence of $E_8$ Wilson lines, they reduce to Seiberg's $E_n$ theories in five dimensions\cite{Seiberg:1996bd,Morrison:1996xf,Ganor:1996pc}.  Compactifying on Riemann surfaces lands us on various $\cN=1$ theories in four dimensions \cite{Heckman:2016xdl,Razamat:2016dpl}.

There is a larger class of ${\cal N} = (1,0)$ theories coming from F-theory constructions that contains the E-string theories as a subclass\cite{DelZotto:2014hpa,Heckman:2013pva,Heckman:2014qba,Heckman:2015bfa}.  In order to pinpoint specific theories on the solution space of bootstrap, we need to know the values of certain physical observables.  One physical observable that has been computed in known six-dimensional theories is the anomaly polynomial \cite{Harvey:1998bx,Intriligator:2000eq,Ohmori:2014pca,Ohmori:2014kda,Intriligator:2014eaa,Mekareeya:2016yal,Shimizu:2017kzs}.  By superconformal symmetry, the anomaly polynomial uniquely fixes both the central charge $C_T$ and flavor central charge $C_J$, which are in turn related to certain OPE coefficients \cite{Cordova:2015fha,Beccaria:2015uta,Beccaria:2015ypa}.  The precise relation between $C_J$ and the 't Hooft anomaly coefficients should appear in \cite{Dumitrescu}, and the relation for $C_T$ was determined in \cite{Beccaria:2015ypa,Yankielowicz:2017xkf,Beccaria:2017dmw}.

Employing numerical bootstrap techniques, we analyze the four-point function of scalar superconformal primaries in the $E_8$ flavor current multiplets.  Based on the results, we propose the following conjecture:
\begin{conj}
\label{Conj:MinCJ}
The rank-one E-string theory has the minimal flavor central charge $C_J = 150$ among all unitary interacting superconformal field theories in six dimensions with an $E_8$ flavor group.
\end{conj}
We emphasize to the reader that the true virtue of this conjecture is {\it not} that we can compute $C_J$ by bootstrap, but rather the fact that if the rank-one E-string theory indeed saturates the bootstrap bound, then the entire OPEs between the flavor current multiplets can be determined (up to signs) by the linear functional method.  This would be invaluable input towards a full solution of the rank-one E-string theory by the conformal bootstrap.  We shall comment on the possibility of solving the higher-rank E-string theories and thereby probing the dual M-theory on AdS${}_7\times{\rm S}^4$/$\bZ_2$.

The organization of this paper is as follows.  Section~\ref{Sec:Rep} reviews the superconformal representation theory of the ${\cal N} = (1,0)$ algebra in six dimensions.  In Sections~\ref{sec:4pt} and~\ref{sec:SB}, we write down the general form of the four-point function involving $1\over2$-BPS scalars in flavor current multiplets that solves the superconformal Ward identities, and determine the superconformal blocks.  Section~\ref{Sec:Flavor} explains how to introduce non-abelian flavor symmetry.  In Section~\ref{Sec:CentralCharges}, we relate the central charge $C_T$ and flavor central charge $C_J$ to certain coefficients in the OPEs between flavor current multiplet scalars.  In Section~\ref{Sec:Semi}, we review the linear functional method which turns the problem of bounding OPE coefficients to a problem in semidefinite programming.  Section~\ref{Sec:Bounds} presents the numerical bounds and their physical implications.  Section~\ref{Sec:Outlook} discusses the future outlook.

\section{Review of superconformal representation theory}
\label{Sec:Rep}

The six-dimensional ${\cal N}= (1,0)$ superconformal algebra is $\mathfrak{osp}(8|2)$, which contains a bosonic subalgebra $\mathfrak{so}(2,6)\times \mathfrak{su}(2)_R$. There are sixteen fermonic generators: eight supercharges $Q^A_\A$ and eight superconformal supercharges $S^\A_A$, where $\A=1,\cdots,4$ and $A=1,2$ are the $\mathfrak{so}(6)$ and $\mathfrak{su}(2)_R$ spinor indices, respectively. Superconformal primaries are operators that are annihilated by all the superconformal supercharges $S^\A_A$.  A highest weight state of $\mathfrak{osp}(8^*|2)$ is a superconformal primary that is also a highest weight state of the maximal compact subalgebra $\mathfrak{so}(2) \times \mathfrak{so}(6) \times \mathfrak{su}(2)_R$. Representations of the superconformal algebra are generated by successively acting the supercharges $Q^A_\A$ and the lowering generators of $\mathfrak{so}(6)\times \mathfrak{su}(2)_R$ on the highest weight states. 
While some descendants of a highest weight state can appear to have zero norm, in unitary theories, they must be decoupled, and the shortened multiplets are referred to as short multiplets.

Each superconformal multiplet can be labeled by the charges $\Delta,h_1,h_2,h_3, J_R$ of its highest weight state under the Cartan generators of $\mathfrak{so}(2) \times \mathfrak{so}(6)\times \mathfrak{su}(2)_R$, where $h_1,h_2,h_3$ are the charges under the subgroup $\mathfrak{so}(2)^3 \subset \mathfrak{so}(6)$.  All the charges are real for unitary representations of the Lorentzian conformal algebra $\mathfrak{so}(2,6) \times \mathfrak{su}(2)_R$. The short representations are classified into ${\cal A}, {\cal B}, {\cal C}, {\cal D}$ types, satisfying the following relations\cite{Minwalla:1997ka,Bhattacharya:2008zy,Buican:2016hpb,Cordova:2016emh},
\ie
\label{ShortDelta}
{\cal A}&:&&\Delta=4J_R + {c_1\over 2} + c_2 +{3c_3\over 2} + 6,
\\
{\cal B}&:&&\Delta=4J_R + {c_1\over 2} + c_2 + 4,&&c_3=0,
\\
{\cal C}&:&&\Delta=4J_R + {c_1\over 2} + 2,&&c_2=c_3=0,
\\
{\cal D}&:&&\Delta=4J_R,&&c_1=c_2=c_3=0,
\fe
where $c_1$, $c_2$ and $c_3$ are the Dynkin labels of $\mathfrak{su}(4)$ which is related to the $h_1$, $h_2$ and $h_3$ by
\ie
h_1={1\over 2}c_1+c_2+{1\over 2}c_3,~~h_2={1\over 2} c_1 + {1\over 2}c_3,~~h_3 = {1\over 2} c_1 - {1\over 2} c_3.
\fe
The ${\cal D}$-type highest weight states are annihilated by the four supercharges with positive R-charge, and are therefore ${1\over 2}$-BPS.  The ${\cal A}$-, ${\cal B}$-, and ${\cal C}$-type multiplets always contain BPS operators, although their highest weight states are not BPS. The long representations satisfy the inequality
\ie\label{eqn:unitarityBound}
{\cal L}:\Delta>4J_R + {c_1\over 2} + c_2 +{3c_3\over 2} + 6.
\fe

Let us denote the multiplets by\footnote{We use $2J_R$ since it is the Dynkin label of $\mathfrak{su}(2)_R$.
}
\ie
{\cal X}[\Delta;c_1,c_2,c_3;2J_R],\quad {\cal X}={\cal L},{\cal A},{\cal B},{\cal C},{\cal D}.
\fe
Due to OPE selection rules, later we only have to consider multiplets whose superconformal primaries are in the symmetric rank-$\ell$ representation of $\mathfrak{so}(6)$. We denote such representations by
\ie
\label{ell6}
{\cal X}[2J_R]_{\Delta,\ell}={\cal X}[\Delta;0,\ell,0;2J_R].
\fe
The $\Delta,\ell$ subscripts for ${\cal D}$-type multiplets and the $\Delta$ subscript for ${\cal B}$-type will be omitted since their values are fixed by \eqref{ShortDelta} and \eqref{ell6}.

\paragraph{Important short multiplets}

We give names to certain special short multiplets, some of which contain conserved currents.

\begin{itemize}

\item Identity multiplet ${\cal D}[0]$: This multiplet contains only the identity operator (vacuum state). 

\item Hypermultiplet ${\cal D}[1]$: contains two complex scalars and one Weyl spinor.

\item Flavor current multiplet ${\cal D}[2]$:  contains conserved currents transforming in the adjoint of a flavor symmetry, and their supertners.

\item Stress tensor multiplet ${\cal B}[0]_{0}$: contains the R-symmetry currents, the stress tensor and their superpartners.

\item Higher spin multiplet ${\cal B}[0]_{\ell}$ for $\ell>0$: contains a spin-$(\ell+2)$ higher spin conserved current and and their superpartners.  These multiplets generally live in a decoupled free subsector\cite{Alba:2015upa,Maldacena:2011jn,Alba:2013yda}.

\end{itemize}

\section{Four-point function of half-BPS operators}\label{sec:4pt}

In this section, we consider the four-point function of the scalar superconformal primaries in the ${1\over 2}$-BPS multiplet ${\cal D}[k]$, and review the constraints from superconformal symmetry\cite{Dolan:2004mu}. The $1\over2$-BPS condition implies that this four-point function uniquely fixes the entire set of four-point functions of the (primary or descendant) operators in ${\cal D}[k]$.\footnote{The superfield for a $1\over 2$-BPS multiplet only depends on four fermionic coordinates (half the number of fermionic coordinates in full superspace). The four-point function of such superfields depends on sixteen fermionic coordinates, which is the same as the number of fermionic generators in the superconformal algebra. Hence the four-point function of the superfields can be obtained by supersymmetrizing the four-point function of the superconformal primaries. There is no extra constraint coming from the crossing symmetry of the four-point functions of superconformal descendants.
}  Although we are interested in ${\cal N} = (1,0)$ in six dimensions, the setup is the same for superconformal field theories in other dimensions where the R-symmetry is $\mathfrak{su}(2)_R$, namely, ${\cal N} = 1$ in five dimensions and ${\cal N} = 3$ in three dimensions.\footnote{Our setup does not apply to ${\cal N} = 2$ in four dimensions.  In particular, such a theory has a protected subsector corresponding to a two-dimensional chiral algebra\cite{Dolan:2004mu,Beem:2014zpa}.} Hence we keep the spacetime dimension general and write it as $d=2(\epsilon+1)$.

The scalar superconformal primaries form a spin-${k\over 2}$ representation of $\mathfrak{su}(2)_R$, and their weight is fixed by the BPS condition $\Delta=\epsilon  k$. The scalars can be written as $\cO_{A_1\cdots A_{k}}(x)$, which is a symmetric rank-$k$ tensor of the fundamental representation of $\mathfrak{su}(2)_R$, $A_i=1,2$. 
We can contract the indices with auxiliary variables $Y^A$ to form an operator $\cO(x,Y)$ that has homogenous degree $(-\epsilon k,k)$. The four point function of $\cO(x,Y)$ is then a homogenous degree $(-4\epsilon k,4k)$ function, and is polynomial in $Y^A$.  Therefore it must take the form
\ie
\label{4pf}
&\vev{\cO(x_1,Y_1)\cO(x_2,Y_2)\cO(x_3,Y_3)\cO(x_4,Y_4)}=\left({(Y_1\cdot Y_2)(Y_3\cdot Y_4)\over x_{12}^{2\epsilon}x_{34}^{2\epsilon}}\right)^{k}G(u,v;w),
\\
&G(u,v;w) = G_{0}(u,v)+G_1(u,v)w^{-1}+\cdots +G_k(u,v)w^{-k},
\fe
where the cross ratios $u$, $v$, and $w$ are defined as\footnote{The variables $Y^A_i$ satisfy the identity $(Y_1\cdot Y_2)(Y_3\cdot Y_4)-(Y_1\cdot Y_3)(Y_2\cdot Y_4)+(Y_1\cdot Y_4)(Y_2\cdot Y_3)=0$.}
\ie
&u={x_{12}^2 x_{34}^2\over x_{13}^2 x_{24}^2},\quad v={x_{14}^2 x_{23}^2\over x_{13}^2 x_{24}^2},\quad w={(Y_1\cdot Y_2)(Y_3\cdot Y_4)\over (Y_1\cdot Y_4)(Y_2\cdot Y_3)},
\\
&x_{12}^2=(x_1-x_2)^2,\quad Y_1\cdot Y_2=Y_1^A Y^B_2\epsilon_{BA}.
\fe
As all four external scalars are identical, the invariance of \eqref{4pf} under $(x_1,Y_1)\leftrightarrow (x_3,Y_3)$ leads to the crossing symmetry constraint
\ie\label{crossing13}
G(u,v;w)=\left({u^\epsilon \over v^\epsilon w}\right)^kG(v,u;w^{-1}).
\fe
Similarly, the invariance of \eqref{4pf} under $(x_1,Y_1)\leftrightarrow (x_2,Y_2)$ leads to the constraint
\ie\label{crossing12}
G(u,v;w)=G\left({u\over v},{1\over v};-{w\over w+1}\right).
\fe

The four-point function is further constrained by the superconformal Ward identities, which we review in Appendix~\ref{App:Ward}.  They were solved in \cite{Dolan:2004mu}, and the solutions are parametrized by $k-2$ functions $b_n(u,v)$,
\ie\label{eqn:WardSolution}
G(u,v;w)=&\sum_{n=0}^{k-2}u^{(n+2)\epsilon}\Delta_{\epsilon}\left[(v+u w^{-1})(1+w^{-1})-w^{-1}\right]\left(1+w^{-1}\right)^nb_n(u,v),
\fe
where the differential operator $\Delta_\epsilon$ is defined as
\ie
\label{DefineDelta}
&D_\epsilon=u{\partial^2\over\partial u^2}+v{\partial^2\over\partial v^2} +(u+v-1){\partial^2\over \partial u\partial v}+(1+\epsilon)\left({\partial\over \partial u}+{\partial\over \partial v}\right),
\\
&\Delta_\epsilon=(D_\epsilon)^{\epsilon-1}u^{\epsilon-1}.
\fe

In even dimensions, $\Delta_\epsilon$ is a well-defined differential operator, and is invariant under crossing.  One approach to solving the crossing equation is to ``factor out'' $(D_\epsilon)^{\epsilon-1}$ and write down a crossing equation for $b_n(u, v)$ (while carefully taking care of the kernel of $(D_\epsilon)^{\epsilon-1}$), as was the approach of \cite{Beem:2015aoa}.  However, in odd dimensions, the differential operator $(D_\epsilon)^{\epsilon-1}$ is defined only formally on the functional space spanned by Jack polynomials with eigenvalues given in \eqref{eqn:DeltaEigenvalue}, and this functional space does not map to itself under crossing $u \leftrightarrow v$.\footnote{We thank Silviu S. Pufu for a discussion on the subtleties of the differential operator $\Delta_\epsilon$.
}  To make our setup easily generalizable to five and three dimensions, we will not study the crossing equation for $b_n(u, v)$, but will instead analyze the crossing equation for $G(u, v; w)$ directly.  See Appendix~\ref{App:bCrossing} for the setup of the crossing equation for $b_n(u, v)$ in the special case of $\epsilon = k = 2$.

The rest of the paper specializes to the case of $k=2$. Then $G(u,v;w)$ is a second degree polynomial in $w^{-1}$.  By matching the coefficients of the monomials in $w$, the crossing equation \eqref{crossing13} can be separated into three equations involving only $u$ and $v$,
\ie
\label{eqn:crossingk=2}
u^{-2\epsilon}G_2(u,v)=v^{-2\epsilon}G_0(v,u),
\\
u^{-2\epsilon}G_1(u,v)=v^{-2\epsilon}G_1(v,u),
\\
u^{-2\epsilon}G_0(u,v)=v^{-2\epsilon}G_2(v,u),
\fe
where $G_i$ are defined in \eqref{4pf}, and the third equation is trivially equivalent to the first equation.  In Appendix~\ref{App:Ward}, we show that the second equation also follows from the first equation as a consequence of the superconformal Ward identities \eqref{eqn:SCW}.  Moreover, the superconformal Ward identities imply an identity \eqref{CrossingIdentity} on the first equation, which is important when we need to identify the independent constraints from the crossing equation in order when applying the linear functional method.

\section{Superconformal blocks}
\label{sec:SB}
The four point function can be expanded in superconformal blocks as
\ie\label{eqn:sBlockExpansion}
G(u,v;w)=\sum_{\mathcal X}\lambda^2_{\mathcal X} {\cal A}^{\cal X}(u,v;w),
\fe
where ${\cal A}^{\cal X}(u,v;w)$ is the superconformal block of the superconformal multiplet $\cal X$. The sum is over the superconformal multiplets allowed in the OPE of two ${\cal D}[2]$. The selection rule is
\ie\label{eqn:(1,0)selectionRole}
\hspace{-0.15in}
&{\cal D}[2]\times{\cal D}[2]
=\sum^{2}_{j=0}{\cal D}[4-2j]+ \sum^{2}_{j=1}\sum_{\ell=0}^\infty{\cal B}[4-2j]_{\ell}+ 
\sum_{\ell=0}^\infty\sum_{\Delta>\ell+2\epsilon(2-j)+6}{\cal L}[0]_{\Delta,\ell},
\hspace{-0.15in}
\fe
as we presently argue.
First, a generalization of \cite{Ferrara:2001uj} shows \eqref{eqn:(1,0)selectionRole} with the possible addition of ${\cal A}[0]_\ell$ and ${\cal C}[0]$.\footnote{Consider the three-point function of two superfields of ${\cal D}[2]$ with a generic superfield $\cal O$ in the harmonic superspace. The bottom component of such three-point function takes the form of equation (3.3) in \cite{Ferrara:2001uj} with the obvious modifications. By the arguments of \cite{Ferrara:2001uj}, $\cal O$ must correspond to either a ${\cal D}$- or ${\cal B}$-type multiplet if $\cal O$ has $2J_R=2$, and a ${\cal D}$-type if $2J_R=4$.
}
However, no consistent superconformal block satisfying the superconformal Ward identities exists for ${\cal A}[0]_\ell$ and ${\cal C}[0]$, thereby proving their absence (see footnote~\ref{ACProof}).
The constraint \eqref{crossing12} imposes an additional selection rule $\ell+J_R \in 2\bZ$ on the intermediate primary operators.\footnote{Note the the bosonic conformal blocks satisfy ${\cal G}_{\Delta,\ell}(u,v)=(-1)^\ell {\cal G}_{\Delta,\ell}(u/v,1/v)$.
}  A superconformal block can be expanded in products of bosonic conformal blocks ${\cal G}_{\Delta,\ell}$ and $\mathfrak{su}(2)_R$ harmonics,
\ie\label{eqn:sBlock}
{\cal A}^{\cal X}(u,v;w)=\sum_{(2J_R,\Delta,\ell)\in{\cal X}} c_{2J_R,\Delta,\ell}P_{J_R}(1+\tfrac{2}{w}) {\cal G}_{\Delta,\ell}(u,v),
\fe
where $P_{J_R}(x)$ are Legendre polynomials.
The summation $\sum_{(2J_R,\Delta,\ell)\in{\cal X}}$ is over all primary operators in the superconformal multiplet $\cal X$ that appear in the OPE, labeled by $(2J_R,\Delta,\ell)$. It is a finite sum as there are only finitely many primary operators contained in each superconformal multiplet. Bosonic conformal blocks are reviewed in Appendix~\ref{App:Bosonic}.

The coefficients $c_{2J_R,\Delta,\ell}$ are fixed by the superconformal Ward identities \eqref{eqn:SCW}.  The superconformal block expansion \eqref{eqn:sBlockExpansion} implies that the functions $b_n(u,v)$ parameterizing solutions to the superconformal Ward identities (see \eqref{eqn:WardSolution}) have expansions
\ie\label{eqn:bExpansion}
&b_n(u,v)=\sum_{\cal X}\lambda^2_{\mathcal X}b_n^{\cal X}(u,v).
\fe
Comparing \eqref{eqn:sBlockExpansion} and \eqref{eqn:sBlock} with \eqref{eqn:WardSolution} gives the relation
\ie\label{eqn:Pab}
&\sum_{J_R=0}^2 P_{J_R}(1+\tfrac{2}{w}){\cal A}^{\cal X}_{2J_R}(u,v)
= u^{2\epsilon}\Delta_{\epsilon}\left[(v+u w^{-1})(1+w^{-1})-w^{-1}\right]b^{\cal X}_0(u,v),
\fe
where ${\cal A}^{\cal X}_{2J_R}(u,v)$ is defined as
\ie\label{eqn:calAdef}
&{\cal A}^{\cal X}_{2J_R}(u,v)= \sum_{(\Delta,\ell)\in {\cal X}|_{2J_R}} c_{2J_R,\Delta,\ell} {\cal G}_{\Delta,\ell}(u,v).
\fe
The relation \eqref{eqn:Pab} can then be written as
\ie\label{eqn:bEquation}
&{\cal A}^{\cal X}_0(u,v)=u^{2\epsilon}
\Delta_{\epsilon}\left({1\over 2}-{1\over 6}u+{1\over 2}v\right)b^{\cal X}(u,v),
\\
&{\cal A}^{\cal X}_2(u,v)={1\over 2}u^{2\epsilon}
\Delta_{\epsilon}(v-1)b^{\cal X}(u,v),
\\
&{\cal A}^{\cal X}_4(u,v)={1\over 6}u^{2\epsilon}
\Delta_{\epsilon}ub^{\cal X}(u,v),
\fe
where we abbreviate $b_0$ as $b$ since there is no other $b_n$.

In the following subsections, we give explicit expressions for the superconformal blocks by solving \eqref{eqn:bEquation}.  The bosonic conformal blocks are normalized such that in the limit of $u = v \ll 1$, the leading term in the $u$ expansion is $u^\Delta$.
%\footnote{We do not include $(-)^\ell$ in our normalization of the conformal blocks.
%}
The superconformal blocks are normalized such that in the same limit, the leading term is $(-)^{J_R} u^\Delta P_{J_R}(1+{2 \over w})$.

\subsection{Long multiplets}

Inside the superconformal multiplet ${\cal L}[0]_{\Delta,\ell}$, there is a unique conformal primary of dimension $\Delta+2$, spin $\ell$, and transforming in the $\mathfrak{su}(2)_R$ representation with $2J_R = 4$\cite{Buican:2016hpb,Cordova:2016emh}.  Thus we can solve for $b$ using the last line of \eqref{eqn:bEquation}:
\ie\label{eqn:bL}
b_{{\cal L}[0]_{\Delta,\ell}}(u,v)=6 c_{4,\Delta,\ell}u^{-1}\Delta_{\epsilon}^{-1}u^{-2\epsilon}{\cal G}_{\Delta+2,\ell}(u,v),
\fe
where
\ie
c_{4,\Delta,\ell} = \frac{(\Delta +\ell ) (\Delta -\ell -2 \epsilon )}{6 (\Delta +\ell -2 \epsilon +2) (\Delta -\ell -4 \epsilon +2)}.
\fe
Using the formulae in Appendix~D of \cite{Chester:2014fya}, we obtain the explicit decomposition of the long multiplet superconformal blocks into bosonic conformal blocks, as follows:
\ie
\label{LongBlock}
& {\cal A}^{{\cal L}[0]_{\Delta,\ell}}_{0} = {\cal G}_{\Delta,\ell}{}
\\
&\hspace{-.2in} +\frac{(\ell -1) \ell  (\Delta -\ell ) (\Delta -\ell -2 \epsilon ) (\Delta -\ell -2 \epsilon +2)}{16 (\ell +\epsilon -1) (\ell +\epsilon ) (\Delta -\ell -4 \epsilon +2) (\Delta -\ell -2 \epsilon +1) (\Delta -\ell -2 \epsilon +3)} {\cal G}_{\Delta +2,\ell -2}{} 
\\
&\quad +\frac{(\Delta +\ell ) (\Delta +\ell +2) (\ell +2 \epsilon ) (\ell +2 \epsilon +1) (\Delta +\ell +2 \epsilon )}{16 (\Delta +\ell +1) (\Delta +\ell +3) (\ell +\epsilon ) (\ell +\epsilon +1) (\Delta +\ell -2 \epsilon +2)} {\cal G}_{\Delta +2,\ell +2}{} 
\\
&\hspace{-.2in} +\frac{(\Delta +1) (\Delta +2) (\Delta -2 \epsilon +2) (\Delta -2 \epsilon +3) (\Delta -\ell ) (\Delta +\ell ) (\Delta +\ell +2)}{256 (\Delta -\epsilon +1) (\Delta -\epsilon +2)^2 (\Delta -\epsilon +3) (\Delta +\ell +1) (\Delta +\ell +3) (\Delta -\ell -4 \epsilon +2)}
\\
&\quad \hspace{.5in} \times\frac{ (\Delta -\ell -2 \epsilon ) (\Delta -\ell -2 \epsilon +2) (\Delta +\ell +2 \epsilon )}{ (\Delta -\ell -2 \epsilon +1) (\Delta -\ell -2 \epsilon +3) (\Delta +\ell -2 \epsilon +2)} {\cal G}_{\Delta +4,\ell}{} 
\\
&\quad +\frac{(\Delta +\ell ) (\Delta -\ell -2 \epsilon ) \left(1-\frac{3 (\epsilon -1) \epsilon  \left(\Delta ^2-2 \Delta  (\epsilon -1)+\ell ^2+2 (\ell +3) \epsilon -2 \epsilon ^2-4\right)}{2 (\Delta -\epsilon ) (\Delta -\epsilon +2) (\ell +\epsilon -1) (\ell +\epsilon +1)}\right)}{12 (\Delta +\ell -2 \epsilon +2) (\Delta -\ell -4 \epsilon +2)} {\cal G}_{\Delta +2,\ell}{},
\\
& {\cal A}^{{\cal L}[0]_{\Delta,\ell}}_{2} = -\frac{\ell  (\Delta -\ell -2 \epsilon )}{2(\ell +\epsilon ) (\Delta -\ell -4 \epsilon +2)} {\cal G}_{\Delta+1,\ell-1}{}
\\
&\quad  -\frac{(\Delta +\ell ) (\ell +2 \epsilon )}{2(\ell +\epsilon ) (\Delta +\ell -2 \epsilon +2)} {\cal G}_{\Delta+1,\ell+1}{}
\\
&\quad -\frac{(\Delta +1) \ell  (\Delta -2 \epsilon +2) (\Delta -\ell )}{32(\Delta -\epsilon +1) (\Delta -\epsilon +2) (\ell +\epsilon ) (\Delta -\ell -4 \epsilon +2)} 
\\
&\quad \quad \times \frac{ (\Delta +\ell ) (\Delta -\ell -2 \epsilon ) (\Delta -\ell -2 \epsilon +2)}{ (\Delta -\ell -2 \epsilon +1) (\Delta -\ell -2 \epsilon +3) (\Delta +\ell -2 \epsilon +2)}{\cal G}_{\Delta+3,\ell-1}{}
\\
&\quad -\frac{(\Delta +1) (\Delta -2 \epsilon +2) (\Delta +\ell ) (\Delta +\ell +2)}{32(\Delta -\epsilon +1) (\Delta -\epsilon +2) (\Delta +\ell +1) (\Delta +\ell +3)} 
\\
&\quad \quad \times\frac{ (\ell +2 \epsilon ) (\Delta -\ell -2 \epsilon ) (\Delta +\ell +2 \epsilon ) G(\Delta +3,\ell +1)}{ (\ell +\epsilon ) (\Delta -\ell -4 \epsilon +2) (\Delta +\ell -2 \epsilon +2)}{\cal G}_{\Delta+3,\ell+1}{},
\\
& {\cal A}^{{\cal L}[0]_{\Delta,\ell}}_{4} = \frac{(\Delta +\ell ) (\Delta -\ell -2 \epsilon )}{6 (\Delta +\ell -2 \epsilon +2) (\Delta -\ell -4 \epsilon +2)}{\cal G}_{\Delta+2,\ell}{}.
\fe

\subsection{Short multiplets}
The superconformal blocks for the short multiplets can be obtained by taking limits of the superconformal block for ${\cal L}[0]_{\Delta,\ell}$, as follows:
\ie
\label{ShortBlocks}
{\cal A}^{{\cal B}[2]_{\ell}}(u,v;w)&={\ell+\epsilon+1\over (\ell+1)(\epsilon-1)}\lim_{\Delta\to \ell+4\epsilon-1}(\Delta-\ell-4\epsilon+1){\cal A}^{{\cal L}[0]_{\Delta,\ell+1}}(u,v;w),
\\
{\cal A}^{{\cal B}[0]_{\ell}}(u,v;w)&=\lim_{\Delta\to \ell+2\epsilon}(\Delta-\ell-2\epsilon){\cal A}^{{\cal L}[0]_{\Delta,\ell}}(u,v;w),
\\
{\cal A}^{{\cal D}[4]}(u,v;w)&={3\epsilon\over (\epsilon-1)(2\epsilon-1)}\lim_{\Delta\to 4\epsilon-2}(\Delta-4\epsilon+2){\cal A}^{{\cal L}[0]_{\Delta,0}}(u,v;w),
\\
{\cal A}^{{\cal D}[2]}(u,v;w)&={1\over (2\epsilon-1)}\lim_{\Delta\to 2\epsilon-1}(\Delta-2\epsilon+1){\cal A}^{{\cal L}[0]_{\Delta,-1}}(u,v;w),
\fe
where the first and third equations follow from the recombination rules at the unitary bound.\footnote{See (4.4) in \cite{Buican:2016hpb} or (2.63) in \cite{Cordova:2016emh}.}
In the second and forth equations, we need to analytically continue the superconformal block ${\cal A}^{{\cal L}[0]_{\Delta,\ell}}$ to $\Delta$ below the unitarity bound \eqref{eqn:unitarityBound}, so the limits should be regarded as mere tricks to generate solutions to the superconformal Ward identities.  One can explicitly check that the superconformal blocks for short multiplets obtained this way indeed have the correct decompositions into bosonic conformal blocks. One can also show that given the content of each multiplet, \eqref{LongBlock} or \eqref{ShortBlocks} is the unique combination of bosonic conformal blocks that solves the superconformal Ward identities.  
In fact, as mentioned earlier, the lack of a solution for ${\cal A}_\ell[0]$ and ${\cal C}[0]$ proves their absence in the selection rule \eqref{eqn:(1,0)selectionRole}.\footnote{\label{ACProof}All the bosonic component fields in ${\cal C}[0]$ are R-symmetry neutral, hence the superconformal Ward identities reduce to 
\ie
\partial_\chi G(u,v;w)|_{w\to\chi}=0, \quad \partial_{\bar\chi} G(u,v;w)|_{w\to\bar\chi}=0,
\fe
which cannot be satisfied by any non-vacuum block. The superconformal block for ${\cal A}[0]_\ell$ must take the form
\ie
\label{A0Form}
{\cal A}^{{\cal A}[0]_\ell}(\rho,\theta;w) = a \left[{\cal G}_{\ell+6,\ell}(\rho,\theta)+{\cal O}(\rho^{\ell+7})\right] + b \left[{\cal G}_{\ell+7,\ell+1}(\rho,\theta) + {\cal O}(\rho^{\ell+8})\right] P_1(1+\tfrac{2}{w}),
\fe
where $\rho$ and $\theta$ are defined by $\chi = \rho e^{i\theta}$ and $\bar\chi = \rho e^{-i\theta}$. By
\ie
&\lim_{\theta\to 0}(\partial_{\chi}+2\partial_{w}){\cal A}^{{\cal A}[0]_\ell}(\rho,\theta;w)\Big|_{w\to \chi}={1\over 2}\left[a(\ell+6)+2b(\ell+3)\right]\rho^{\ell+5} +{\cal O}(\rho^{\ell+6}),
\\
&\lim_{\theta\to {\pi\over 2}}(\partial_{\chi}+2\partial_{w}){\cal A}^{{\cal A}[0]_\ell}(\rho,\theta;w)\Big|_{w\to \chi}
\\
& \hspace{.5in} = {3\over2}\left[{a(\ell+4)+2b(\ell+3)\over (\ell+2)(\ell+4)}\sin\frac{\pi\ell}{ 2}-i{a(\ell+6)+2b(\ell+1)\over (\ell+1)(\ell+3)}\cos\frac{\pi\ell}{ 2}\right]\rho^{\ell+5} +{\cal O}(\rho^{\ell+6}),
\fe
it is clear that \eqref{A0Form} cannot satisfy the superconformal Ward identities unless $a = b = 0$.
}

The superconformal block for ${\cal B}[2]_{\ell}$ is given by
\ie
\label{B2Block}
& {\cal A}^{{\cal B}[2]_{\ell}}_{0} = \frac{\ell  \epsilon}{2(2 \epsilon +1) (\ell +\epsilon )}{\cal G}_{\ell +4 \epsilon +1, \ell- 1}{} +\frac{ (\ell +2 \epsilon ) (\ell +4 \epsilon )}{6 (\ell +\epsilon ) (\ell +3 \epsilon +1)} {\cal G}_{\ell +4 \epsilon +1, \ell+ 1}{}
\\
&\quad +\frac{\epsilon  (\ell +2 \epsilon ) (\ell +2 \epsilon +1)^2 (\ell +2 \epsilon +2) (\ell +4 \epsilon ) (\ell +4 \epsilon +1)}{8(\ell +1) (2 \epsilon +1) (\ell +3 \epsilon +1)^2 (\ell +3 \epsilon +2) (2 \ell +4 \epsilon +1) (2 \ell +4 \epsilon +3)} {\cal G}_{\ell +4 \epsilon +3, \ell+ 1}{},
\\
& {\cal A}^{{\cal B}[2]_{\ell}}_{2} = - {\cal G}_{\ell+ 4 \epsilon, \ell}{} -\frac{ \epsilon  (\ell +2 \epsilon ) (\ell +2 \epsilon +1) (\ell +4 \epsilon ) }{4(2 \epsilon +1) (\ell +\epsilon +1)
   (\ell +3 \epsilon ) (\ell +3 \epsilon +1)} {\cal G}_{\ell +4 \epsilon +2, \ell}{} 
\\
&\quad -\frac{(\ell +2 \epsilon ) (\ell +2 \epsilon +1)^3 (\ell +4 \epsilon )}{4 (\ell +1) (\ell +\epsilon +1) (\ell +3
   \epsilon +1) (2 \ell +4 \epsilon +1) (2 \ell +4 \epsilon +3)} {\cal G}_{\ell +4 \epsilon +2, \ell+ 2}{} ,
\\
& {\cal A}^{{\cal B}[2]_{\ell}}_{4} = \frac{\ell +2\epsilon}{3 (\ell +1)} {\cal G}_{\ell+ 4\epsilon+ 1, \ell+ 1}{}.
\fe
The superconformal block for ${\cal B}[0]_{\ell}$ is given by
\ie\label{eqn:B[0]SB}
&{\cal A}^{{\cal B}[0]_{\ell}}_{0}={\cal G}_{\ell+2\epsilon,\ell}{}+\frac{(\ell +2 \epsilon )^2 (\ell +2 \epsilon +1) }{4 (\ell +1) (2 \ell +2 \epsilon +1) (2 \ell +2 \epsilon +3)}{\cal G}_{\ell+2\epsilon+2,\ell+2}{},
\\
&{\cal A}^{{\cal B}[0]_{\ell}}_{2}=-\frac{\ell +2 \epsilon }{2(\ell +1)}{\cal G}_{\ell+2\epsilon+1,\ell+1}{}.
\fe
The superconformal block for ${\cal D}[4]$ is given by
\ie
\label{D4Block}
{\cal A}^{{\cal D}[4]}_{0} &=\frac{2 \epsilon ^2 }{3(4\epsilon + 1)(3\epsilon + 1)}{\cal G}_{4 \epsilon +2,0}{},
\\
{\cal A}^{{\cal D}[4]}_{2} &= -\frac{2 \epsilon}{4 \epsilon +1}{\cal G}_{4 \epsilon +1,1}{} ,
\\
{\cal A}^{{\cal D}[4]}_{4} &= {\cal G}_{4\epsilon, 0}{}.
\fe
The superconformal block for ${\cal D}[2]$ is given by
\ie\label{eqn:D[2]SB}
&{\cal A}^{{\cal D}[2]}_{0}={\epsilon\over 2\epsilon+1}{\cal G}_{2\epsilon+1,1}{},
\\
&{\cal A}^{{\cal D}[2]}_{2}=-{\cal G}_{2\epsilon,0}{}.
\fe
The superconformal block for ${\cal D}[0]$ is given by
\ie
&{\cal A}^{{\cal D}[0]}_{0} = 1.
\fe
A different derivation of the blocks using the superconformal Casimir equations appears in \cite{Bobev:2017jhk}.\footnote{The paper \cite{Bobev:2017jhk} points out typos in \eqref{B2Block} and \eqref{D4Block} in the early versions of this paper.  However, the correct formulae for the blocks were used in the actual bootstrap implementation in all versions.
}

\section{Flavor symmetry}
\label{Sec:Flavor}

We want to consider theories with non-abelian flavor symmetry.  Since flavor currents are contained in the ${\cal D}[2]$ multiplets, the superconformal primaries $\cO_a(x_i,Y_i)$ transform in the adjoint representation of the flavor symmetry group $G_F$, where $a$ is the adjoint index. The four-point function of $\cO_a(x_i,Y_i)$ takes the form
\ie\label{eqn:4pfWithFlavor}
&\vev{\cO_a(x_1,Y_1)\cO_b(x_2,Y_2)\cO_c(x_3,Y_3)\cO_d(x_4,Y_4)}={(Y_1\cdot Y_2)^2(Y_3\cdot Y_4)^2\over x_{12}^{4\epsilon}x_{34}^{4\epsilon}}G_{abcd}(u,v;w),
\fe
and $G_{abcd}(u,v;w)$ admits a decomposition into superconformal blocks as in Section~\ref{sec:SB}.  The operators that appear in the OPE of $\cO_a(x_1,Y_1)$ and $\cO_b(x_2,Y_2)$ transform in the tensor product representation ${\bf adj}\otimes{\bf adj}$, which can further be decomposed into irreducible representations ${\cal R}_i$.  The decomposition of $G_{abcd}(u,v;w)$ takes the form
\ie\label{eqn:flavorG}
& G_{abcd}(u,v;w) = \sum_{{\cal R}_i\in {\bf adj}\otimes{\bf adj}}P^{abcd}_iG_{i}(u,v;w),
\\
& G_{i}(u,v;w) =\sum_{\mathcal X}\lambda^2_{{\mathcal X},i} {\cal A}^{\cal X}(u,v;w),
\fe
where $P^{abcd}_i$ is the projection matrix that projects onto the contributions of operators in the OPE that transform in the representation ${\cal R}_i$. They satisfy \cite{cvitanovic2008group}
\ie
P^{abcd}_iP^{dcef}_j=\delta_{ij}P^{abef}_i,\quad P^{abba}_i={\rm dim}({\cal R}_i).
\fe
The projection matrices of the trivial representation and the adjoint representation are
\ie
P_{\bf 1}^{abcd}={1\over {\rm dim}(G_F)}\delta^{ab}\delta^{cd},\quad P_{\bf adj}^{abcd}={1\over \psi^2 h^{\vee}}f^{abe}f^{ecd},
\fe
where $h^{\vee}$ is the dual Coxeter number and $\psi^2=2$ is the length squared of the longest root of the flavor group.

The identity operator and the stress tensor multiplet ${\cal B}[0]_0$ can only transform in the trivial representation $\bf 1$ of the flavor group, while the flavor current multiplet ${\cal D}[2]$ can only be in the adjoint representation $\bf adj$.  Their OPE coefficients satisfy
\ie \label{FlavorNorm}
\lambda^2_{{\cal D}[0],i}={\rm dim}(G_F)\delta_{i,{\bf 1}},\quad \lambda^2_{{\cal B}[0]_0,i}=\lambda^2_{{\cal B}[0]}{\rm dim}(G_F)\delta_{i,{\bf 1}},\quad \lambda^2_{{\cal D}[2],i}=\lambda^2_{{\cal D}[2]}\delta_{i,{\bf adj}}.
\fe
In Section~\ref{Sec:CentralCharges}, we will relate the coefficients $\lambda^2_{{\cal B}[0]_0}$ and $\lambda^2_{{\cal D}[2]}$ to the central charge $C_T$ and flavor central charge $C_J$, which are in turn related to the anomaly coefficients and can be determined through other methods.

Because all four external scalars are identical, the four-point function \eqref{eqn:4pfWithFlavor} is invariant under $(x_1,Y_1,a)\leftrightarrow (x_3,Y_3,c)$, leading to the crossing symmetry constraint
\ie\label{crossingF13}
F_{i}{}^jG_j(u,v;w)={u^{2\epsilon} \over v^{2\epsilon} w^2}G_i(v,u;w^{-1}),
\fe
where the crossing matrix $F_i{}^j$ is defined as
\ie
F_i{}^j = {1\over {\rm dim}({\cal R}_i)}P_i^{dabc}P_j^{abcd}.
\fe
Similarly, the invariance under $(x_1,Y_1,a)\leftrightarrow (x_2,Y_2,b)$ leads to the constraint
\ie\label{crossingF12}
G_i(u,v;w)=(-1)^{|{\cal R}_i|}G_i\left({u\over v},{1\over v};-{w\over w+1}\right),
\fe
where $|{\cal R}_i| = 0$ for ${\cal R}_i$ appearing in the symmetric tensor product of two adjoint representations, and $|{\cal R}_i| = 1$ for ${\cal R}_i$ appearing in the anti-symmetric tensor product. The constraint \eqref{crossingF12} amounts to imposing the selection rule $\ell+J_R+|{\cal R}_i| \in 2\bZ$ on the intermediate primary operators.

We will be interested in the SU(2) and $E_8$ flavor groups.  The $\text{\bf adj} \otimes \text{\bf adj}$ decompositions and crossing matrices are summarized in Table~\ref{Tab:F}.\footnote{We compute the crossing matrices following the methods explained in \cite{cvitanovic2008group}.}
\begin{table}[H]
\centering
\begin{tabular}{|c|c|c|c|c|}
\hline
$G_F$ & $h^{\vee}$ & {\bf adj} $\otimes_S$ {\bf adj} & {\bf adj} $\otimes_A$ {\bf adj} & $F$
\\\hline\hline
SU(2) & 2 & {\bf 1} + {\bf 5} & {\bf 3} & $\begin{pmatrix}
 \frac{1}{3} & \frac{5}{3} & 1 \\
 \frac{1}{3} & \frac{1}{6} & -\frac{1}{2} \\
 \frac{1}{3} & -\frac{5}{6} & \frac{1}{2} \\
\end{pmatrix}$
\\
$E_8$ & 30 & {\bf 1} + {\bf 3875} + {\bf 27000} & {\bf 248} + {\bf 30380} & $\begin{pmatrix}
 \frac{1}{248} & \frac{125}{8} & \frac{3375}{31} & 1 & \frac{245}{2} \\
 \frac{1}{248} & -\frac{3}{8} & \frac{27}{31} & \frac{1}{5} & -\frac{7}{10} \\
 \frac{1}{248} & \frac{1}{8} & \frac{23}{62} & -\frac{1}{30} & -\frac{7}{15} \\
 \frac{1}{248} & \frac{25}{8} & -\frac{225}{62} & \frac{1}{2} & 0 \\
 \frac{1}{248} & -\frac{5}{56} & -\frac{90}{217} & 0 & \frac{1}{2} 
\end{pmatrix}$
\\\hline
\end{tabular}
\caption{The decomposition of {\bf adj} $\otimes$ {\bf adj} into irreducible representations and the crossing matrices for SU(2) and $E_8$ flavor groups. The basis of representations in the crossing matrix are in the order shown in columns {\bf adj} $\otimes_S$ {\bf adj} and then {\bf adj} $\otimes_A$ {\bf adj}.}
\label{Tab:F}
\end{table}

\section{Central charges}
\label{Sec:CentralCharges}

In this section, we review the definitions of the central charge $C_T$ and the flavor central charge $C_J$, and derive their relations to the OPE coefficients $\lambda_{{\cal B}[0]_{0}}^2$, $\lambda_{{\cal D}[2]}^2$.

\subsection{Central charge $C_T$}
\label{Sec:CT}

Conformal symmetry fixes the two-point function of the stress tensor up to an overall coefficient.  Since the stress tensor has a canonical normalization, this coefficient is physical and is referred to in the literature is as the central charge $C_T$.  More precisely\cite{Osborn:1993cr},
\ie\label{eqn:TTtpf}
&\vev{T_{\m\n}(x)T_{\sigma\rho}(0)} = {C_T\over V_{\widehat {\rm S}^{d-1}}^2}{{\cal I}_{\m\n,\sigma\rho}(x)\over x^{2d}} ,
\fe
where $V_{\widehat {\rm S}^{d-1}}=2\pi^{d\over 2}/ \Gamma\left(d\over 2\right)$ is the volume of a unit $(d-1)$-sphere, and the conformal structure ${\cal I}_{\m\n,\sigma\rho}(x)$ is given by
\ie\label{eqn:structureI}
&{\cal I}_{\m\n,\sigma\rho}(x)={1\over 2}\left[I_{\mu\sigma}(x)I_{\nu\rho}(x)+I_{\mu\rho}(x)I_{\nu\sigma}(x)\right]-{1\over d}\delta_{\m\n}\delta_{\sigma\rho},
\\
&I_{\mu\nu}(x)=\delta_{\mu\nu}-2{x_\mu x_\nu\over x^2}.
\fe
In Appendix~\ref{App:CT2OPE}, we review how the contribution of the stress tensor multiplet to the four-point function of identical scalars is fully determined by the value of $C_T$.  Assuming that there is a unique flavor-singlet stress tensor multiplet ${\cal B}[0]_{0}$, the relation between the OPE coefficient $\lambda_{{\cal B}[0]_{0}}$ and the central charge $C_T$ is
\ie\label{eqn:lambdaB0toCT}
\lambda_{{\cal B}[0]_{0}}^2 = {4(2\epsilon+2)(2\epsilon+3)\over 2\epsilon+1}{1\over C_T}.
\fe

To later compare with numerical bounds, we present here the values of $C_T$ for six-dimensional superconformal field theories of interest, by relating $C_T$ to a Weyl anomaly coefficient.  The Weyl anomaly in six-dimensional conformal field theories takes the form\cite{Bonora:1985cq,Deser:1993yx,Bastianelli:2000hi}
\ie\label{eqn:weylAnomaly}
{\cal A}_{6d}=(4\pi)^3\vev{T^\m_\m}= -aE_6+c_1 I_1 + c_2 I_2 + c_3 I_3 +{\text{scheme dependent}},
\fe
where $E_6$ is the Euler density and $I_{1,2,3}$ are certain Weyl invariants.  $I_3$ is normalized as $I_3 = C_{\m\n\sigma\rho} \nabla^2 C^{\m\n\sigma\rho} + \dotsb$, $C_{\m\n\sigma\rho}$ being the Weyl tensor (see \cite{Bastianelli:2000hi} for the precise definition of $I_3$). The $a$-coefficient appears in the stress tensor four-point function, $c_1$ and $c_2$ in the stress tensor three-point function, and $c_3$ in the stress tensor two-point function.  The relation between $c_3$ and $C_T$ is
\ie
C_T = 3024 c_3.
\fe

In theories with supersymmetry, the Weyl anomaly coefficients are linearly related to the 't Hooft anomaly coefficients\cite{Cordova:2015fha,Beccaria:2015uta,Beccaria:2015ypa}, which appear in the anomaly polynomial involving gravitational and R-symmetry anomalies (see \cite{Beccaria:2015ypa} for precise definitions and normalizations)
\ie\label{eqn:tHooftPoly}
{\cal I}_8={1\over 4!}\left(\A c_2(R)^2+\beta c_2(R)p_1(T)+\gamma p_1(T)^2+\delta p_2(T)\right).
\fe
In \cite{Beccaria:2015ypa}, the authors proposed that the coefficients appearing in the linear relations can be fixed by computing the values of $\A$, $\B$, $\C$, $\D$ and $a$, $c_1$, $c_2$, $c_3$ in free theories, {\it e.g.}, the free hypermultiplet, the free tensor multiplet, and a class of non-unitary free theories.  The relation between $c_3$ and $\A$, $\B$, $\C$, $\D$ was determined up to an unfixed parameter $\xi$,
\ie
\label{c3Relation}
&c_3 = {1\over 9}\A-{1\over 14}(5+3\xi)\B+{1\over 21}(2-3\xi)\C-{1\over 12}(2+3\xi)\D.
\fe
The value of $\xi$ can be further fixed by considering a superconformal vector multiplet $V^{(1,0)}$, which has the same field content as the flavor current multiplet, but whose component fields have higher-derivative kinetic terms.  More explicitly, the multiplet consists of a four-derivative vector, a three-derivative Weyl fermion, and three standard two-derivative scalars.  The anomaly coefficients are\cite{Beccaria:2015ypa}
\ie
(\A, \B, \C, \D) = (-1, -{1\over2}, -{7\over240}, {1\over60}).
\fe
Thus the constant $\xi$ can be determined by
\ie
\xi = {C_T(V^{(1,0)}) \over 324} - {26 \over 45}.
\fe

Since the theory is free, the $C_T$ of $V^{(1,0)}$ is simply the sum of that of its component fields.  The $C_T$ of a free scalar is known from\cite{Osborn:1993cr},
\ie
C_T = {6 \over 5} \quad \text{(standard scalar)},
\fe
and that of a free four-derivative vector was computed in \cite{Giombi:2016fct,Osborn:2016bev} to be
\ie
C_T = -90 \quad \text{(four-derivative vector)}.
\fe 
In \cite{Beccaria:2017dmw}, the authors computed the $C_T$ for a three-derivative Weyl fermion by studying the partition function on S$^1\times{\mathbb H}^{5}$, and found
\ie
C_T = - {72 \over 5} \quad \text{(three-derivative Weyl fermion)}.
\fe
In Appendix~\ref{App:Vec}, we verify this answer by explicitly constructing the stress tensor for the three-derivative fermion and computing its two-point function.  Thus
\ie
C_T(V^{(1,0)}) = 3 \times {6 \over 5} - {72 \over 5} - 90 = - {504 \over 5},
\fe
and they concluded that
\ie
\xi = - {8 \over 9},
\fe
which corroborates with what was first found in \cite{Yankielowicz:2017xkf} via a different method.\footnote{We thank Matteo Beccaria, Arkady A. Tseytlin, and Yang Zhou for sharing this result before publication.}  In \cite{Beccaria:2017dmw}, the conformal anomaly coefficients for an infinite family of free, non-unitary, higher-derivative ${\cal N} = (1,0)$ superconformal multiplets were also computed, and indeed found to satisfy the linear relation \eqref{c3Relation} with this value of $\xi$.

There are various techniques for inferring the values of 't Hooft anomaly coefficients in superconformal field theories, even when the theory is strongly interacting and direct handles are lacking.  For instance, if a construction within string theory or M theory exists, the 't Hooft anomaly coefficients can be computed by anomaly inflow\cite{Harvey:1998bx,Ohmori:2014pca}.  Another approach is anomaly matching by going onto the tensor branch or the Higgs branch \cite{Intriligator:2000eq,Ohmori:2014kda,Intriligator:2014eaa,Shimizu:2017kzs}.

In the following, we present the values of $C_T$ for the free hypermultiplet and the E-string theories.

\paragraph{Free hypermultiplet} The $C_T$ for each free scalar $\phi$ and each free Dirac spinor $\psi$ are\cite{Osborn:1993cr}
\ie\label{eqn:CTfree}
C^{\phi}_T={(2\epsilon+2)\over (2\epsilon+1)},\quad C^{\psi}_T=2^{\lfloor\epsilon\rfloor}(2\epsilon+2).
\fe
Thus the $C_T$ for a free hypermultiplet is
\ie
\label{FreeHyperCT}
C^{\text{\tiny hyper}}_T = 
4C^{\phi}_T+{1\over 2}C^{\psi}_T={84\over 5}.
\fe

\paragraph{E-string theories} 

The rank-$N$ E-string theory is realized by stacking $N$ M5 branes inside an end-of-the-world M9 brane \cite{Ganor:1996mu,Seiberg:1996vs}.  The flavor symmetry is $E_8$ for rank-one and $E_8 \times {\rm SU(2)}$ for higher ranks.  The 't Hooft anomaly coefficients and the conformal anomaly coefficient $c_3$ are given by (including the free hypermultiplet describing the center-of-mass degrees of freedom parallel to the M9 brane)
\ie
&\A=N(4N^2+6N+3),\quad \B=-{N\over 2}(6N+5),\quad \C={7N\over 8},\quad \delta=-{N\over 2},
\\
&c_3 = {4\over 9}N^3+{7\over 6}N^2+{11\over 12}N,\quad C_T=84N(16N^2+42N+33).
\fe
The minimal central charge is achieved in the $N=1$ case, which after decoupling the free hypermultiplet is
\ie
C_T = 7644  - {84 \over 5 } ={ 38136\over 5}.
\fe

\subsection{Flavor central charge $C_J$}

We can perform a similar analysis for the flavor currents $J^a_\mu$, which are canonically normalized in the following way.  In radial quantization, the non-abelian charge of a state on the cylinder which corresponds to an operator inserted at the origin $x^\m=0$ is measured by
\ie
\label{QfromJ}
Q^a=\int_{{\rm S}^{d-1}} J^a_{\m}(x)\hat r^\m dS,
\fe
where $\hat r^\m=x^\m/|x|$ is the radial unit vector, and the integral is over an ${\rm S}^{d-1}$ surrounding the origin.  If we consider a state $\ket{J_\m^b}$ that corresponds to the current $J_\m^b$, then the non-abelian charge of this state is given by the structure constants,
\ie
Q^a\ket{J_\m^b}=if^{ab}{}_c\ket{J_\m^c}.
\fe
We can normalize the structure constants by 
\ie
{1 \over \psi^2 h^\vee} f^{ab}{}_c f^{dc}{}_b = \delta^{ad},
\fe
where $h^{\vee}$ is the dual Coxeter number and $\psi^2 = 2$ is the length squared of the longest root of the flavor group.  This then endows the currents with a normalization.

Conformal symmetry constrains the two point function of the flavor currents $J^a_\mu$ up to an overall coefficient, which is called the flavor central charge $C_J$\cite{Osborn:1993cr},
\ie\label{eqn:JJTPF}
\vev{J^a_\mu(x)J^b_\nu(0)}={C_J\over V_{\widehat  {\rm S}^{d-1}}^2}{\delta_{ab}I_{\mu\nu}(x)\over x^{2(d-1)}}.
\fe
The contribution of the flavor current multiplet to the four-point function of identical scalars is fully determined by the value of $C_J$.  In Appendix~\ref{App:CJ2OPE}, we derive the relation between the OPE coefficient $\lambda_{{\cal D}[2]}$ and the central charge $C_J$,
\ie\label{eqn:lambdaD2toCJ}
\lambda^2_{{\cal D}[2]}
= {2(2\epsilon+1)\over 2\epsilon} {\psi^2 h^\vee\over C_J}.
\fe

Similar to the central charge $C_T$, the flavor central charge $C_J$ can be linearly related to 't Hooft anomaly coefficients\cite{Dumitrescu}. We list the values of $C_J$ for the theories of interest.

\paragraph{Free hypermultiplet}   The flavor central charge of a single free hypermultiplet can be determined by \eqref{eqn:lambdaD2toCJ} and \eqref{HyperOPE}, giving
\ie
 C_J = {5\over 2}.
\fe

\paragraph{E-string theories}  The $C_J$ of the $E_8$ flavor group of E-string theories is
\ie
C_J = 60N^2 + 90N.
\fe
For rank one, $C_J = 150$.

\section{Semidefinite programming}
\label{Sec:Semi}

We proceed by employing the linear functional method \cite{Rattazzi:2008pe} to exploit the crossing symmetry constraint \eqref{crossing13} (setting $\epsilon = k = 2$), as well as the non-negativity of the coefficients in the superconformal block expansion \eqref{eqn:sBlockExpansion}, where ${\cal X}$ is summed over the multiplets \eqref{eqn:(1,0)selectionRole} allowed by selection rules.  To keep the discussion simple, we only display formulae for $U(1)$ flavor symmetry.  Also recall from that $G(u, v; w)$ has an expansion in $w^{-1}$ as shown in \eqref{4pf}.  Putting these together, we have
\ie
G(u, v; w) &= \left( u^2 \over v^2 w \right)^2 G(v, u; w^{-1}),
\\
G(u, v; w)&= \sum_{\mathcal X} \lambda^2_{\mathcal X} {\cal A}^{\cal X}(u,v;w) ,
\\
G(u, v; w)&= G_0(u, v) + G_1(u, v) w^{-1} + G_2(u, v) w^{-2},
\fe
where each superconformal block ${\cal A}^{\cal X}(u,v;w)$ also has an expansion in $w^{-1}$ that terminates at quadratic order,\footnote{Notice that $\widetilde A^{\cal X}_i(u,v)$ are different from the $ {\cal A}^{\cal X}_i(u,v)$ defined in \eqref{eqn:calAdef}, which are the coefficients in the expansion of superconformal blocks ${\cal A}^{\cal X}(u,v; w)$ in Legendre polynomials rather than in monomials in $w^{-1}$.}
\ie
{\cal A}^{\cal X}(u,v;w) = \widetilde A^{\cal X}_0(u,v) +  \widetilde A^{\cal X}_1(u,v) w^{-1} + \widetilde A^{\cal X}_2(u,v)w^{-2}.
\fe
The precise formulae for these superconformal blocks are detailed in Section~\ref{sec:SB}.

As explained in the final paragraph of Section~\ref{sec:4pt}, the superconformal Ward identities imply that the independent constraints from crossing symmetry are contained in the equation
\ie
v^4 G_2(u, v) = u^4 G_0(v, u).
\fe
Putting things together compactly, the constraints we need to analyze are\footnote{Recall from \eqref{FlavorNorm} that when the flavor group is non-abelian, the normalization is $\lambda^2_{{\cal D}[0],i}={\rm dim}(G_F)\delta_{i,{\bf 1}}$.  
}
\ie
\label{constraints}
& 0 = \sum_{\mathcal X \in {\cal I} \cup \{{\cal D}[0]\}} \lambda^2_{\mathcal X} {\cal K}^{\cal X}(u,v),\quad {\cal K}^{\cal X}(u,v)\equiv v^4 \widetilde A^{\cal X}_2(u,v) - u^4 \widetilde A^{\cal X}_0(v,u),
\\
& \lambda_{{\cal D}[0]}^2 = 1, \quad \lambda_{\cal X}^2 \geq 0 \text{ for } {\cal X} \in {\cal I},
\fe
where ${\cal I}$, the putative spectrum of superconformal multiplets with the identity multiplet excluded, contains a subset of
\ie
{\cal L}[0]_{\Delta,\ell},\quad{\cal B}[2]_{\ell},\quad{\cal B}[0]_{\ell},\quad{\cal D}[4],\quad{\cal D}[2].
\fe
It is a subset because there are further restrictions on the set of ${\cal X}$ over which we sum:
\begin{itemize}
\item  With abelian flavor symmetry, there is a further selection rule that requires $\ell + J_R$ to be even.
\item  With non-abelian flavor symmetry, the selection rule allows symmetric representations in ${\bf adj} \times {\bf adj}$ for $\ell + J_R$ even and anti-symmetric ones for $\ell + J_R$ odd.
\item ${\cal D}[0]$ only appears in the trivial representation of the flavor group.
\item  ${\cal D}[2]$ can only appear in the adjoint representation of the flavor group since these multiplets contain flavor currents (hence ${\cal D}[2]$ are absent for abelian flavor).
\item  In interacting theories with a unique stress tensor, ${\cal B}[0]_0$ only exists in the trivial representation, and ${\cal B}[0]_\ell$ for $\ell > 0$ do not exist since these multiplets contain higher spin conserved currents.\footnote{Later when we mention ``interacting theories'', we always assume that the stress tensor is unique.}\footnote{We thank the JHEP referee for pointing out a mistake in our draft, where we wrongly assumed that even in free theories, ${\cal B}[0]_\ell$ for all $\ell$ can only transform in the trivial or adjoint representation of the flavor group.
}
\end{itemize}

Our goal is to put bounds on the central charges $C_T$ and $C_J$, which are inversely proportional to $\lambda^2_{{\cal B}[0]_{0}}$ and  $\lambda^2_{{\cal D}[2]}$ via \eqref{eqn:lambdaB0toCT} and \eqref{eqn:lambdaD2toCJ}.  We presently explain how to put a universal lower bound on $C_T$, or equivalently an upper bound on $\lambda^2_{{\cal B}[0]_{0}}$, using the linear functional method.  Simple modifications of the following setup allow us to further bound theories to within a finite region in the $C_T^{-1}-C_J^{-1}$ plane.

Consider the space of linear functionals on functions of $u, v$.  Suppose we can find a linear functional $\A$ that satisfies
\ie
\label{FunctionalConstraints}
\A[{\cal K}^{{\cal D}[0]} ]= -1, \quad \A[{\cal K}^{{\cal X}}]\geq 0 \text{ for } {\cal X} \in {\cal I},
\fe
then these constraints combined with the constraints \eqref{constraints} imply an upper bound on $\lambda^2_{{\cal D}[2]}$,
\ie
\label{FunctionalBound}
\lambda^2_{{\cal D}[2]} &= {\lambda^2_{{\cal D}[2]} \over \sum_{{\cal X} \in {\cal I}}  \lambda^2_{\cal X} \A[{\cal K}^{\cal X}]}\leq {1 \over \A[{\cal K}^{{\cal D}[2]}]}.
\fe
The optimal upper bound is obtained by maximizing $\A[{\cal K}^{{\cal D}[2]}]$ within the space of linear functionals satisfying \eqref{FunctionalConstraints}.  The resulting functional is referred to as the {\it extremal functional}, which we denote by $\A_E$\cite{ElShowk:2012hu}.  Thus the linear functional method turns the problem of putting an upper bound on $\lambda^2_{{\cal D}[2]}$ to a problem in semidefinite programming.

Generically, there is a unique four-point function saturating \eqref{FunctionalBound}, called the {\it extremal four-point function}\cite{ElShowk:2012hu,El-Showk:2014dwa}.  This four-point function satisfies
\ie
0 = \sum_{{\cal X} \in {\cal I} \setminus \{{\cal D}[2] \}} \lambda^2_{\cal X} \, \A_E[{\cal K}^{{\cal X}}],
\fe
which, given \eqref{FunctionalConstraints}, means that the long multiplets that can contribute to this extremal four-point function must have $\Delta, \ell$ at which $\A_E[{\cal K}^{{\cal L}[0]_{\Delta,\ell}}]$ vanishes.

In practice, we can only perform the above minimization procedure within a finite-dimensional subspace of linear functionals, with the constraints \eqref{FunctionalConstraints} imposed on a finite number of multiplets.  We achieve the latter by restricting to multiplets with spins no larger than a certain maximum $\ell_{max}$, and estimate how the bound weakens with increasing $\ell_{max}$.  Empirically we find that the amount of weakening is roughly inversely proportional to $\ell_{max}$, and so we can estimate the errors by extrapolations.  This issue is examined further in Appendix~\ref{App:Numerics}.  As for truncating the linear functionals, a convenient subspace is given by the following.  Define variables $z, \bar z$ by
\ie
u = z \bar z,\quad v = (1-z)(1-\bar z),
\fe
such that crossing $u \leftrightarrow v$ amounts to $(z,\bar z) \leftrightarrow (1-z,1-\bar z)$.  Consider the expansion of linear functionals in the basis of taking derivatives with respect to $\partial_z$ and $\partial_{\bar z}$ and evaluating at the crossing symmetric point $z = \bar z = {1\over2}$.  Our subspace is simply the truncation of these derivatives to having total degree no larger than $\Lambda$, namely,
\ie
\A = \sum_{m, n = 0}^\Lambda \A_{m,n} \partial_z^m \partial_{\bar z}^n |_{z = \bar z = {1\over2}}.
\fe

Bosonic conformal blocks and their derivatives evaluated at the crossing symmetric point are computed by utilizing the recursive representation\cite{Penedones:2015aga}, the diagonal limit\cite{ElShowk:2012ht,Hogervorst:2013kva}, and a recursion relation on transverse derivatives\cite{ElShowk:2012ht} that follows from the conformal Casimir equation.  The computations are described in Appendix~\ref{App:Bosonic}.  We use the SDPB package\cite{Simmons-Duffin:2015qma} to perform the semidefinite programming procedure.  Details on the numerical implementations are discussed in Appendix~\ref{App:Numerics}.

\section{Results}
\label{Sec:Bounds}

\subsection{Free hypermultiplet: a check}

In the semidefinite programming approach to constraining superconformal field theories, free theories differ from interacting theories by the presence of multiplets that contain higher spin conserved currents, ${\cal B}[0]_{\ell}$ with $\ell > 0$.  This means that the functional $\A$ acted on these multiplets must also be non-negative, leading to weaker constraints than the interacting case.  \hspace{.05in}

\begin{figure}[h]
\centering
\subfloat{
\includegraphics[width=.47\textwidth]{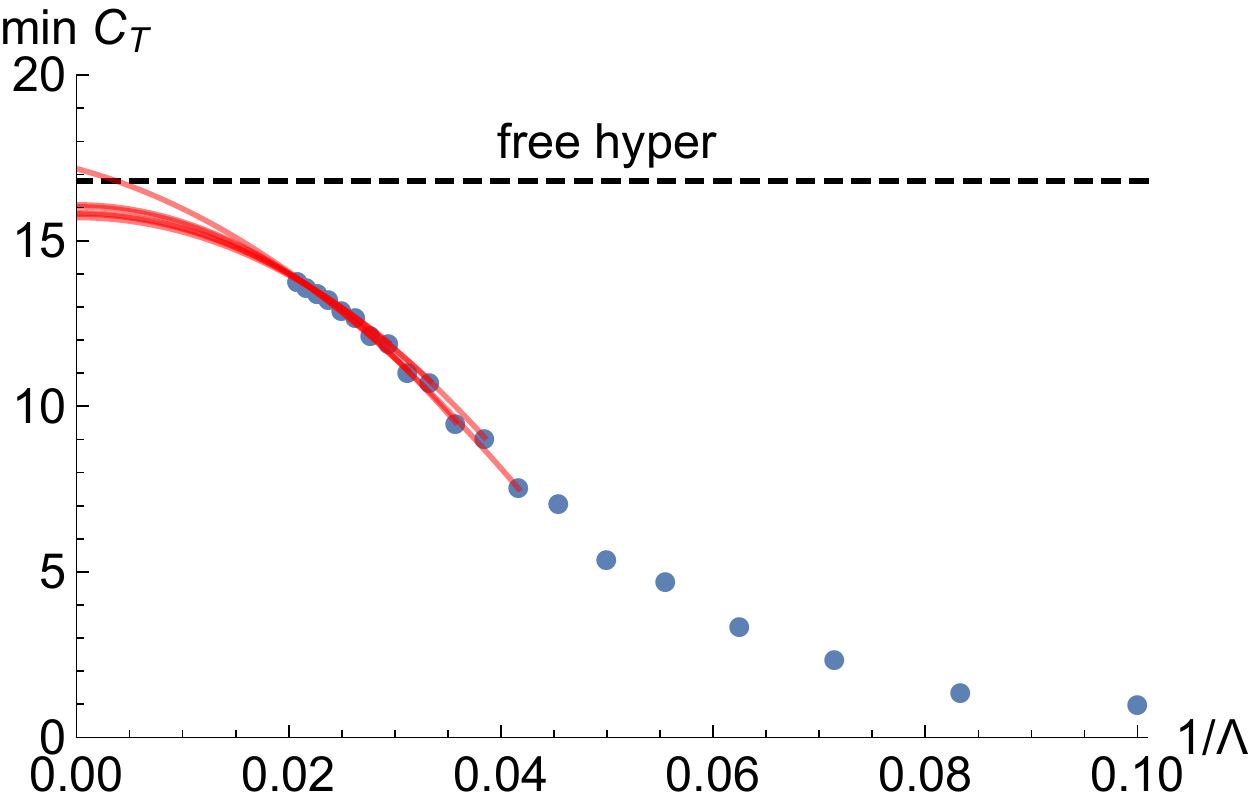}
}
\quad
\subfloat{
\includegraphics[width=.47\textwidth]{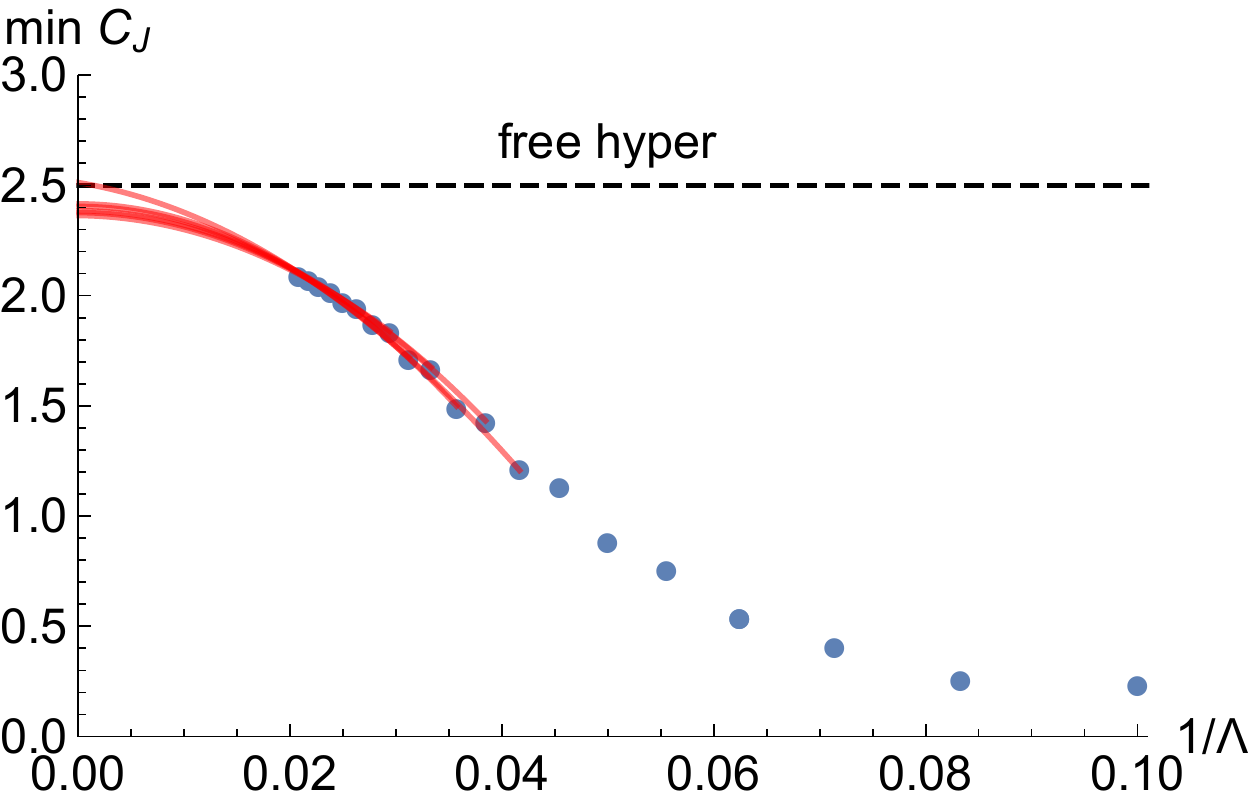}
}
\caption{The lower bounds on $C_T$ and $C_J$ at different derivative orders $\Lambda$, assuming SU(2) flavor group and allowing higher spin conserved currents in the trivial or adjoint representation.  Also shown are the values for a free hypermultiplet, $C_T = {84 \over 5}$ and $C_J = {5\over 2}$.   Also shown are the extrapolations to $\Lambda \to \infty$ using the ansatz \eqref{GapAnsatz}, for $\Lambda \in 4\bZ$ and $\Lambda \in 4\bZ+2$, separately.}
\label{Fig:FreeHyperCTCJ}
\end{figure}

\hspace{.05in}  A single free hypermultiplet has SU(2) flavor symmetry.  In particular, the SO(4) that rotates the four real scalars is the combination of the flavor SU(2) and R-symmetry SU(2)$_R$.  The superconformal primaries of the ${\cal D}[2]$ multiplets are scalar bilinears, and their four-point function can be computed explicitly by Wick contractions.  We refer the reader to Appendix~\ref{App:FreeHyper} for the explicit form of this four-point function and its decomposition into superconformal blocks.  An important property is the absence of ${\cal B}[0]_\ell$ in the {\bf 5} representation, an additional condition that we impose in the bootstrap analysis.  We also note that the long multiplets appearing in the {\bf 1} channel have lowest scaling dimension $\Delta = 8$, and in the {\bf 5} channel have lowest $\Delta = 10$.

\begin{figure}[H]
\centering
\subfloat{
\includegraphics[width=.47\textwidth]{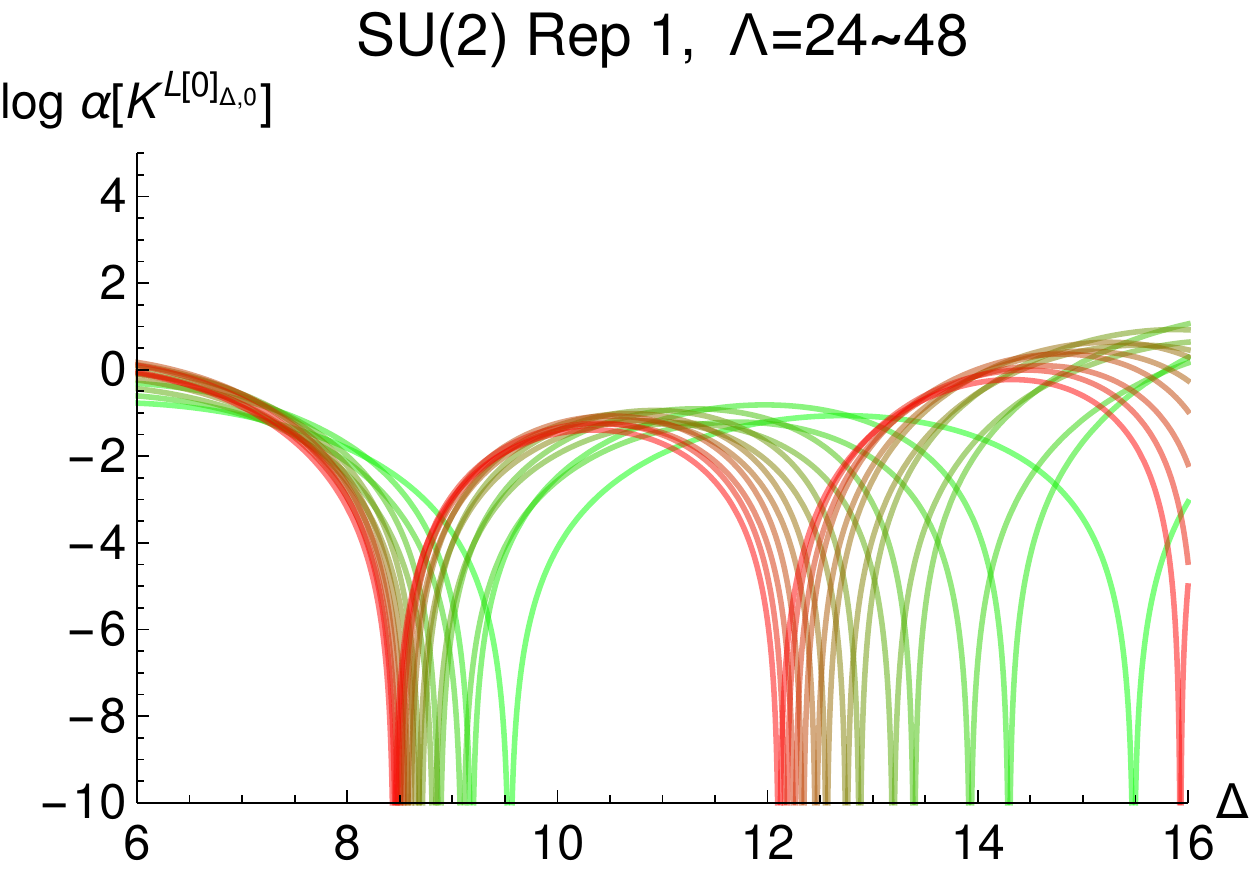}
}
\quad
\subfloat{
\includegraphics[width=.47\textwidth]{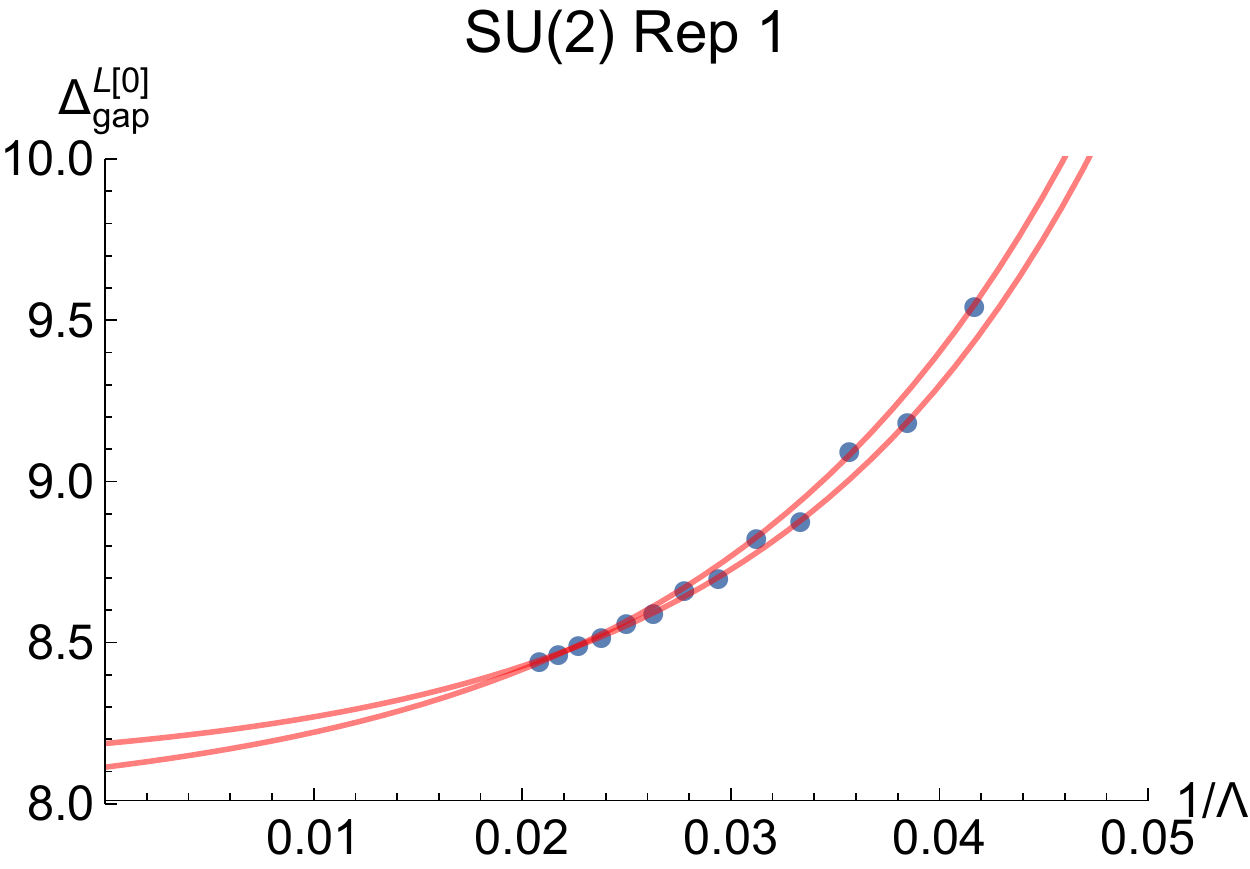}
}
\\
\subfloat{
\includegraphics[width=.47\textwidth]{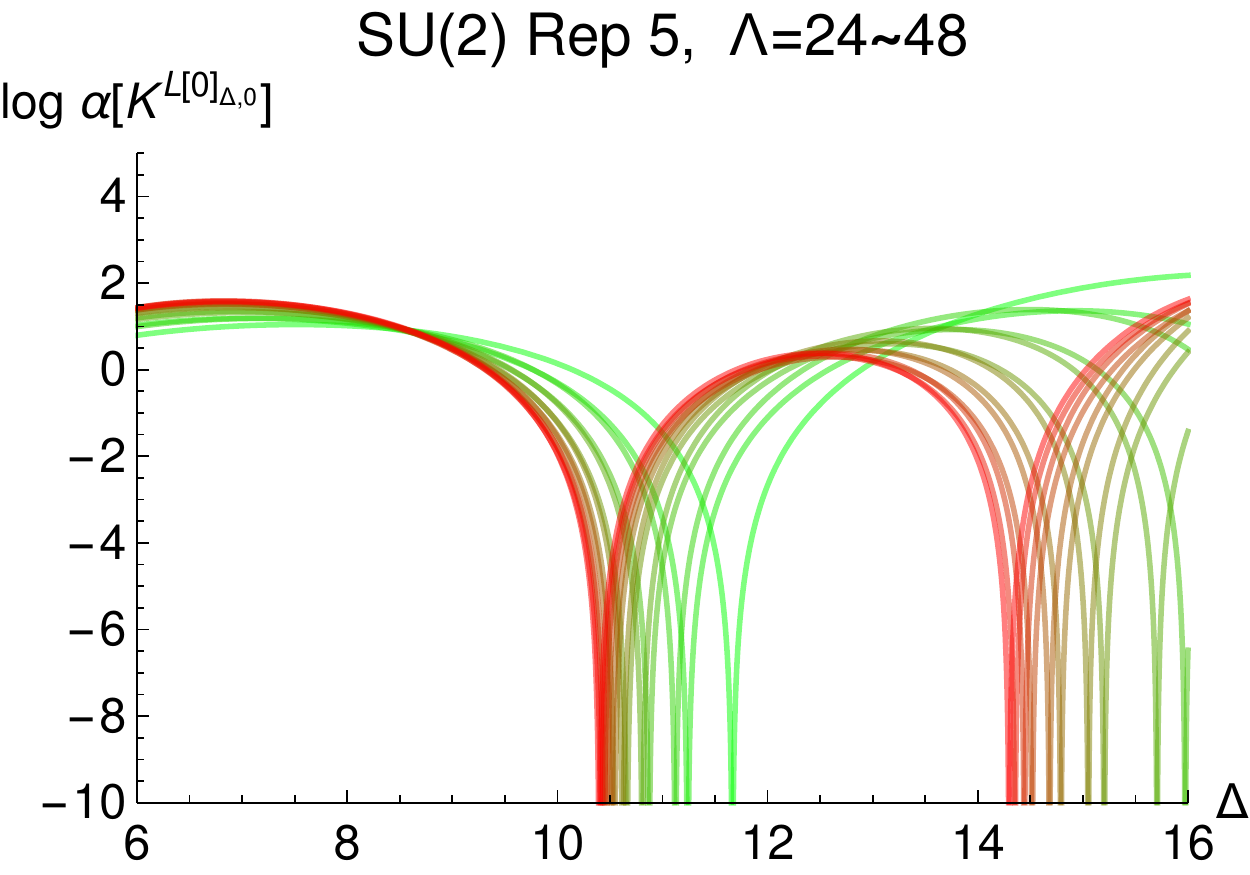}
}
\quad
\subfloat{
\includegraphics[width=.47\textwidth]{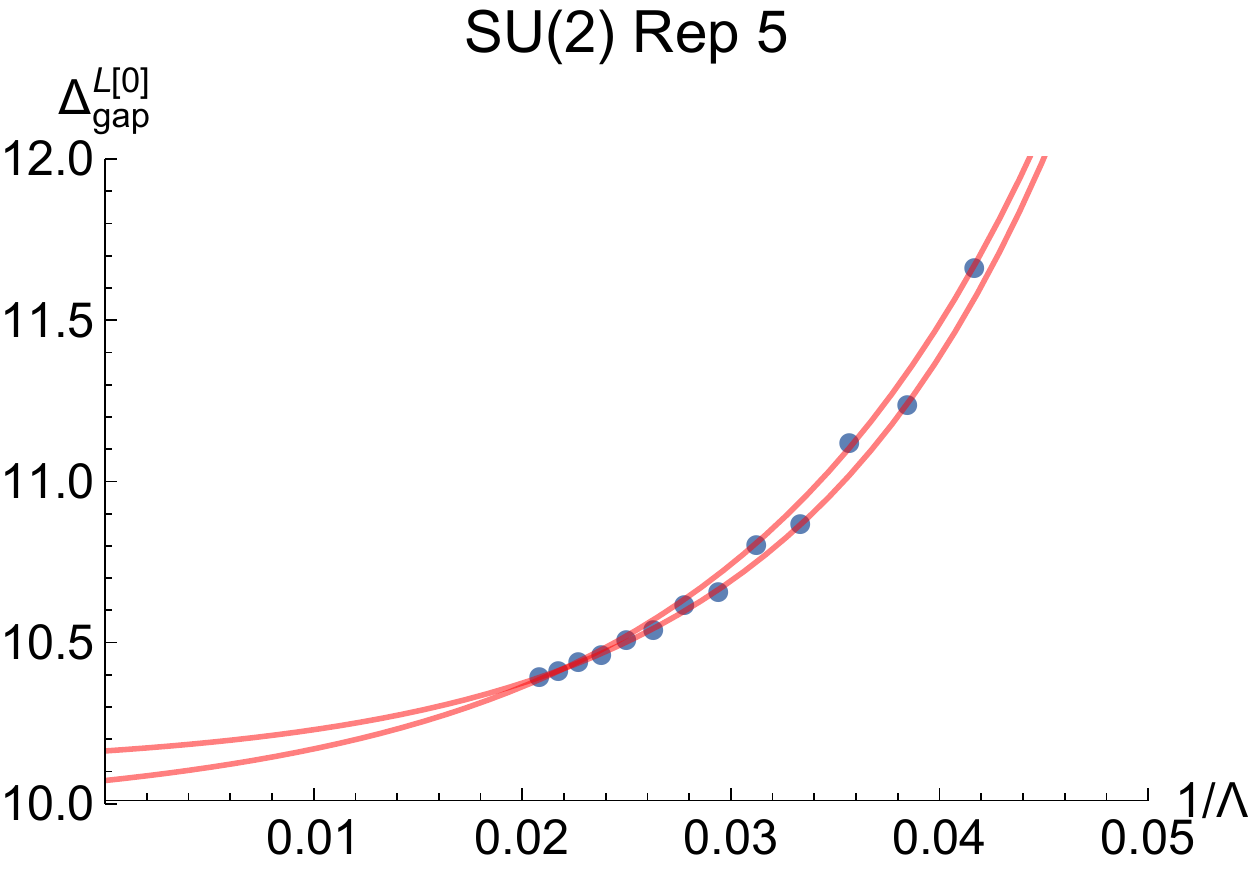}
}
\caption{{\bf Left:}  The extremal functional optimizing the lower bound on $C_J$ acted on the contribution of the spin-zero long multiplet to the crossing equation, $\A_E[{\cal K}^{{\cal L}[0]_{\Delta,0}}]$, in the {\bf 1} and {\bf 5} channels of the SU(2) flavor, plotted in logarithmic scale.  Increasing derivative orders $\Lambda = 24, 26, \dotsc, 48$ are shown from green to red.    
{\bf Right:}  The gap (lowest scaling dimension) in the spectrum of long multiplets in each channel at different $\Lambda$.  Also shown are the extrapolations to $\Lambda \to \infty$ using the ansatz \eqref{ExtrapAnsatz}, for $\Lambda \in 4\bZ$ and $\Lambda \in 4\bZ+2$, separately.
}
\label{Fig:FreeHyperFunctional}
\end{figure}

Assuming SU(2) flavor symmetry and the existence of higher spin conserved currents in the trivial {\bf 1} or adjoint {\bf 3} representation, Figure~\ref{Fig:FreeHyperCTCJ} shows the universal lower bounds on $C_T$ and $C_J$ at various derivative orders $\Lambda$, as well as extrapolations to $\Lambda \to \infty$ using the quadratic ansatz
\ie
\label{ExtrapAnsatz}
{\rm min} \, C_{T/J} = a + {b\over\Lambda} + {c\over\Lambda^2}, \quad b < 0, \quad \Lambda \geq 24, 28, 32.
\fe
We see that both ${\rm min} \, C_T$ and ${\rm min} \, C_J$ tend towards the values for a single free hypermultiplet.  The left side of Figure~\ref{Fig:FreeHyperFunctional} shows the extremal functional 
optimizing the lower bound on $C_J$ 
acted on the contribution of the spin-zero long multiplet to the crossing equation, $\A_E[{\cal K}^{{\cal L}[0]_{\Delta,0}}]$, in the {\bf 1} and {\bf 5} channels of the SU(2) flavor.  We can read off the low-lying spectrum of long multiplets from the zeroes.\footnote{The results are almost identical to those using the extremal functional obtained by minimizing $C_T$, $\A_E^{{\cal B}[0]_{0}}[{\cal K}^{{\cal L}[0]_{\Delta,0}}]$.
}  The right side of Figure~\ref{Fig:FreeHyperFunctional} shows how the lowest $\Delta$ in each channel varies with increasing $\Lambda$ and tends towards $\Delta = 8$ and $\Delta = 10$.  Also shown are extrapolations to infinite $\Lambda$ using the ansatz
\ie
\label{GapAnsatz}
\Delta_{gap}^{{\cal L}[0]} = a + b \exp{c \over \Lambda}, \quad \Lambda \geq 24.
\fe
Due to the oscillatory behavior of the data points, we perform separate extrapolations for $\Lambda \in 4\bZ$ and $\Lambda \in 4\bZ+2$, for both ${\rm min} \, C_{T/J}$ and $\Delta_{gap}^{{\cal L}[0]}$.
These results suggest that a free hypermultiplet saturates the lower bounds on both $C_T$ and $C_J$.

\subsection{E-string theories}

\begin{figure}[H]
\centering
\subfloat{
\includegraphics[width=.47\textwidth]{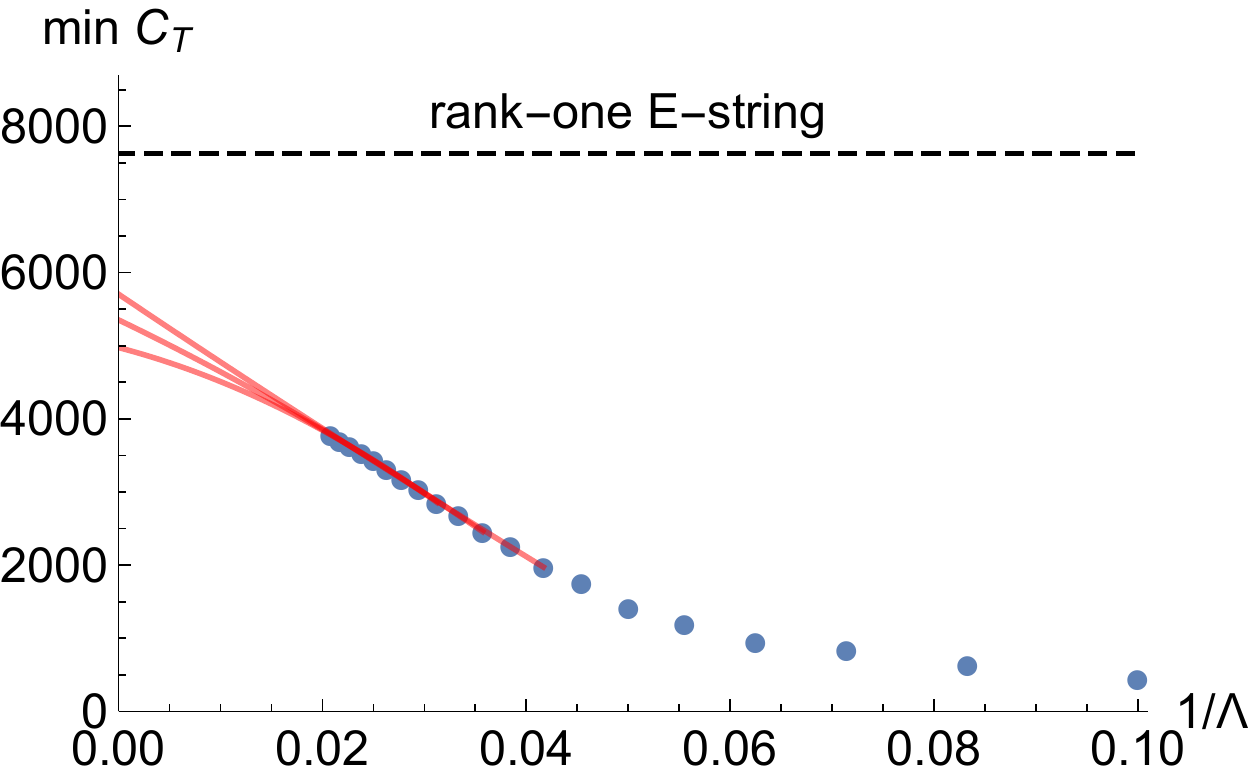}
}
\quad
\subfloat{
\includegraphics[width=.47\textwidth]{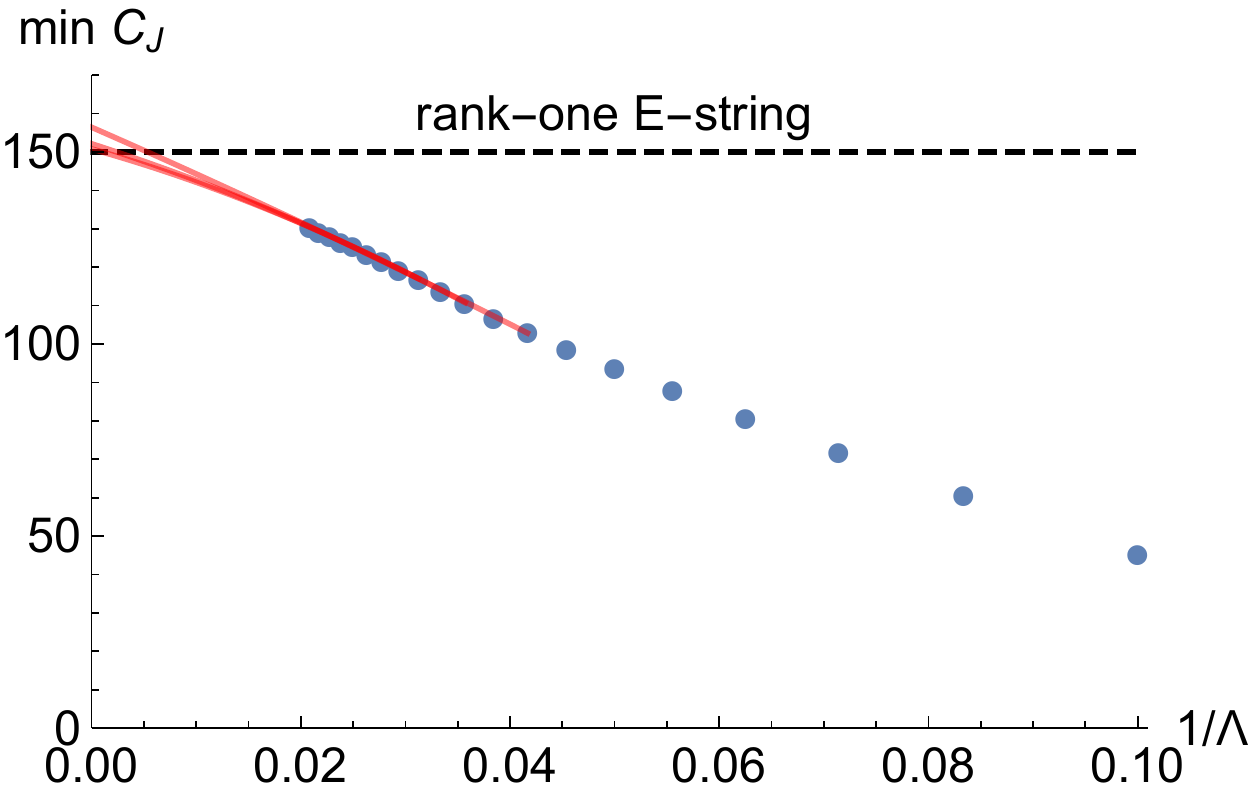}
}
\caption{The lower bounds on $C_T$ and $C_J$ at different derivative orders $\Lambda$ for interacting theories with $E_8$ flavor group.  Also shown are the extrapolations to infinite derivative order using the quadratic ansatz \eqref{ExtrapAnsatz}, as well as the values in the rank-one E-string theory. }
\label{Fig:CTCJ}
\end{figure}

Let us now turn our attention to the E-string theories.  We first present universal lower bounds on $C_T$ and $C_J$ for theories whose flavor group contains $E_8$ as a subgroup.  Figure~\ref{Fig:CTCJ} shows the bounds on $C_T$ and $C_J$ at different derivative orders $\Lambda$, and extrapolations to infinite $\Lambda$ using the quadratic ansatz \eqref{ExtrapAnsatz}.

%\newpage

%

\noindent Table~\ref{Tab:Bounds} summarizes the results of the extrapolations, as well as the $C_T$ and $C_J$ values in the rank-one E-string theory.  Notice that the extrapolated lower bound on $C_J$ sits close to the rank-one E-string value, while that on $C_T$ is still some distance away.  The former observation motivates Conjecture~\ref{Conj:MinCJ} stated in the introduction.

\vspace{.1in}

\begin{table}[H]
\centering
\subfloat{}
\\
\subfloat{
\begin{tabular}{|c|c|c|}
\hline
& min $C_T$ & min $C_J$
\\\hline\hline
$\Lambda = 48$ & $3.78 \times 10^3$ & $1.30 \times 10^2$
\\\hline
Extrapolations & &
\\
$\Lambda \geq 24$ & $5.71 \times 10^3$ & $1.56 \times 10^2$
\\
$\Lambda \geq 28$ & $5.35 \times 10^3$ & $1.52 \times 10^2$
\\
$\Lambda \geq 32$ & $4.98 \times 10^3$ & $1.51 \times 10^2$
\\\hline
Rank-one E-string & ${38136 \over 5} \approx 7.63 \times 10^3$ & 150
\\\hline
\end{tabular}
}
\caption{The lower bounds on $C_T$ and $C_J$ for interacting theories with $E_8$ flavor symmetry.  Presented are the bounds at the highest derivative order computed $(\Lambda=48)$, as well as the extrapolations to infinite $\Lambda$ using the quadratic ansatz \eqref{ExtrapAnsatz}.
}
\label{Tab:Bounds}
\end{table}

%

%\newpage

\begin{figure}[h]
\centering
\subfloat{
\includegraphics[width=.8\textwidth]{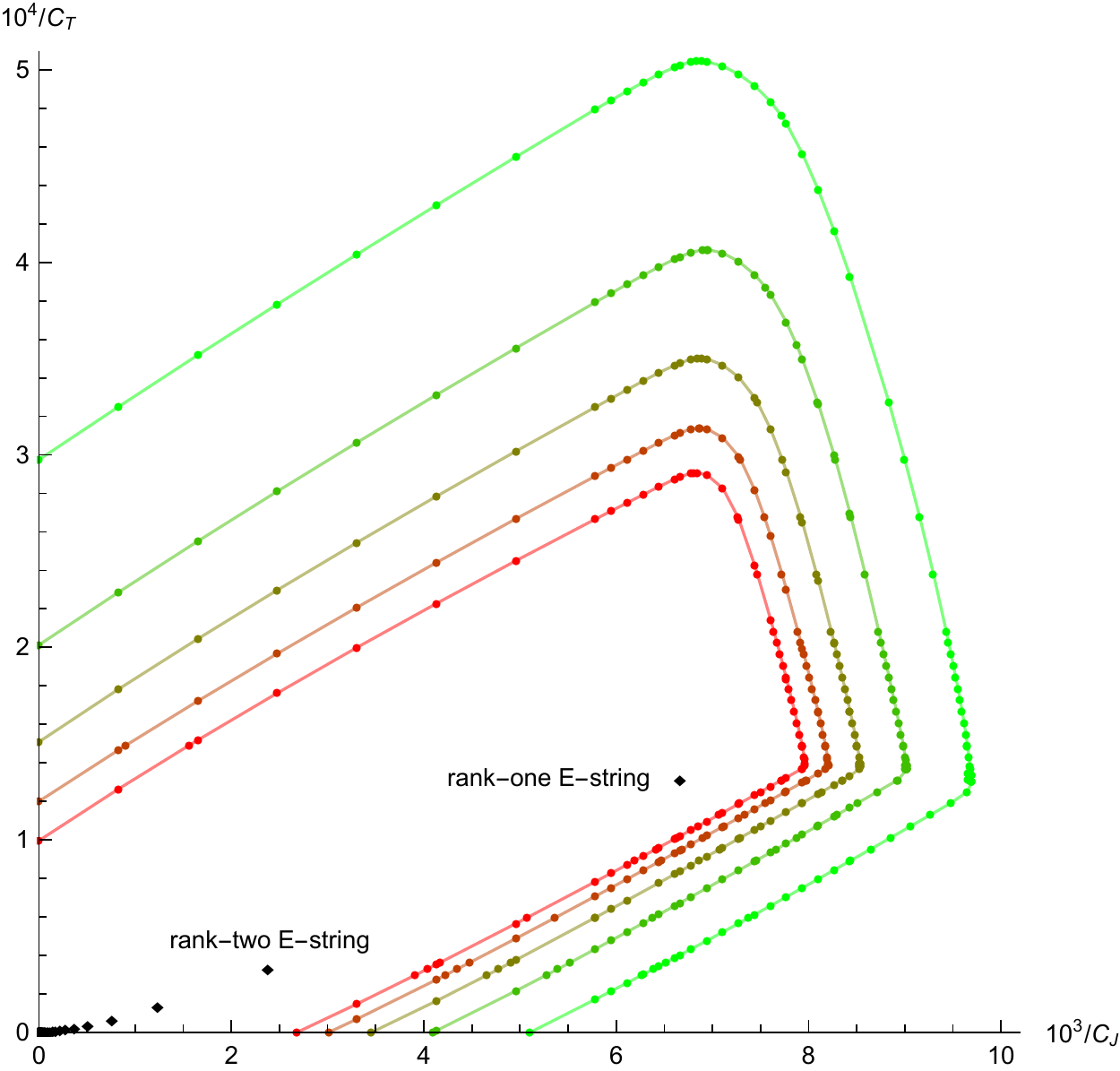}
}
\caption{The allowed region in the $C_T^{-1}-C_J^{-1}$ plane for interacting theories with $E_8$ flavor group, at derivative orders $\Lambda = 24, 28, \dotsc, 40$, shown from green to red.  Also plotted are the points corresponding to the E-string theories.}
\label{Fig:CTCJE8}
\end{figure}

\begin{figure}[h]
\centering
\subfloat{
\includegraphics[width=.47\textwidth]{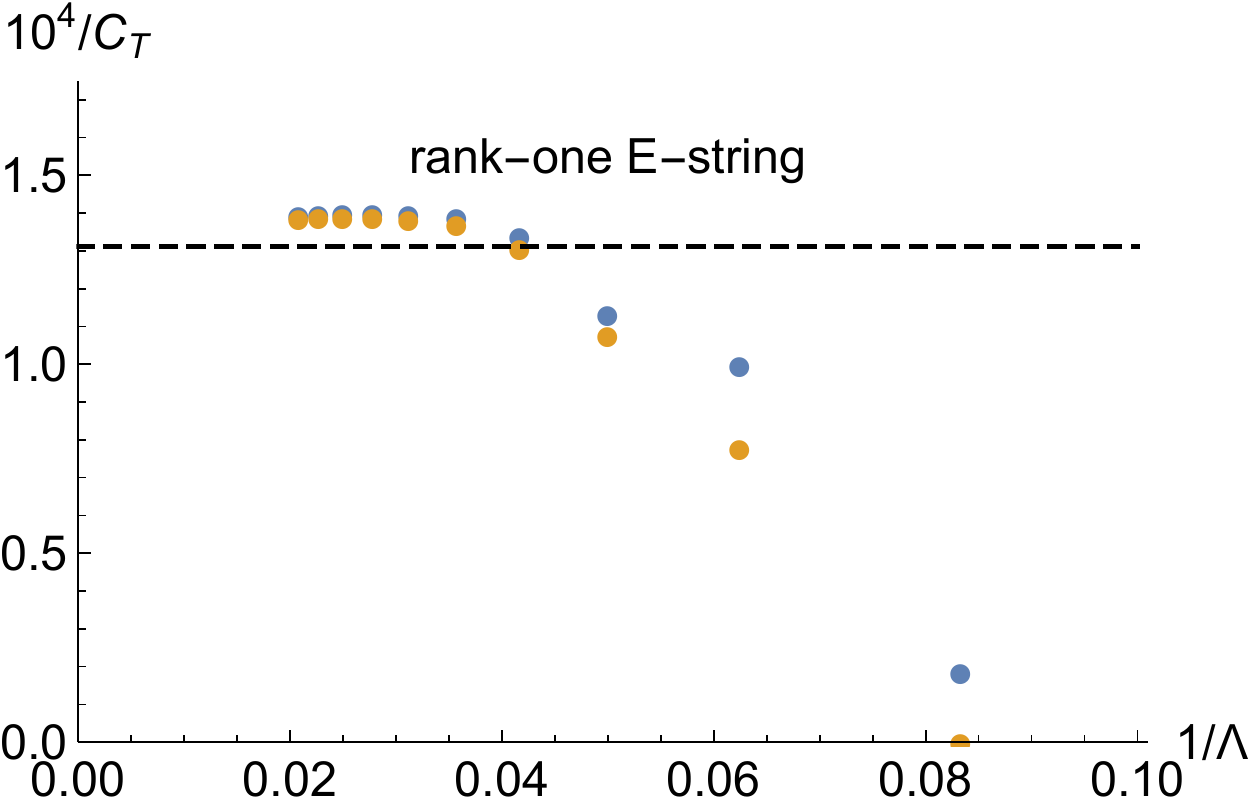}
}
\quad
\subfloat{
\includegraphics[width=.47\textwidth]{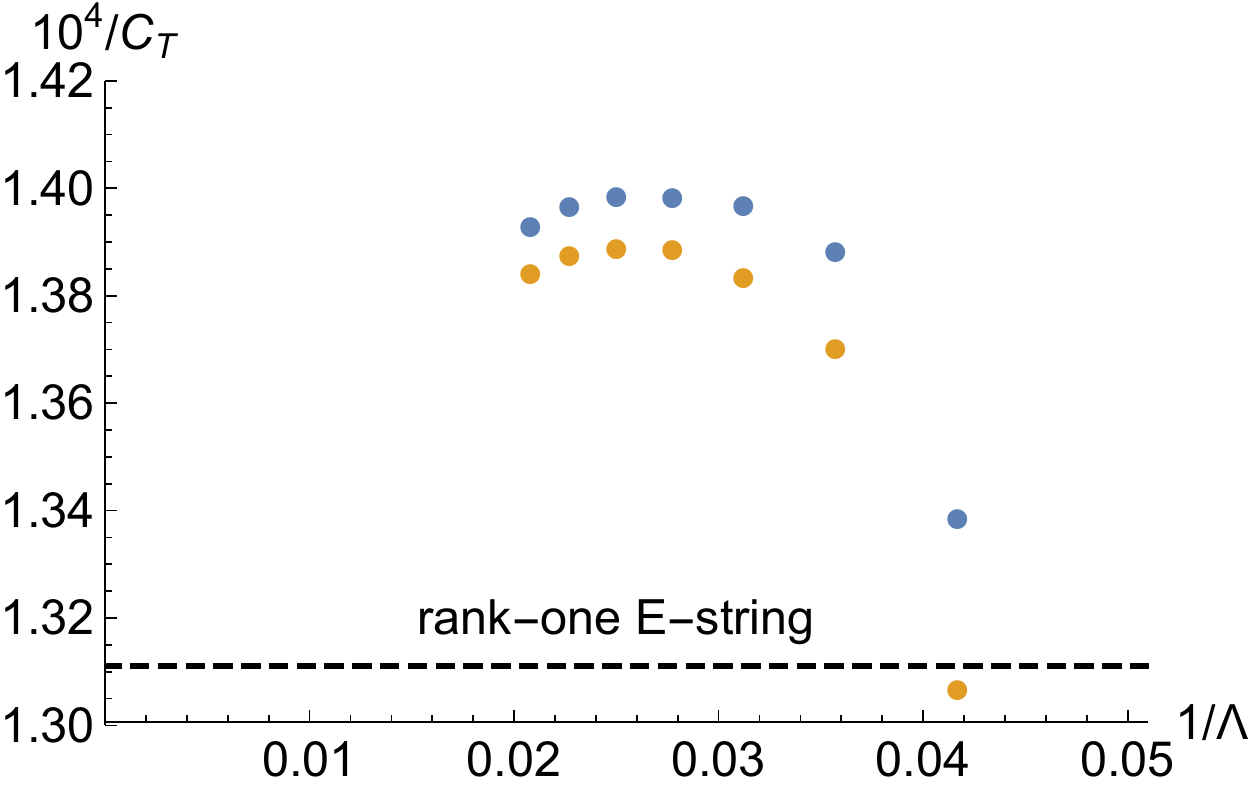}
}
\caption{{\bf Left:} The upper and lowers bounds on the inverse of the central charge $C_T^{-1}$ when the value of the flavor central charge $C_J$ is set close to saturating the lower bound, $C_J = (1+10^{-4}) \, {\rm min}\,C_J$, at different derivative orders $\Lambda$, for interacting theories with $E_8$ flavor group. Also shown is the value for the rank-one E-string theory.
{\bf Right:}  The same plot zoomed in on high derivative orders, showing a trend that the value of $C_T$ at min $C_J$ potential approaches the rank-one E-string value as $\Lambda \to \infty$.
}
\label{Fig:CTMinCJ}
\end{figure}

To supply further evidence for Conjecture~\ref{Conj:MinCJ}, we perform a full survey of the range of allowed $(C_J, C_T)$.  Figure~\ref{Fig:CTCJE8} shows the allowed region in the $C_T^{-1}-C_J^{-1}$ plane for derivative orders $\Lambda = 24, 28, \dotsc, 40$.  Notice that the point of minimal $C_J$ has a value of $C_T$ that sits close to the value of $C_T$ in the rank-one E-string theory.  To quantify this observation more precisely, we show in Figure~\ref{Fig:CTMinCJ} how the value of $C_T$ at min $C_J$ tends to the rank-one E-string value with increasing derivative order.  The value appears to be rather stable between derivative orders 24 and 48, and although it is somewhat smaller than the rank-one E-string value, a closer examination shows a trend of potential convergence to the rank-one E-string at higher derivative orders.\footnote{The deviation of $C_T$ at min $C_J$ from the rank-one E-string value ($\sim 7\%$) is larger than the estimated error due to the truncation on spins ($\lesssim 2\%$).  See Appendix~\ref{App:Numerics}.}

\begin{figure}[h]
\centering
\subfloat{
\includegraphics[width=.47\textwidth]{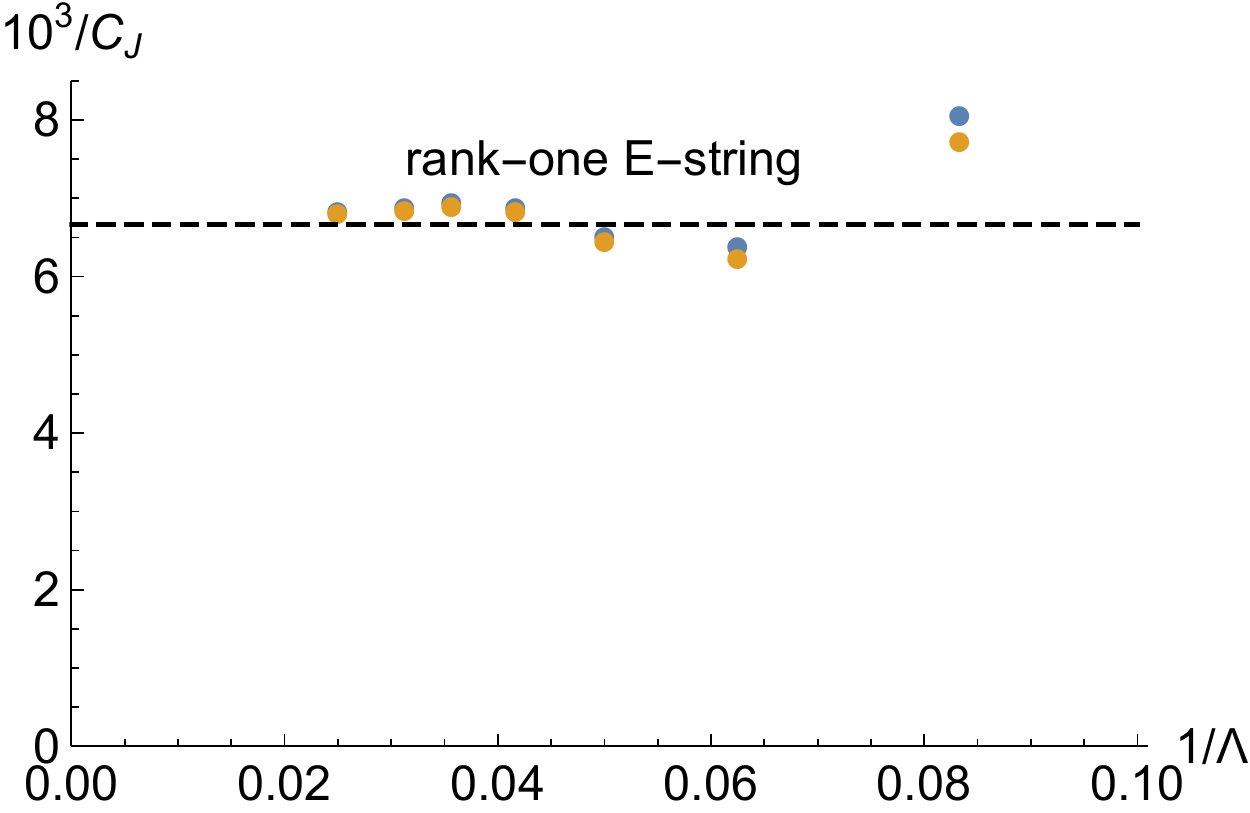}
}
\caption{The upper and lowers bounds on the inverse of the flavor central charge $C_J^{-1}$ when the value of the central charge $C_T$ is set close to saturating the lower bound, $C_T = (1+10^{-4}) \, {\rm min}\,C_T$, at different derivative orders $\Lambda$, for interacting theories with $E_8$ flavor group.  Also shown is the value for the rank-one E-string theory.
}
\label{Fig:CJMinCT}
\end{figure}

While our data do not permit a reliable extrapolation of the entire allowed region to infinite derivative order, we comment on some of the features.  First, given any two unitary solutions to crossing, $G_1(u, v; w)$ and $G_2(u, v; w)$, we can construct a family of unitary solutions $\A G_1(u, v; w) + (1-\A) G_2(u, v; w)$ for $0 \leq \A \leq 1$ that populate the line segment between the two points corresponding to $G_1(u, v; w)$ and $G_2(u, v; w)$ on the $C_T^{-1}-C_J^{-1}$ plane.  This means that the allowed region is convex.\footnote{Unitary solutions to crossing that populate the boundary of the allowed region can be explicitly constructed using the extremal functional method.
}
Second, there seem to be two kinks, one corresponding to the rank-one E-string theory, and another with a $C_J$ value close to that of the rank-one E-string, but with a smaller $C_T$.\footnote{We do not know what to make of the proximity of $C_J$ at min $C_T$ to the rank-one E-string value, as shown in Figure~\ref{Fig:CJMinCT}, or are aware of any candidate theory that sits at this second kink; one logical possibility is that min $C_T$ changes trend at very high derivative orders and becomes saturated by the rank-one E-string theory.
}  A third feature is that the lower boundary appears to approach the locus of points corresponding to the higher rank E-string theories. We discuss the last feature more in Section \ref{Sec:Outlook}.

A further check of Conjecture~\ref{Conj:MinCJ} is the following.  The Higgs branch of the rank-one E-string theory is the one-instanton moduli space of the flavor group $E_8$, which is isomorphic to the minimal nilpotent orbit of $E_8$ \cite{Ganor:1996mu,Gaiotto:2008nz,Beem:2013sza}.  The minimal nilpotent orbit can be defined by quadratic polynomial equations in the complexified $\mathfrak{e}_8$ Lie algebra. More explicitly, for ${\bf r}\in \mathfrak{e}_8$, the defining equation for the minimal nilpotent orbit is
\ie
({\bf r}\otimes{\bf r})\big|_{{\bf 1}\oplus{\bf 3875}}=0.
\fe
The Higgs branch chiral ring is isomorphic to the coordinate ring of the Higgs branch\cite{Seiberg:1994aj,Argyres:1996eh,Gaiotto:2008nz,Argyres:2012fu}.  The latter admits a description as the polynomial ring generated by the $E_8$ moment maps (the superconformal primaries of the ${\cal D}[2]$ multiplets), quotient by the Joseph ideal generated by the superconformal primaries of the ${\cal D}[4]$ multiplets in the representations $\bf 1$ and $\bf 3875$\cite{Gaiotto:2008nz,Beem:2013sza}. In other words, in the rank-one E-string theory, the ${\cal D}[4]$ multiplets in the representations $\bf 1$ and $\bf 3875$ do not appear in the OPE of two ${\cal D}[2]$ multiplets, while those in the $\bf 27000$ do.\footnote{We thank Yifan Wang for explaining this fact to us.}  In accordance with this expectation, Table \ref{Tab:Joseph} shows the extremal functional optimizing the lower bound on $C_J$ acted on the contributions of the ${\cal D}[4]$ multiplets to the crossing equation in each channel.
\begin{table}[H]
\centering
\subfloat{}
\\
\subfloat{
\begin{tabular}{|c|c|}
\hline
Representation & $\A_E[{\cal K}^{{\cal D}[4]}]$
\\\hline\hline
{\bf 1} & $1.1 \times 10^1$
\\
{\bf 3875} & $5.7 \times 10^2$
\\
{\bf 27000} & $1.1 \times 10^{-11}$
\\\hline
\end{tabular}
}
\caption{The extremal functional optimizing the lower bound on $C_J$ for interacting theories with $E_8$ flavor group, at derivative order $\Lambda = 48$, acted on the contributions of the ${\cal D}[4]$ multiplets to the crossing equation in each channel.
}
\label{Tab:Joseph}
\end{table}

\begin{figure}[H]
\centering
\subfloat{
\includegraphics[width=.47\textwidth]{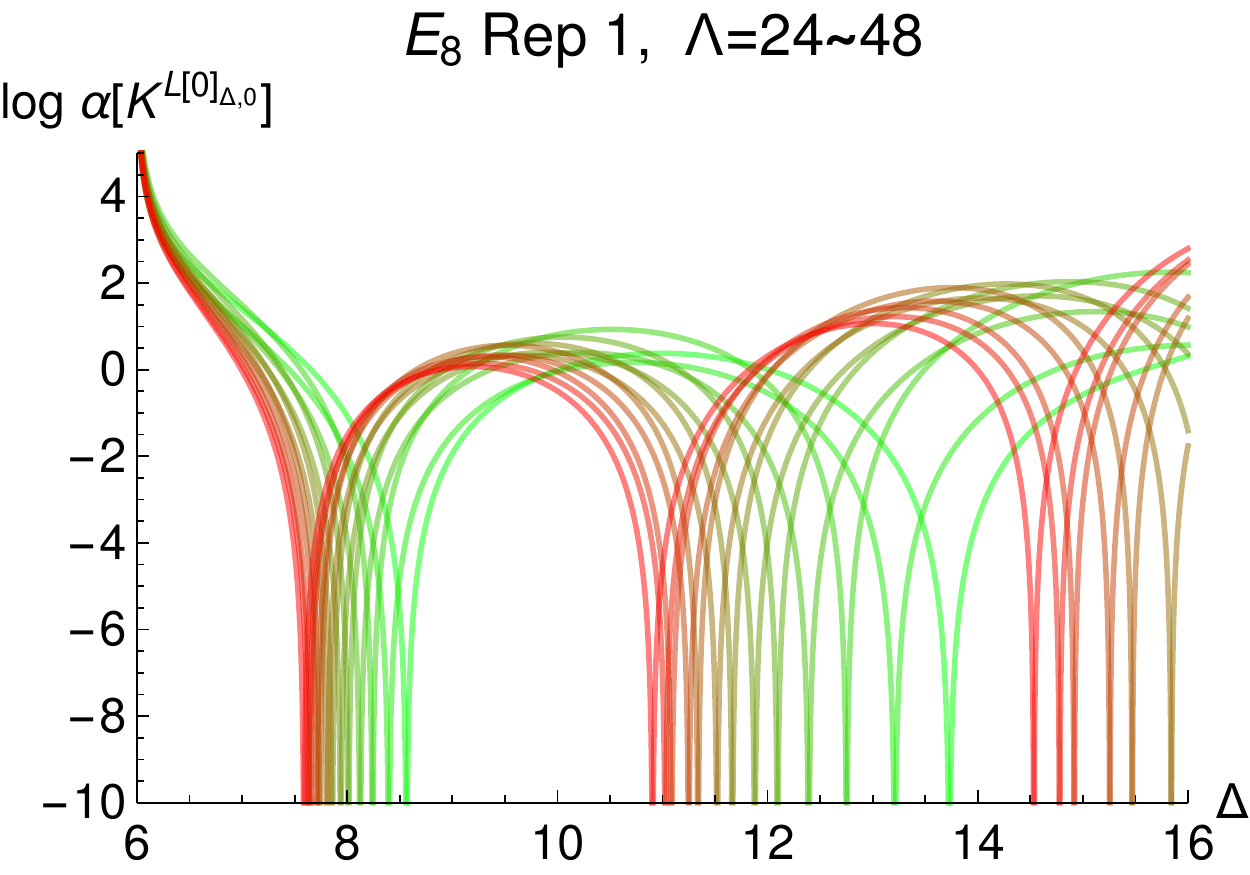}
}
\quad
\subfloat{
\includegraphics[width=.47\textwidth]{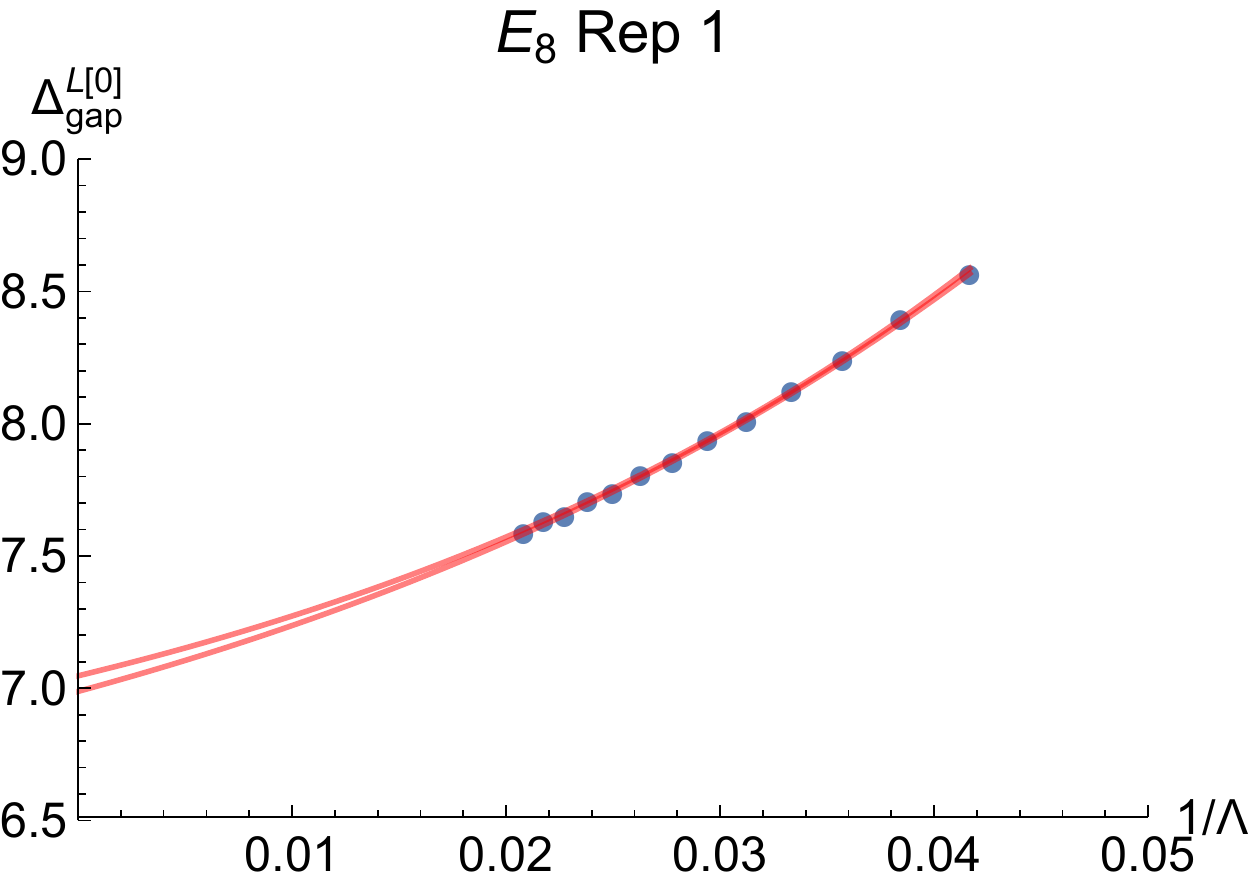}
}
\\
\subfloat{
\includegraphics[width=.47\textwidth]{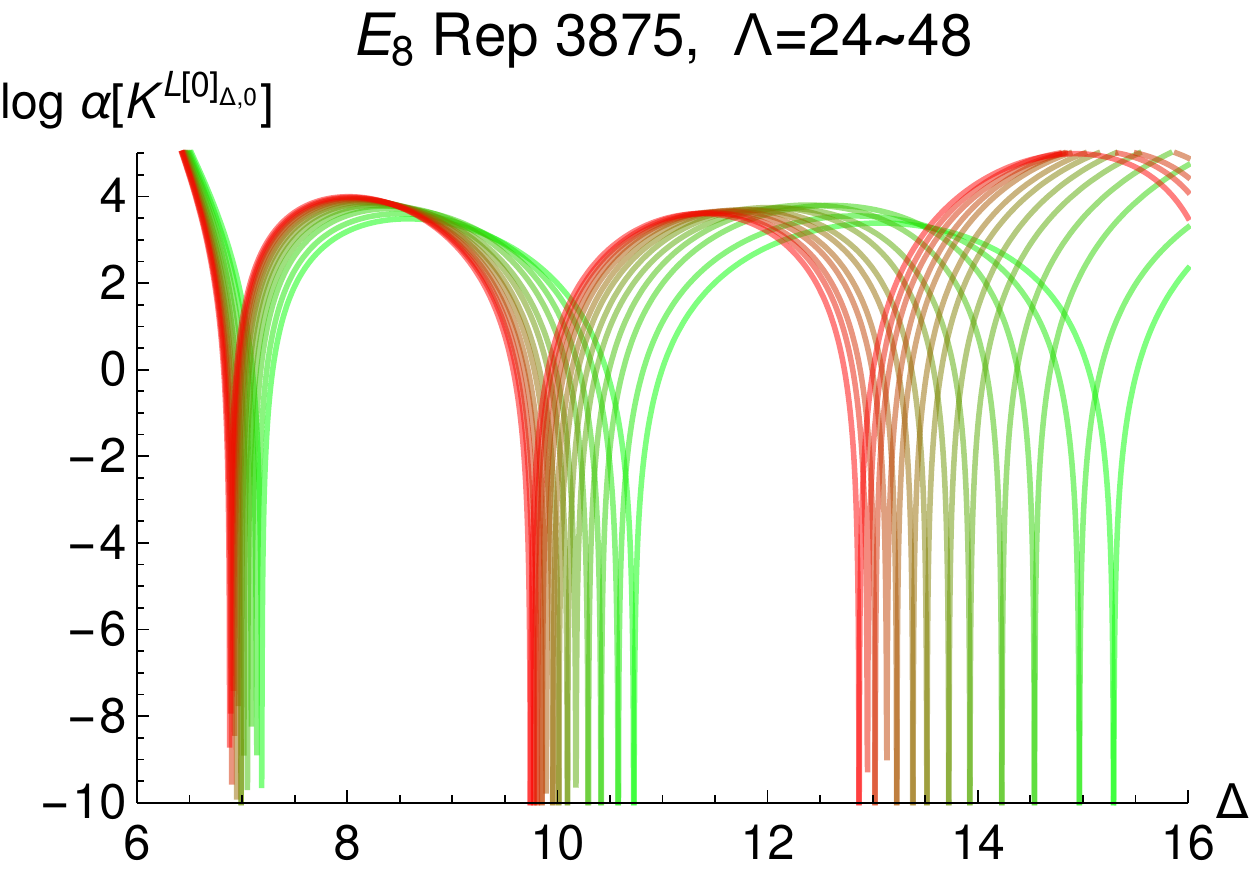}
}
\quad
\subfloat{
\includegraphics[width=.47\textwidth]{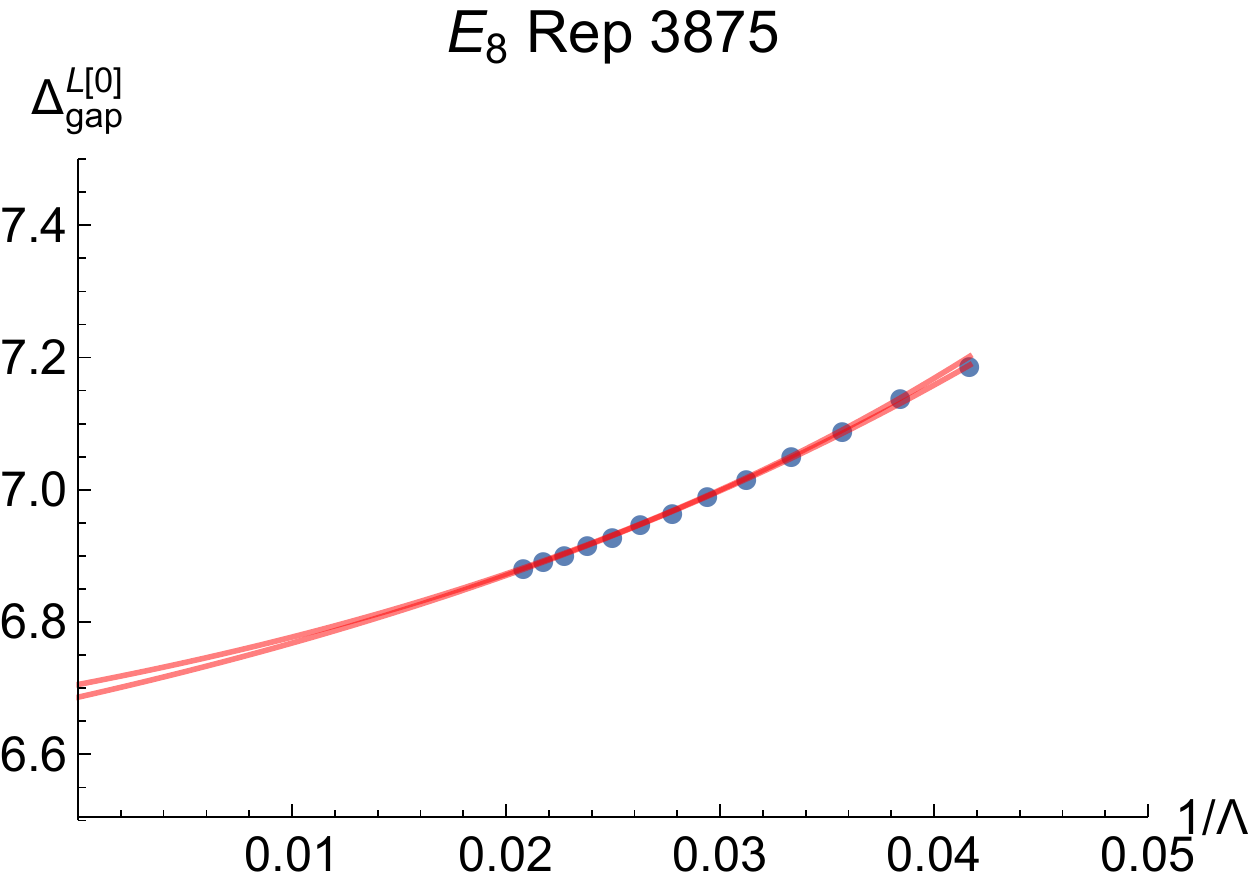}
}
\\
\subfloat{
\includegraphics[width=.47\textwidth]{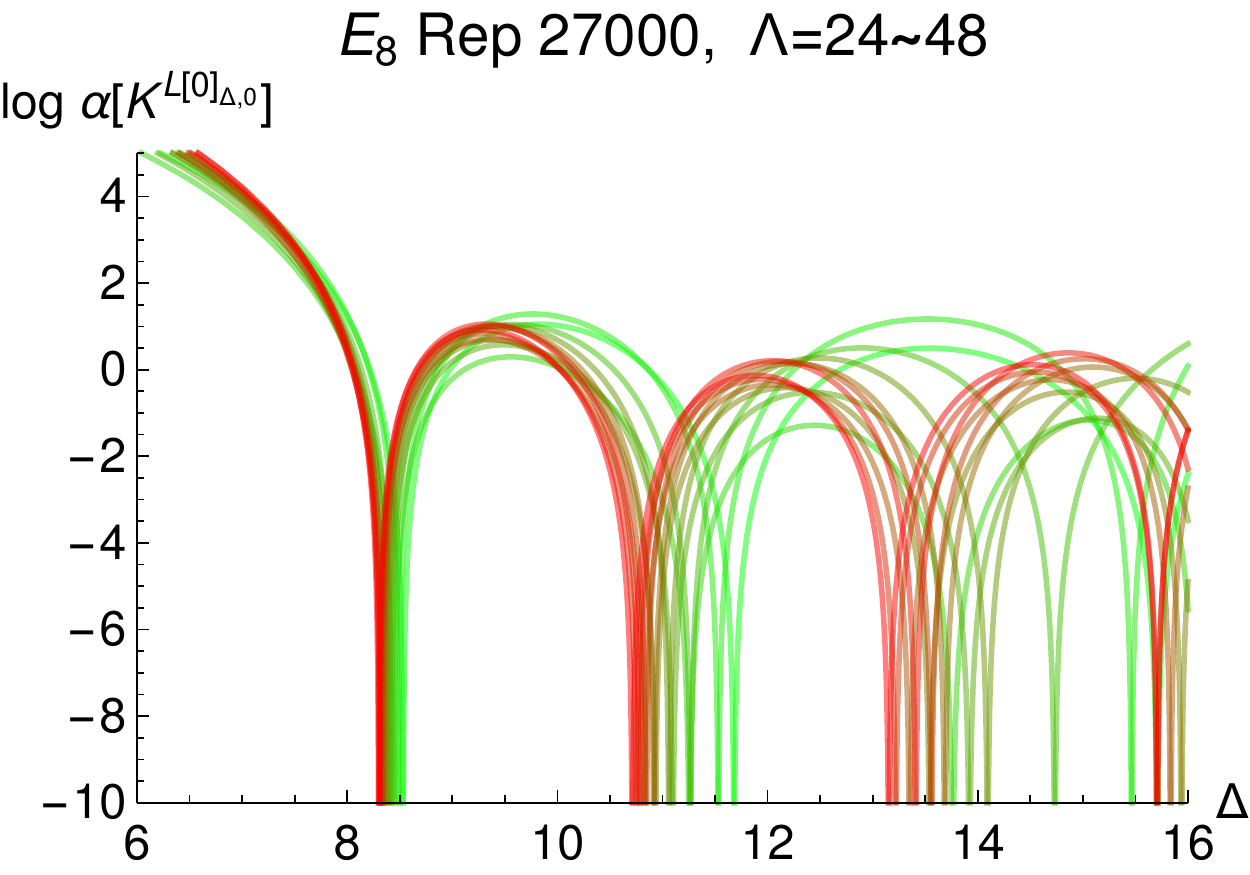}
}
\quad
\subfloat{
\includegraphics[width=.47\textwidth]{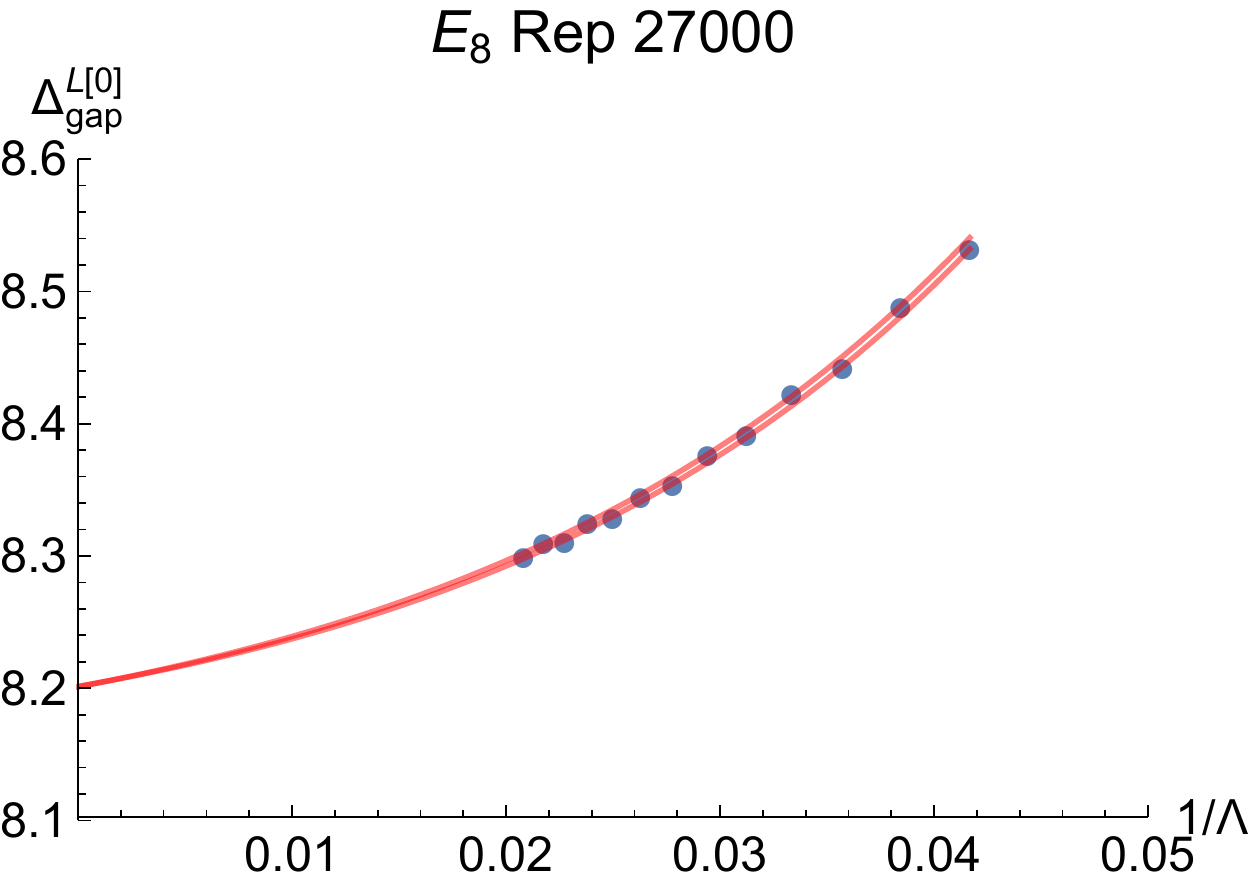}
}
\caption{{\bf Left:} The extremal functional optimizing the lower bound on $C_J$, acted on the contribution of the spin-zero long multiplet to the crossing equation, $\A_E[{\cal K}^{{\cal L}[0]_{\Delta,0}}(u,v)]$, in the {\bf 1}, {\bf 3875}, {\bf 27000} channels of $E_8$, plotted in logarithmic scale.  Increasing derivative orders $\Lambda = 24, 26, \dotsc, 48$ are shown from green to red.  {\bf Right:} The gap (lowest scaling dimension) in each channel at different $\Lambda$, and an extrapolation to $\Lambda \to \infty$ using the ansatz \eqref{GapAnsatz}, for $\Lambda \in 4\bZ$ and $\Lambda \in 4\bZ+2$, separately.}
\label{Fig:E8Functional}
\end{figure}

Assuming that Conjecture~\ref{Conj:MinCJ} is true, we can determine various physical properties of the rank-one E-string theory, such as the spectrum of long multiplets.  The left side of Figure~\ref{Fig:E8Functional} shows the extremal functional acted on the contribution of the spin-zero long multiplet to the crossing equation, namely, $\A_E[{\cal K}^{{\cal L}[0]_{\Delta,0}}(u,v)]$, in the {\bf 1}, {\bf 3875}, {\bf 27000} channels of $E_8$.  The right side shows how the lowest $\Delta$ in each channel varies with increasing derivative order $\Lambda$, as well as an extrapolation to infinite $\Lambda$ using the ansatz \eqref{GapAnsatz}, for $\Lambda \in 4\bZ$ and $\Lambda \in 4\bZ+2$, separately.
The results motivate the next conjecture.
\begin{conj}
In the ${\cal D}[2] \times {\cal D}[2]$ OPEs of flavor current multiplets in the rank-one E-string theory, the lightest long multiplet transforms in the {\bf 3875} of $E_8$.  The estimated gaps in the scaling dimensions of long multiplets in the {\bf 1}, {\bf 3875}, {\bf 27000} in ${\cal D}[2] \times {\cal D}[2]$
are given in Table~\ref{Tab:E8Gap}.
\end{conj}

\begin{table}[H]
\centering
\begin{tabular}{|c|c|}
\hline
Rep of $E_8$ & Extrapolated $\Delta_{gap}^{{\cal L}[0]}$
\\\hline\hline
\bf 1 & 
7.0
\\
\bf 3875 & 
6.7
\\
\bf 27000 & 
8.2
\\\hline
\end{tabular}
\caption{The estimated gaps in the scaling dimensions of long multiplets transforming in the {\bf 1}, {\bf 3875}, {\bf 27000} of $E_8$ in the OPEs of flavor current multiplets, in the rank-one E-string theory.}
\label{Tab:E8Gap}
\end{table}

\section{Outlook}
\label{Sec:Outlook}

\begin{figure}[h]
\centering
\subfloat{
\includegraphics[width=.47\textwidth]{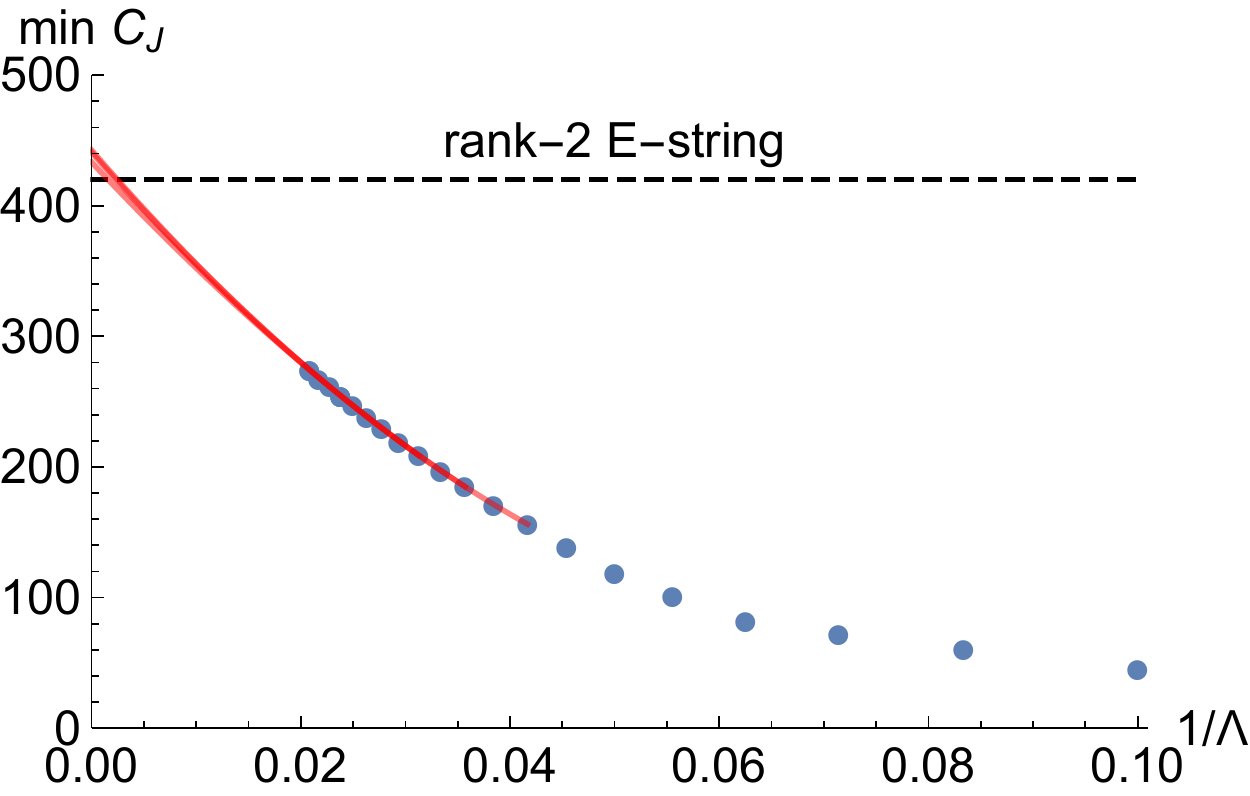}
}
\caption{The lower bounds on $C_J$ at different derivative orders $\Lambda$, for interacting theories with $E_8$ flavor group and assuming $C_T={151956 \over 5}$, which is the value in the rank-two E-string theory.  Also shown is an extrapolation to infinite derivative order using the quadratic ansatz \eqref{ExtrapAnsatz} with $\Lambda \geq 24$.}
\label{Fig:CJE82}
\end{figure}

Based on our observations on Figure~\ref{Fig:CTCJE8}, we put forward an optimistic conjecture.
\begin{conj}\label{conj:higherN}
The E-string theories of all ranks sit at the boundary of the space of unitary solutions to crossing.
\end{conj}
As a piece of supporting evidence, Figure~\ref{Fig:CJE82} shows the lower bound on $C_J$ assuming the value of $C_T={151956 \over 5}$ in the rank-two E-string theory, where we see that the extrapolated $C_J$ sits close to the rank-two E-string value $C_J=420$.  There is actually more we can do.  For $N>1$, the E-string theories have a larger flavor group $E_8\times{\rm SU(2)}$, and the SU(2) flavor central charge is given by
\ie
C_J={5\over 2}\left(16N^3+6N^2-21N-1\right).
\fe
This additional input may be necessary to put the higher-rank E-string theories on the boundary of the space of unitary solutions to crossing.

If Conjecture~\ref{conj:higherN}
is true, then the conformal bootstrap can potentially solve the E-string theories of arbitrary rank $N$.  We can then consider the large $N$ regime, and study the dual M-theory on AdS${}_7\times{\rm S}^4$/$\bZ_2$ beyond the supergravity limit.  On the M-theory side, the low energy excitations consist of a supergravity multiplet in the eleven-dimensional bulk and an $\cN=1$ $E_8$ vector multiplet supported on a ten-dimensional locus, AdS${}_7\times{\rm S}^3$ that is fixed by $\bZ_2$.  With enough computational power, we can collect information about the non-BPS spectra in the E-string theories of large $N$, filter out the operators dual to multi-particle excitations of the bulk supergravity and $E_8$ vector multiplets, and determine for instance the scaling dimension of the operator that corresponds to the first M-brane excitation.\footnote{Such an operator is analogous to the Konishi operator in ${\cal N}=4$ SYM, whose dimension to leading order at large $N$ is $2g_{\rm YM}^{1/2}N^{1/4}$ at strong coupling\cite{Gubser:2002tv,Gromov:2009zb} and ${3g_{YM}^2 N/4\pi^2}$ at weak coupling\cite{Anselmi:1996mq}.
}  The scaling dimension of this operator should behave as
\ie
\Delta = a N^b
\fe
to leading order at large $N$.  The knowledge of $a$ and $b$ would be an important step towards understanding the quantum nature of M-branes. 

We are also exploring other flavor groups.  For instance, the Sp(4)$_R$ R-symmetry in ${\cal N} = (2,0)$ theories breaks up into R-symmetry and flavor symmetry parts, Sp(2)$_R$ $\times$ Sp(2), when interpreted as ${\cal N} = (1,0)$ theories. For the $A_{N-1}$ theory, which is the infrared fixed point of the world-volume theory on a stack of $N$ M5 branes, the central charge and flavor central charge are
\ie
C_T=84(4N^3-3N-1),\quad C_J={5\over 2}(4N^3-3N-1).
\fe
Other ${\cal N} = (1,0)$ theories include the large class of theories constructed in F-theory \cite{DelZotto:2014hpa,Heckman:2013pva,Heckman:2014qba,Heckman:2015bfa}, whose $C_T$ and $C_J$ can be computed by using the anomaly polynomials given in \cite{Ohmori:2014kda,Mekareeya:2016yal}.  Finally, a particularly interesting example is a conjectural theory that has SU(3) flavor symmetry, and whose Higgs branch is given by the one-instanton moduli space of SU(3), recently proposed in \cite{Shimizu:2017kzs}.  It has central charge and flavor central charge
\ie
C_T={19488\over 5},\quad C_J={195\over 2}.
\fe
This theory does not seem to appear in the F-theoretic ``classification" of ${\cal N} = (1,0)$ theories\cite{DelZotto:2014hpa,Heckman:2013pva,Heckman:2014qba,Heckman:2015bfa}.\footnote{We thank Tom Rudelius for a discussion on this point.} The conformal bootstrap can provide evidence for the existence or non-existence of this theory.

The system of equations studied in this paper has straightforward generalizations to superconformal field theories in lower spacetime dimensions, ${\cal N} = 1$ in five and ${\cal N} = 3$ in three dimensions, which have SU(2)$_R$ R-symmetry\cite{Chang}.  The $C_T$ of such theories can be computed by taking the second derivative of the squashed three- or five-sphere partition function with respect to the squashing parameter\cite{Hama:2011ea,Imamura:2011wg,Closset:2012ru,Nishioka:2013gza,Imamura:2012xg,Lockhart:2012vp,Imamura:2012bm,Kim:2012qf,Spiridonov:2012de,Alday:2014rxa,Alday:2014bta,Bobev:2017asb}.\footnote{We thank Hee-Cheol Kim for a discussion on this point.}  In five dimensions, there is another distinguished class of superconformal field theories -- Seiberg's $E_n$ theories \cite{Seiberg:1996bd,Morrison:1996xf}.  If an analog of Conjecture~\ref{conj:higherN} is true for these theories, then we can study the type I' string theory on a warped product of AdS$_6$ and S$^4$ \cite{Brandhuber:1999np}. In three dimensions, the Chern-Simons-Matter theories provide many examples of ${\cal N} = 3$ superconformal field theories\cite{Kao:1995gf,Kapustin:1999ha,Gaiotto:2007qi}.  It would be interesting if the conformal bootstrap predicts new ${\cal N} = 3$ theories that do not admit Chern-Simons-Matter constructions.

\section*{Acknowledgments}

We are grateful to Matteo Beccaria, Clay C\'ordova, Thomas T. Dumitrescu, Ori Ganor, Hee-Cheol Kim, Petr Kravchuk, Luca V. Iliesiu, Kenneth Intriligator, Silviu S. Pufu, Tom Rudelius, Arkady A. Tseytlin, Yifan Wang, Xi Yin, and Yang Zhou for insightful discussions, and to Yifan Wang for comments on the first draft.  CC thanks the California Institute of Technology and Harvard University, and YL thanks Yale University, the Center for Quantum
Mathematics and Physics at University of California, Davis, the University of Amsterdam, and Princeton University for hospitality during the course of this work.  The numerical computations were performed on the Harvard Odyssey cluster, the UC Davis Peloton cluster, and the Caltech high energy theory group cluster.  We thank Tony Bartolotta for technical support with the Caltech cluster. YL is supported by the Sherman Fairchild Foundation and the U.S. Department of Energy, Office of Science, Office of High Energy Physics, under Award Number DE-SC0011632.  Lastly, the discoveries in this paper would not have been possible without John's foundational work on string theory, and we wish John a very happy birthday!

\appendix

\section{Bosonic conformal blocks}
\label{App:Bosonic}

This appendix reviews properties of bosonic conformal blocks for the four-point function of scalar primaries with scaling dimensions $\Delta_1, \Delta_2, \Delta_3, \Delta_4$ in $d = 2\epsilon + 2$ spacetime dimensions.   The conformal blocks depend on the external scaling dimensions only through the differences $\Delta_{12} \equiv \Delta_1 - \Delta_2$ and $\Delta_{34} \equiv \Delta_3 - \Delta_4$, and will be denoted by ${\cal G}^{\Delta_{12},\Delta_{34}}_{\Delta,\ell}$.  In Section~\ref{App:Jack}, we keep $\Delta_{12}$ and $\Delta_{34}$ arbitrary since blocks with nonzero $\Delta_{34}$ will be needed in Appendix~\ref{App:bCrossing}, but for later sections we set $\Delta_{12} = \Delta_{34} = 0$.  For notationally simplicity, we abbrevaite
\ie
{\cal G}^{0,0}_{\Delta,\ell} \to {\cal G}_{\Delta,\ell}.
\fe

The standard conformal cross ratios $u, v$ are defined in terms of the positions of operators as
\ie
&u={x_{12}^2 x_{34}^2\over x_{13}^2 x_{24}^2},\quad v={x_{14}^2 x_{23}^2\over x_{13}^2 x_{24}^2}, \quad x_{12}^2=(x_1-x_2)^2.
\fe
We also introduce the variables $z$, $\bar z$ and $\chi$, $\bar \chi$ as alternative ways to parameterize the cross ratios,\footnote{The reader should be careful when comparing with \cite{Dolan:2004mu}, as we have swapped what they called $z$ and $\chi$.
}
\ie
&u=z\bar z={\chi \bar \chi\over(1+\chi)(1+\bar \chi)}, \quad &&v=(1-z)(1-\bar z)={1\over(1+\chi)(1+\bar \chi)}.
\fe
Radial coordinates $r$ and $\eta$, defined as\cite{Hogervorst:2013sma}
\ie
r (\eta + i \sqrt{1-\eta^2}) = {z \over (1 + \sqrt{1-z})^2}, \quad r (\eta - i \sqrt{1-\eta^2}) = {\bar z \over (1 + \sqrt{1-\bar z})^2},
\fe
will be the variables in which we expand the conformal block in the recursive representation.

\subsection{Expansion in Jack polynomials}
\label{App:Jack}

The conformal block can be expanded in Jack polynomials\cite{Dolan:2003hv},
\ie
{\cal G}^{\Delta_{12},\Delta_{34}}_{\Delta,\ell}({z},\bar{z})=\sum_{m,n\ge0}r_{mn}(\Delta_{12},\Delta_{34},\Delta,\ell)P^{(\epsilon)}_{{1\over 2}(\Delta+\ell)+m,{1\over 2}(\Delta-\ell)+n}({z},\bar {z}),
\fe
where the expansion coefficients $r_{mn}$ are given by
\ie
r_{mn} &= \left({1\over 2}(\Delta+\ell-\Delta_{12})\right)_m\left({1\over 2}(\Delta+\ell+\Delta_{34})\right)_m
\\
& \times\left({1\over 2}(\Delta-\ell-\Delta_{12})-\epsilon\right)_n\left({1\over 2}(\Delta-\ell+\Delta_{34})-\epsilon\right)_n\widehat r_{mn},
\fe
and $\widehat r_{mn}$ are defined recursively via
\ie
&(m(m+\Delta+\ell -1)+n(n+\Delta-\ell-2\epsilon-1)) \widehat r_{mn}
\\
& \hspace{.5in} ={\ell+m-n-1+2\epsilon\over \ell+m-n-1+\epsilon}\widehat r_{m-1,n}+{\ell+m-n+1\over \ell+m-n+1+\epsilon}\widehat r_{m,n-1},
\fe
with the initial condition $r_{00} = 1$.

Jack polynomials can be defined in terms of Gegenbauer polynomials
\ie
P^{(\epsilon)}_{\lambda_1,\lambda_2}({z},\bar {z})={(\lambda_1-\lambda_2)!\over (2\epsilon)_{\lambda_1-\lambda_2}}({z}\bar{z})^{{1\over 2}(\lambda_1+\lambda_2)}C_{\lambda_1-\lambda_2}^{(\epsilon)}\left({z}+\bar{z}\over 2({z}\bar{z})^{1/2}\right),
\fe
which satisfy the orthogonality condition
\ie
\int^1_{-1}C_{m}^{(\epsilon)}(x)C_{n}^{(\epsilon)}(x)(1-x^2)^{\epsilon-1/2}dx=\delta_{m,n}{ 2^{1-2\epsilon}\pi\Gamma(m+2\epsilon)\over m!(m+\epsilon)\Gamma(\epsilon)^2}.
\fe
Jack polynomials are eigenfunctions of the differential operator $\Delta_\epsilon$ defined in \eqref{DefineDelta}, with eigenvalues
\ie
\label{eqn:DeltaEigenvalue}
\Delta_\epsilon P^{(\epsilon)}_{\lambda_1,\lambda_2}({z},\bar{z}) = {\cal E}^{(\epsilon)}_{\lambda_1,\lambda_2}P^{(\epsilon)}_{\lambda_1,\lambda_2}({z},\bar{z}),\quad {\cal E}^{(\epsilon)}_{\lambda_1,\lambda_2} = (\lambda_1+1+\epsilon)_{\epsilon-1}(\lambda_2+1)_{\epsilon-1}.
\fe
They also satisfy the relations
\ie\label{eqn:jackRelations}
& ({z}\bar{z})^{n} P^{(\epsilon)}_{\lambda_1,\lambda_2}({z},\bar{z}) = P^{(\epsilon)}_{\lambda_1+n,\lambda_2+n}({z},\bar{z}),
\\
&({z}+\bar{z})P^{(\epsilon)}_{\lambda_1,\lambda_2}({z},\bar{z})={\lambda_1-\lambda_2+2\epsilon\over \lambda_1-\lambda_2+\epsilon}P^{(\epsilon)}_{\lambda_1+1,\lambda_2}({z},\bar{z})+{\lambda_1-\lambda_2\over \lambda_1-\lambda_2+\epsilon}P^{(\epsilon)}_{\lambda_1,\lambda_2+1}({z},\bar{z}),
\\
&P^{(\epsilon)}_{\lambda_1,\lambda_2}({z},\bar{z})=P^{(\epsilon)}_{\lambda_2-\epsilon,\lambda_1+\epsilon}({z},\bar{z}).
\fe

\subsection{Recursive representation}
\label{App:Recur}

From now on we only consider the conformal blocks for the four-point function of identical scalar primaries, and set $\Delta_{12} = \Delta_{34} = 0$.

\begin{table}[H]
\centering
\begin{tabular}{|c|c|c|c|}
\hline
$n_i$ & $\Delta_i$ & $\ell_i$ & range of $k$
\\\hline\hline
$k$ & $1-\ell-k$ & $\ell+k$ & $k = 1, 2, \dotsc$
\\
$2k$ & $1+\epsilon-k$ & $\ell$ & $k = 1, 2, \dotsc$
\\
$k$ & $1+\ell+2\epsilon-k$ & $\ell-k$ & $k = 1, 2, \dotsc, \ell$
\\\hline
\end{tabular}
\caption{Three families of degenerate primaries that can appear in the OPE of two scalars, labeled by their scaling dimension $\Delta_i$, spin $\ell_i$, and level $n_i$ of the first null descendant.}
\label{tab:deg}
\end{table}

When the scaling dimension of the internal primary is taken to values where a descendant becomes null, the conformal block encounters a simple pole whose residue is again another conformal block.  This fact was first used in \cite{Zamolodchikov:1985ie,Zamolodchikov:1995aa} to write down a recursion formula for Virasoro blocks.  The generalization to higher dimensions was obtained in \cite{Penedones:2015aga}, where the authors found that when the external operators are scalars, the degenerate primaries come in three classes, as we list in Table~\ref{tab:deg}.  Then the conformal blocks admit the following recursive representation
\ie
& {\cal G}_{\Delta,\ell} (r, \eta) = (-)^\ell (4r)^{\Delta} h_{\Delta,\ell} (r, \eta),
\\
& h_{\Delta,\ell} (r, \eta) = \tilde h_\ell (r, \eta) + \sum_{i=1}^3 \sum_k {c_i(k) \over \Delta - \Delta_i(k)} r^{n_i(k)} h_{\Delta_i(k) + n_i(k),\ell_i(k)}(r, \eta),
\\
& \tilde h_\ell (r, \eta) = {\ell! \over (2\epsilon)_\ell} {(-1)^\ell C^{(\epsilon)}_\ell(\eta) \over (1-r^2)^\epsilon (1+r^2+2r\eta)^{{1 \over 2}} (1+r^2-2r\eta)^{{1 \over 2}}},
\fe
where $C^{(\epsilon)}_\ell(\eta)$ is the Gegenbauer polynomial.  The coefficients $c_i(k)$ for the three types of degenerate weights are
\ie
c_1(k) &= -{4^k k (-1)^k \over (k!)^2} {(\ell+2\epsilon)_k ({1\over2}(1-k))_k ({1\over2}(1-k))_k \over (\ell+\epsilon)_k},
\\
c_2(k) &= -{4^{2k} k (-1)^k \over (k!)^2} {(\epsilon-k)_{2k} \over (\ell+\epsilon-k)_{2k} (\ell+\epsilon+1-k)_{2k}} 
\\
& \hspace{.5in} \times \left( {1 \over 2} (1-k+\ell+\epsilon) \right)_k\left( {1 \over 2} (1-k+\ell+\epsilon) \right)_k
\\
& \hspace{.5in} \times
 \left( {1 \over 2} (1-k+\ell+\epsilon) \right)_k\left( {1 \over 2} (1-k+\ell+\epsilon) \right)_k,
\\
c_3(k) &= -{4^k k (-1)^k \over (k!)^2} {(\ell+1-k)_k ({1\over2}(1-k))_k ({1\over2}(1-k))_k \over (\ell+\epsilon+1-k)_k}.
\fe

The virtue of this recursive representation is not only its computational efficiency.  Firstly, the expansion in $r$ converges better than the $z$ expansion, as $r = 3-2\sqrt2 \approx 0.17$ at the crossing symmetric point.  Secondly, to a fixed order in $r$, the truncated conformal block with the $(4r)^\Delta$ prefactor stripped off is a rational function of $\Delta$, whose poles are at values of $\Delta$ below the unitarity bound.  This latter fact is crucial because semidefinite programming is much more efficient when the inputs are polynomials (for the sake of imposing non-negativity, we can strip off manifestly positive factors from the truncated conformal block); in fact, the SDPB package\cite{Simmons-Duffin:2015qma} only allows polynomial input.

For the purpose of computing derivatives of conformal blocks evaluated at the crossing symmetric point, we find it most efficient to -- instead of implementing the above recursion relation -- expand closed form expressions for conformal blocks in the diagonal limit $\bar z \to z$ to a fixed order in $r$ ($\eta = 1$ on the diagonal), take the diagonal derivatives at the crossing symmetric point, and then apply a further recursion relation to obtain the transverse derivatives\cite{ElShowk:2012ht}.  The closed form expressions and the recursion on transverse derivatives are reviewed in the next two sections.

\subsection{Diagonal limit}

When all external scalars have the same scaling dimension, the conformal blocks admit closed form expressions in the diagonal limit $\bar z \to z$, defined via a recursion relation \cite{ElShowk:2012ht}
\ie
& (\ell+d-3)(2\Delta+2-d) {\cal G}_{\Delta, \ell}(z, z)
\\
&= (d-2)(\Delta+\ell-1){\cal G}_{\Delta,\ell-2}(z, z) + {2-z\over2z} (2\ell+d-4) (\Delta-d+2) {\cal G}_{\Delta+1,\ell-1}(z, z)
\\
&~~~- {\Delta(2\ell+d-4)(\Delta+2-d)(\Delta+3-d)(\Delta-\ell-d+4)^2 \over 16(\Delta+1-{d\over2})(\Delta-{d\over2}+2)(\ell-\Delta+d-5)(\ell-\Delta+d-3)} {\cal G}_{\Delta+2,\ell-2}(z, z),
\fe
starting with seeds
\ie
& {\cal G}_{\Delta,0}(z, z) = \left( z^2 \over 1-z \right)^{\Delta\over2} {}_3F_2 \left( \tfrac{\Delta}{2}, \tfrac{\Delta}{2}, \tfrac{\Delta}{2}-\epsilon; \tfrac{\Delta+1}{2}, \Delta-\epsilon; {z^2 \over 4(z-1)} \right),
\\
& {\cal G}_{\Delta,1}(z, z) = {2-z\over2z} \left( z^2 \over 1-z \right)^{\Delta+1\over2} {}_3F_2 \left( \tfrac{\Delta+1}{2}, \tfrac{\Delta+1}{2}, \tfrac{\Delta+1}{2}-\epsilon; \tfrac{\Delta}{2}+1, \Delta-\epsilon; {z^2 \over 4(z-1)} \right).
\fe

\subsection{Recursion on transverse derivatives}

Define
\ie
z = {1 + a + \sqrt b \over 2}, \quad \bar z = {1 + a - \sqrt b \over 2},
\fe
and denote $\partial_a^m \partial_b^n {\cal G}_{\Delta, \ell}|_{a = b = 0}$ by $h_{m,n}$.  Given the diagonal limit of the conformal block, we can compute $h_{m,0}$ for all $m \geq 0$.  The transverse derivatives can then be obtained by the following recursion relation\cite{ElShowk:2012ht},
\ie
\hspace{.5in} & \hspace{-.5in} 2(d+2n-3) h_{m,n}
\\
& \hspace{-.5in} =2m(d+2n-3) [-h_{m-1,n} + (m-1)h_{m-2,n} + (m-1)(m-2)h_{m-3,n}]
\\
&- h_{m+2,n-1} + (d-m-4n+4) h_{m+1,n-1}
\\
&+ [2C_{\Delta,\ell} + 2d(m+n-1) + m^2 + 8mn - 9m + 4n^2 - 6n + 2] h_{m,n-1}
\\
&+ m[d(m-2n+1) + m^2 + 12mn - 15m + 12n^2 - 30n + 20] h_{m-1,n-1}
\\
&+ (n-1)[h_{m+2,n-2}-(d-3m-4n+4)h_{m+1,n-2}].
\fe

\section{Superconformal Ward identities}
\label{App:Ward}

The superconformal Ward identities read\cite{Dolan:2004mu}
\ie\label{eqn:SCW}
(\partial_{\chi}+\epsilon\partial_{w})G(u,v;w)\Big|_{w \to \chi} = (\partial_{\bar\chi}+\epsilon\partial_{w})G(u,v;w)\Big|_{w \to \bar\chi}=0,
\fe
where the variables $\chi$ and $\bar \chi$ are related to $u$ and $v$ by
\ie
&u={\chi \bar \chi\over(1+\chi)(1+\bar \chi)},\quad v={1\over(1+\chi)(1+\bar \chi)}.
\fe

We presently show in the case of $k=2$ that the second equation in \eqref{eqn:crossingk=2} follows from the first as a consequence of the first superconformal Ward identity in \eqref{eqn:SCW}, which explicitly reads
\ie
\chi^2\partial_\chi G_0(u,v)+\chi\partial_\chi G_1(u,v)+\partial_\chi G_2(u,v)=\epsilon G_1(u,v)+{2\epsilon\over \chi} G_2(u,v).
\fe
It can be rewritten as
\ie
\partial_\chi  (\chi \bar \chi)^{-\epsilon}G_1(u,v)
%&=z^{-\epsilon-1}\bar z^{-\epsilon}\left[z\partial_zG_1(u,v)-\epsilon G_1(u,v)\right]
%\\
%&=2\epsilon z^{-\epsilon-2}\bar z^{-\epsilon} G_2(u,v)-z^{-\epsilon+1}\bar z^{-\epsilon}\partial_z G_0(u,v)-z^{-\epsilon-1}\bar z^{-\epsilon}\partial_z G_2(u,v)
%\\
&=-\chi^{-\epsilon+1}\bar \chi^{-\epsilon}\partial_\chi G_0(u,v)-\chi^{\epsilon-1}\bar \chi^{\epsilon}\partial_\chi(\chi\bar \chi)^{-2\epsilon}G_2(u,v).
\fe
Applying $u\leftrightarrow v$ or equivalently $(\chi,\bar \chi) \leftrightarrow (\chi^{-1},\bar \chi^{-1})$, the above equation becomes
\ie
\partial_\chi  (\chi\bar \chi)^{\epsilon}G_1(v,u)&=-\chi^{\epsilon-1}\bar \chi^{\epsilon}\partial_\chi G_0(v,u)-\chi^{-\epsilon+1}\bar \chi^{-\epsilon}\partial_\chi(\chi\bar \chi)^{2\epsilon}G_2(v,u).
\fe
The difference of the two equations gives
\ie\label{eqn:Wardz}
&\partial_\chi\left[(\chi\bar \chi)^{-\epsilon}G_1(u,v) - (\chi\bar \chi)^{\epsilon}G_1(v,u)\right] 
\\
& \hspace{-.15in} = -\chi^{-\epsilon+1}\bar \chi^{-\epsilon}\partial_\chi\left[ G_0(u,v)-(\chi\bar \chi)^{2\epsilon}G_2(v,u)\right]-\chi^{\epsilon-1}\bar \chi^{\epsilon}\partial_\chi\left[(\chi\bar \chi)^{-2\epsilon}G_2(u,v)-G_0(v,u)\right].
\fe
Similarly, with $\chi$ replaced by $\bar \chi$, we have
\ie\label{eqn:Wardzbar}
&\partial_{\bar \chi}\left[(\chi\bar \chi)^{-\epsilon}G_1(u,v) - (\chi\bar \chi)^{\epsilon}G_1(v,u)\right] 
\\
& \hspace{-.15in} = - \chi^{-\epsilon}\bar \chi^{-\epsilon+1}\partial_{\bar \chi}\left[ G_0(u,v)-(\chi\bar \chi)^{2\epsilon}G_2(v,u)\right]- \chi^{\epsilon}\bar \chi^{\epsilon-1}\partial_{\bar \chi}\left[(\chi\bar \chi)^{-2\epsilon}G_2(u,v)-G_0(v,u)\right].
\fe
From \eqref{eqn:Wardz} and \eqref{eqn:Wardzbar}, we see that the first and third equations of \eqref{eqn:crossingk=2} imply the second equation of \eqref{eqn:crossingk=2} up to a constant.  This constant can be fixed by considering the case of $u=v$. 

Let us define $K(u,v)=v^{2\epsilon}G_0(u,v)-u ^{2\epsilon}G_2(v,u)$. The compatibility between \eqref{eqn:Wardz} and \eqref{eqn:Wardzbar} gives the identity
\ie
\label{CrossingIdentity}
&\partial_{\bar \chi}\left[\chi^{-\epsilon+1}\bar \chi^{-\epsilon}\partial_\chi v^{-2\epsilon}K(u,v)-\chi^{\epsilon-1}\bar \chi^{\epsilon}\partial_\chi u^{-2\epsilon}K(v,u)\right]
\\
&=\partial_\chi\left[\chi^{-\epsilon}\bar \chi^{-\epsilon+1}\partial_{\bar \chi}v^{-2\epsilon}K(u,v)- \chi^{\epsilon}\bar \chi^{\epsilon-1}\partial_{\bar \chi}u^{-2\epsilon}K(v,u)\right],
\fe
which is important when we want to identity the independent constraints from the crossing equation.

\section{Crossing equation for $b(u,v)$}
\label{App:bCrossing}

Specializing to $\epsilon=k=2$, let us substitute the solution \eqref{eqn:WardSolution} into the superconformal Ward identity into the crossing equation \eqref{crossing13},
\ie\label{eqn:(2,0)crossing}
&D_2 (1-z-zw^{-1})(1-\bar z-\bar zw^{-1})\left[{z}\bar{z} b({z},\bar{z})-(1-{z})(1-\bar{z}) b(1-{z},1-\bar{z})\right]=0.
\fe
Defining $H({z},\bar{z})={z}\bar{z} b({z},\bar{z})-(1-{z})(1-\bar{z}) b(1-{z},1-\bar{z})$, the above equation is equivalent to
\ie\label{eqn:crossingEqH}
D_2H({z},\bar{z})=0,\quad D_2({z}+\bar{z})H({z},\bar{z})=0,\quad D_2{z}\bar{z} H({z},\bar{z})=0.
\fe
The general solution to the first equation is\footnote{There may appear to be another class of solutions $P^{(2)}_{-2,n+2}(z,\bar z)$, but they are related to $P^{(2)}_{n,0}(z,\bar z)$ by \eqref{eqn:jackRelations}.}
\ie
H({z},\bar{z})=\sum_n a_{n}P^{(2)}_{n,0}({z},\bar{z}).
\fe
We also have
\ie
D_2{z}\bar{z} H&=\sum_{n}a_{n}(n+3)P^{(2)}_{n,0}({z},\bar{z}), \quad D_2({z}+\bar{z}) H=\sum_{n}a_{n}nP^{(2)}_{n-1,0}({z},\bar{z}).
\fe
Using the fact that $P^{(2)}_{n,0}({z},\bar{z})$'s are orthogonal polynomials for non-negative integers $n$, one can argue that \eqref{eqn:crossingEqH} has no non-trivial solution if we restrict to such $n$. However, the orthogonality condition fails if we allow $n$ to take negative integer values, and indeed \eqref{eqn:crossingEqH} has an unique solution
\ie
H(z,\bar z)={1\over (z-\bar z)^3}.
\fe 
Therefore, the original crossing equation is equivalent to
\ie
\label{bCrossing}
{z}\bar{z} b({z},\bar{z})-(1-{z})(1-\bar{z}) b(1-{z},1-\bar{z})={c\over (z-\bar z)^3},
\fe
where $c$ is an unphysical constant.  For example, the generalized free field solution \eqref{eqn:Ggff} corresponds to (up to the unphysical term)
\ie
b_{\text{\tiny gff}}(u,v)={1\over 3}u^{-1}(1+u^{-3}+v^{-3}),
\fe
which solves \eqref{bCrossing} with $c = 0$.

A function $b(u, v)$ that gives rise to a physical four-point function $G(u, v; w)$ (via \eqref{eqn:WardSolution}) also admits a decomposition into blocks with non-negative coefficients.  The blocks $b^{\cal X}(u,v)$ for the superconformal multiplets \eqref{eqn:(1,0)selectionRole} can be expressed (up to the unphysical term on the RHS of \eqref{bCrossing}) in terms of bosonic conformal blocks with $\Delta_{12} = 0$ and $\Delta_{34} = -2$,
\ie
b^{{\cal L}[0]_{\Delta,\ell}}(u,v)&=  \left({1\over 2}(\Delta+\ell)\right)_{-1}\left({1\over 2}(\Delta-\ell)-2\right)_{-1}u^{-5}{\cal G}^{0,-2}_{\Delta+2,\ell}(u,v),
\\
b^{{\cal B}[2]_{\ell}}(u,v)&={2(\ell+4)\over (\ell+1)}\left(\ell+5\right)_{-1}u^{-5}{\cal G}^{0,-2}_{\ell+9,\ell+1}(u,v),
\\
b^{{\cal B}[0]_{\ell}}(u,v)&= -{1\over \ell+1} u^{-5}{\cal G}^{0,-2}_{\ell+6,\ell}(u,v),
\\
b^{{\cal D}[4]}(u,v)&=2u^{-5}{\cal G}^{0,-2}_{8,0}(u,v),
\\
b^{{\cal D}[2]}(u,v)&=-{2\over 3}u^{-5}{\cal G}^{0,-2}_{5,-1}(u,v),
\\
b^{{\cal D}[0]}(u,v)&={1\over 3}u^{-4},
\fe
where the unphysical ${\cal G}^{0,-2}_{5,-1}(u, v)$ is formally defined by its expansion into Jack polynomials.  Explicitly, $b^{{\cal D}[2]}(u,v)$ can be written as
\ie
&b^{{\cal D}[2]}(u,v)=-{4({z}+\bar{z}-{z}\bar{z})\over {z}^3\bar{z}^3({z}-\bar{z})^2} +{4(\log{z}-\log\bar{z})\over  {z}\bar{z}({z}-\bar{z})^3}
\\
&\quad + {4\bar{z}^2(\bar{z}-3{z}+3{z}^2-{z}^3\bar{z})\log(1-{z})-4{z}^2({z}-3\bar{z}+3\bar{z}^2-{z}\bar{z}^3)\log(1-\bar{z})\over  {z}^4\bar{z}^4({z}-\bar{z})^3},
\fe
which has a branch point at the origin of the $z$-plane, and the monodromy around it is
\ie
b^{{\cal D}[2]}(u,v)\Big|_{(z,\bar z)\to (e^{2\pi i}z,e^{-2\pi i}\bar z)}=b^{{\cal D}[2]}(u,v)+{16i\pi\over  z\bar z({z}-\bar{z})^3}.
\fe
This monodromy can be absorbed into a shift of the constant $c$,
\ie
c \to c + 32i\pi n.
\fe
We can therefore restrict to the zeroth sheet, where $b^{{\cal D}[2]}(u,v)$ along with other $b^{\cal X}(u, v)$ are all real functions in $z, \bar z$.  Moreover, on this sheet, $b^{\cal X}(u, v)$ are regular as $\bar z \to z$, whereas the term on the right hand side of \eqref{bCrossing} is not. Hence, the constant $c$ must vanish for a solution to \eqref{bCrossing} to also admit an expansion into blocks.

\section{Relating central charges to OPE coefficients}
\label{App:Central2OPE}

\subsection{$C_T$ to $\lambda_{{\cal B}[0]_{0}}^2$}
\label{App:CT2OPE}

Conformal symmetry fixes the three-point function of the stress tensor with two identical scalars ${\cal O}$ to be of the form\cite{Osborn:1993cr}
\ie\label{eqn:TOO}
\vev{T_{\m\n}(x_1){\cal O}(x_2){\cal O}(x_3)}={C_{{\cal O}{\cal O}T}\over x_{12}^d x_{23}^{2\Delta-d}x_{13}^d}t_{\mu\nu}(X_{23}),
\fe
where the conformal structure $t_{\m\n}$ is given by
\ie\label{eqn:structureT}
t_{\m\n}(X)={X_\m X_\n\over X^2} - {1\over d}\delta_{\m\n},\quad X^\m_{23}={x^\m_{21}\over x_{21}^2}-{x^\m_{31}\over x_{31}^2},\quad X_{23}^2 = {x_{23}^2\over x_{21}^2 x_{31}^2}.
\fe
The OPE coefficient $C_{{\cal O}{\cal O}T}$ is fixed by the conformal Ward identity to be \cite{Poland:2010wg,Penedones:2016voo}
\ie\label{eqn:COOTcoef}
C_{{\cal O}{\cal O}T}=-{d\Delta\over (d-1)  V_{\widehat{\rm S}^{d-1}}},\quad  V_{\widehat{\rm S}^{d-1}}={2\pi^{d\over 2}\over\Gamma\left(d\over 2\right)}.
\fe

For later use, we note that the tensor structures $I_{\m\n}(x)$, $X^\m$, ${\cal I}_{\m\n,\sigma\rho}$ and $t_{\m\n}$, defined in \eqref{eqn:structureI} and \eqref{eqn:structureT}, satisfy the identities
\ie\label{eqn:ConStrIds}
&{\cal I}_{\m\n,\A\B}(x){\cal I}_{\A\B,\sigma\rho}(x) = {1\over 2}(\delta_{\m\sigma}\delta_{\n\rho}+\delta_{\m\rho}\delta_{\n\sigma})-{1\over d}\delta_{\m\n}\delta_{\sigma\rho},
\\
&I_{\m\A}(x_{13})X_{23}^\A={x_{23}^2\over x_{12}^2}X_{21}^\m,
\\
&{\cal I}_{\m\n,\sigma\rho}(x_{13})t_{\sigma\rho}(X_{23})=t_{\sigma\rho}(X_{12}).
\fe

From the three-point function \eqref{eqn:TOO}, and using the identities \eqref{eqn:ConStrIds}, we can deduce that the OPE of two identical scalars contains
\ie\label{eqn:OPEoot}
{\cal O}(x_1){\cal O}(x_2)\sim {1\over x_{12}^{2\Delta}}+{C_{{\cal O}{\cal O}T}\over C_T}{ V_{\widehat{\rm S}^{d-1}}^2\over  x_{12}^{2\Delta-d}}t_{\m\n}(x_{12})T^{\m\n}(x_2).
\fe
Now, consider the four-point function of four identical scalars $\cal O$, which reduces to a sum over three-point functions by taking the OPE \eqref{eqn:OPEoot} of the operators ${\cal O}(x_1)$ and ${\cal O}(x_2)$.  Using the formula \eqref{eqn:TOO} for the three-point function, we obtain
\ie
&\vev{{\cal O}(x_1){\cal O}(x_2){\cal O}(x_3){\cal O}(x_4)}
\\
&\sim {1\over x_{12}^{2\Delta}x_{34}^{2\Delta}}\left[1+{C_{{\cal O}{\cal O}T}^2\over C_T}{1\over  x_{12}^{-d}}t_{\m\n}(x_{12}){ V_{\widehat{\rm S}^{d-1}}^2\over x_{23}^d x_{34}^{-d}x_{24}^d}t_{\mu\nu}(X_{34})\right]
\\
&={1\over x_{12}^{2\Delta}x_{34}^{2\Delta}}\left[1+{C_{{\cal O}{\cal O}T}^2\over C_T}{ V_{\widehat{\rm S}^{d-1}}^2\over  x_{12}^{-d} x_{23}^d x_{34}^{-d}x_{24}^d}\left({(x_{12}^2 x_{24}^2+x_{14}^2 x_{23}^2-x_{13}^2 x_{24}^2 - x_{12}^2 x_{23}^2)^2\over 4x_{12}^2 x_{34}^2x_{23}^2 x_{24}^2} - {1\over d}\right)\right],
\fe
which can be written in terms of the cross ratios $u$ and $v$ as
\ie\label{eqn:OOOOexp}
\vev{{\cal O}(x_1){\cal O}(x_2){\cal O}(x_3){\cal O}(x_4)}\sim{1\over x_{12}^{2\Delta}x_{34}^{2\Delta}}\left[1+{C_{{\cal O}{\cal O}T}^2\over C_T}{ V_{\widehat{\rm S}^{d-1}}^2\over  u^{-{d\over 2}}v^{d\over 2}}\left({(u+v-1)^2\over 4u v} - {1\over d}\right)\right].
\fe
Comparing \eqref{eqn:OOOOexp} with the conformal block expansion, we determine the coefficient that sits in front of the bosonic stress-tensor block ${\cal G}_{d,2}(u,v)$,
\ie\label{eqn:CTin4ptfn}
\vev{{\cal O}(x_1){\cal O}(x_2){\cal O}(x_3){\cal O}(x_4)}&\supset {\cal G}_{0,0}(u,v)+{d\over (d-1)}{\Delta^2\over C_T}{\cal G}_{d,2}(u,v).
\fe
The bosonic conformal block ${\cal G}_{d,2}(u,v)$ sits inside the ${\cal B}[0]_{0}$ superconformal block with the coefficient given in \eqref{eqn:B[0]SB}.  We thus obtain the relation \eqref{eqn:lambdaB0toCT} between the OPE coefficient $\lambda_{{\cal B}[0]_{0}}$ and the central charge $C_T$.

\subsection{$C_J$ to $\lambda_{{\cal D}[2]}^2$}
\label{App:CJ2OPE}

Consider the three-point function of one flavor current with two scalars transforming in representation ${\mathcal R}$ of the flavor group. Conformal symmetry fixes this three-point function to be\footnote{Acting the charge \eqref{QfromJ} on the scalar ${\cal O}^j(0)$ gives
\ie
Q^a{\cal O}^j(0) = i(T^a_{\mathcal R})^j{}_k{\cal O}^k(0),
\fe
which fixes the overall coefficient of the three-point function \eqref{eqn:JOO}.}
\ie\label{eqn:JOO}
\vev{J_\m^a(x_1){\cal O}^i(x_2){\cal O}^j(x_3)} =i g^{ik}(T^a_{\mathcal R})^j{}_k{1\over  V_{\widehat{\rm S}^{d-1}}} {X_{23}^\m \over x_{13}^{d-2} x_{12}^{d-2}x^{2\Delta-d+2}_{23}},
\fe
where $i, j$ are the indices for representation $\cal R$, $T^a_{\mathcal R}$ are the generators of the flavor group in the representation $\cal R$, and the two point functions of the scalars are normalized as $\vev{{\cal O}^i(x_1){\cal O}^j(x_2)}=g^{ij}/x_{12}^{2\Delta}$.  We are particularly interested in external scalars that transform in the adjoint representation, in which case $(T^a)^b{}_c = f^{ab}{}_c $.  

From the three-point function \eqref{eqn:JOO}, and using the identities \eqref{eqn:ConStrIds}, we obtain the OPE of two scalars in the adjoint representation,
\ie\label{eqn:OPEooj}
{\cal O}^a(x_1){\cal O}^b(x_2)\sim{\delta^{ab}\over x_{12}^{2\Delta}}-if^{ab}{}_c{ V_{\widehat{\rm S}^{d-1}}\over C_J} {x_{12}^\m \over x^{2\Delta-d+2}_{12}}J^c_\m(x_2).
\fe
Now consider the four-point function of four scalars ${\cal O}^a$.  Using the OPE \eqref{eqn:OPEooj} and the three-point function \eqref{eqn:JOO}, we find
\ie
&\vev{{\cal O}^a(x_1){\cal O}^b(x_2){\cal O}^c(x_3){\cal O}^d(x_4)}
\\
&\sim{1\over x^{2\Delta}_{12} x^{2\Delta}_{34}}\left[\delta^{ab}\delta^{cd}-{f^{abe}f^{ecd} \over C_J}  {x_{12} X_{34}\over x^{-d+2}_{12} x_{24}^{d-2} x_{23}^{d-2}x^{-d+2}_{34}}\right]
\\
& \hspace{-.1in} ={1\over x_{12}^{2\Delta} x_{34}^{2\Delta}}\left[ {\rm dim}(G_F) P_{\bf 1}^{abcd}-{\psi^2 h^{\vee}P_{\bf adj}^{abcd} \over 2C_J}  {x^{d-2}_{12}x^{d-2}_{34}\over  x_{24}^{d} x_{23}^{d}}\left((x_{12}^2-x_{13}^2) x_{24}^2 - (x_{12}^2-x_{14}^2)  x_{23}^2\right)\right]
\fe
which can be expressed in terms of the cross ratios $u$ and $v$ as
\ie\label{eqn:OOOOexpJ}
& \vev{{\cal O}^a(x_1){\cal O}^b(x_2){\cal O}^c(x_3){\cal O}^d(x_4)}
\\
& \hspace{.5in} \sim{1\over x_{12}^{2\Delta} x_{34}^{2\Delta}} \left[ {\rm dim}(G_F)P_{\bf 1}^{abcd}-{\psi^2 h^{\vee} P_{\bf adj}^{abcd}\over 2C_J}{u^{{d\over 2}-1}\over v^{{d\over 2}}}\left(u+v-1\right)\right].
\fe
By comparing \eqref{eqn:OOOOexpJ} with the conformal block expansion, we can determine the coefficient sitting in front of the bosonic conformal block ${\cal G}_{d-1,1}(u,v)$ of the flavor current,
\ie
\vev{{\cal O}^a(x_1){\cal O}^b(x_2){\cal O}^c(x_3){\cal O}^d(x_4)}\supset {\rm dim}(G_F)P_{\bf 1}^{abcd} {\cal G}_{0,0}(u,v)+{\psi^2 h^{\vee}\over C_J} P_{\bf adj}^{abcd}{\cal G}_{d-1,1}(u,v).
\fe
The bosonic conformal block ${\cal G}_{d-1,1}(u,v)$ sits inside the ${\cal D}[2]$ superconformal block with the coefficient given in \eqref{eqn:D[2]SB}.  We thus obtain the relation \eqref{eqn:lambdaD2toCJ} between the OPE coefficient $\lambda_{D[2]}$ and the flavor central charge $C_J$.

\section{The central charge $C_T$ of the three-derivative fermion}
\label{App:Vec}

The $C_T$ of a free three-derivative Weyl fermion was recently computed in \cite{Beccaria:2017dmw} as the second derivative of the partition function on S$^1\times{\mathbb H}^{5}$ with respect to the S$^1$ radius.  In this appendix, we verify their answer by explicitly constructing the stress tensor for a three-derivative Dirac fermion, and computing its two-point function. The $C_T$ of a Weyl fermion is simply half that of a Dirac fermion.  Since the three-derivative Dirac fermion exists in arbitrary $d$ spacetime dimensions, we keep $d = 2\epsilon + 2$ general.  The two-point function of a free Dirac fermion with scaling dimension $\Delta_\psi$ is
\ie
\vev{\psi(x_1) \bar\psi(x_2)}={{\not\!x}_{12}\over x^{2 \Delta_\psi+1}_{12}},
\fe
where $\not\!x = x^\mu \Gamma_\mu$, and $\Gamma^\mu$ are $2^{\lfloor\epsilon\rfloor+1} \times 2^{\lfloor\epsilon\rfloor+1}$ matrices obeying the Clifford algebra $\{ \Gamma^\mu, \Gamma^\nu \} = 2\delta^{\mu\nu} 1\!\!1$.  For a three-derivative fermion, $\Delta_\psi=\epsilon-{1\over2}$.

Our approach is to work in flat space, write down the most general symmetric traceless  spin-two primary operator of scaling dimension $d$, imposed current conservation, and identify the stress tensor by demanding that it has the correct OPE with the fundamental fermion \cite{Iliesiu:2015qra},
\ie
\label{FermionOPE}
T_{\m\n}(x) \psi(0) \sim& - {\Delta_\psi \over (d-1) V_{\widehat{\rm S}^{d-1}}} { \D_{\m\n} x^2 - d x_\m x_\n \over |x|^{d+2} } \psi(0) + {d \over 2 V_{\widehat{\rm S}^{d-1}}} {x_{(\underline{\m}} x^\rho \Gamma_{\rho\underline{\nu})} \psi(0) \over |x|^{d+2}} +  \dotsb.
\fe
Let us first list all the symmetric traceless spin-two operators of scaling dimension $d$ constructed as fermion bilinears,
\ie
&T^1_{\m\n}=\bar \psi{\not\!\partial}\partial_{\m}\partial_{\n}\psi - {1\over d}\delta_{\m\n}\bar \psi{\not\!\partial}\partial^2\psi,&& T^2_{\m\n}=\partial_{(\m}\bar \psi{\not\!\partial}\partial_{\n)}\psi - {1\over d}\delta_{\m\n}\partial_{\rho}\bar \psi{\not\!\partial}\partial_{\rho}\psi,
\\
&T^3_{\m\n} = \partial_{\m}\partial_{\n}\bar \psi{\not\!\partial}\psi - {1\over d}\delta_{\m\n}\partial^2\bar \psi{\not\!\partial}\psi,& &T^4_{\m\n} = \bar\psi \Gamma_{(\m}\partial_{\n)}\partial^2\psi - {1\over d}\delta_{\m\n}\bar\psi {\not\!\partial}\partial^2\psi,
\\
&T^5_{\m\n} = \partial_\rho\bar\psi \Gamma_{(\m}\partial_{\n)}\partial_\rho\psi - {1\over d}\delta_{\m\n}\partial_\rho\bar\psi {\not\!\partial}\partial_\rho\psi,&&T^6_{\m\n} = \partial^2\bar\psi \Gamma_{(\m}\partial_{\n)}\psi - {1\over d}\delta_{\m\n}\partial^2\bar\psi {\not\!\partial}\psi,
\\
&T^7_{\m\n}= \partial_{(\m}\bar\psi \Gamma_{\n)}\partial^2\psi - {1\over d}\delta_{\m\n}\partial_{\rho}\bar\psi \Gamma_{\rho}\partial^2\psi,&&T^8_{\m\n} =\partial_\rho\partial_{(\m}\bar\psi \Gamma_{\n)}\partial_\rho\psi - {1\over d}\delta_{\m\n}  \partial_\rho\partial_{\sigma}\bar\psi \Gamma_{\sigma}\partial_\rho\psi,
\\
&T^9_{\m\n}=\partial^2\partial_{(\m}\bar\psi \Gamma_{\n)}\psi - {1\over d}\delta_{\m\n}\partial^2\partial_{\rho}\bar\psi \Gamma_{\rho}\psi,&&T^{10}_{\m\n}=\partial_\rho\bar\psi \Gamma_\rho\partial_\m\partial_\n \psi- {1\over d}\delta_{\m\n}\partial_{\rho}\bar\psi \Gamma_{\rho}\partial^2\psi,
\\
&T^{11}_{\m\n}=\partial_{(\m }\partial_\rho\bar\psi \Gamma_\rho\partial_{\n)}\psi - {1\over d}\delta_{\m\n}\partial_\sigma\partial_{\rho}\bar\psi \Gamma_{\rho}\partial_\sigma\psi,&&T^{12}_{\m\n}=\partial_\m\partial_\n \partial_\rho\bar\psi \Gamma_\rho\psi - {1\over d}\delta_{\m\n}\partial^2\partial_{\rho}\bar\psi \Gamma_{\rho}\psi,
\\
&T^{13}_{\m\n}=\partial_\rho \bar \psi \partial_\sigma\partial_{(\underline{\m}}\Gamma_{\underline{\n})\rho\sigma}\psi, && T^{14}_{\m\n}=\partial_\rho\partial_{(\underline{\m}} \bar \psi \partial_\sigma\Gamma_{\underline{\n})\rho\sigma}\psi,
\fe
Eleven linearly independent combinations out of the fourteen $T^i_{\m\n}$ are descendants (total derivatives),
\ie\label{eqn:desT}
&T^1_{\m\n}+T^2_{\m\n},\quad T^2_{\m\n}+T^3_{\m\n},\quad T^4_{\m\n}+T^5_{\m\n},\quad T^5_{\m\n}+T^6_{\m\n},\quad T^7_{\m\n}+T^8_{\m\n}, 
\\
&  T^8_{\m\n}+T^9_{\m\n},\quad T^{10}_{\m\n}+T^{11}_{\m\n}, \quad T^{11}_{\m\n}+T^{12}_{\m\n},\quad T^5_{\m\n}+T^8_{\m\n},\quad T^2_{\m\n}+T^{11}_{\m\n}, \quad T^{13}_{\m\n} + T^{14}_{\m\n}.
\fe
Hence there are three linearly independent combinations of $T^i_{\m\n}$ that are primary operators, which by conformal symmetry must have vanishing two-point function with all the descendant operators \eqref{eqn:desT}.  To find the correct linear combinations, we consider the two-point functions involving all fourteen $T^i_{\m\n}$,
\ie\label{eqn:TiTj}
\vev{T^i_{\m\n}(x)T^j_{\rho\sigma}(0)}.
\fe
We outline the intermediate steps for this computation.  First, we compute the four-point functions,
\ie\label{eqn:temp4PT}
 \vev{:\bar\psi(x_1)\Gamma_\m\psi(x_2)::\bar\psi(x_3)\Gamma_\n\psi(x_4):} %= \vev{:\bar\psi(x_1)\Gamma\Gamma_\m\psi(x_2)::\bar\psi(x_3)\Gamma\Gamma_\n\psi(x_4):}
%\\
% = {\Tr(\Gamma_\m{\not\!x}_{23}\Gamma_\n {\not\!x}_{14}) \over x^{2\epsilon}_{23}x^{2\epsilon}_{14}} 
&= 2^{\lfloor\epsilon\rfloor+1} {\left(2x_{14,(\m}x_{23,\n)} - \delta_{\m\n}(x_{14}\cdot x_{23})\right)\over x^{2\epsilon}_{23}x^{2\epsilon}_{14}},
\\
 \vev{:\bar\psi(x_1)\Gamma_\m\psi(x_2)::\bar\psi(x_3)\Gamma_{\n\rho\sigma}\psi(x_4):} %= {\Tr(\Gamma_\m{\not\!x}_{23}\Gamma_{\n\rho\sigma} {\not\!x}_{14}) \over x^{2\epsilon}_{23}x^{2\epsilon}_{14}}
%= (2^{\epsilon+1}) {
%2(x_{23,\A} \D^{\A\m}_{\n\rho} x_{14,\sigma} - x_{23,\A} \D^{\A\m}_{\n\sigma} x_{14,\rho} + x_{23,\A} \D^{\A\m}_{\rho\sigma} x_{14,\n})
%\over x^{2\epsilon}_{23}x^{2\epsilon}_{14}}
%\\
%& \quad = (2^{\epsilon+1}) {
%(x_{23,\n} \D_{\m\rho} x_{14,\sigma} - x_{23,\rho} \D_{\m\n} x_{14,\sigma} - x_{23,\n} \D_{\m\sigma} x_{14,\rho} + x_{23,\sigma} \D_{\m\n} x_{14,\rho} + x_{23,\rho} \D_{\m\sigma} x_{14,\n} - x_{23,\sigma} \D_{\m\rho} x_{14,\n})
%\over x^{2\epsilon}_{23}x^{2\epsilon}_{14}}
&= 2^{\lfloor\epsilon\rfloor+1} {
6 \D_{\m[\n} x_{14,\rho} x_{23,\sigma]}
\over x^{2\epsilon}_{23}x^{2\epsilon}_{14}},
\\
\vev{:\bar\psi(x_1)\Gamma_{\n_1\rho_1\sigma_1}\psi(x_2)::\bar\psi(x_3)\Gamma_{\n_2\rho_2\sigma_2}\psi(x_4):} &= {\Tr(\Gamma_{\n_1\rho_1\sigma_1}{\not\!x}_{23}\Gamma_{\n_2\rho_2\sigma_2} {\not\!x}_{14}) \over x^{2\epsilon}_{23}x^{2\epsilon}_{14}}.
\fe
Then the two-point functions \eqref{eqn:TiTj} can be obtained by taking derivatives on \eqref{eqn:temp4PT}, followed by the limit $x_1, x_2 \to x$ and $x_3, x_4 \to 0$.  For $i,j=13,14$, it is convenient to define
\ie
K_{\n_1\n_2} &= {\partial^4 \over \partial x_{1,\rho_1} \partial x_{2,\sigma_1} \partial x_{3,\rho_2} \partial x_{4,\sigma_2}}{\Tr(\Gamma_{\n_1\rho_1\sigma_1}{\not\!x}_{23}\Gamma_{\n_2\rho_2\sigma_2} {\not\!x}_{14}) \over x^{2\epsilon}_{23}x^{2\epsilon}_{14}}
\\
%&=-{1\over (2\epsilon-2)^2}{\partial^6 \over \partial x_{1,\rho_1} \partial x_{2,\sigma_1} \partial x_{2,\rho_2} \partial x_{1,\sigma_2} \partial x_{1,\tau_2} \partial x_{2,\tau_1}}{\Tr(\Gamma_{\n_1\rho_1\sigma_1}\Gamma_{\tau_1}\Gamma_{\n_2\rho_2\sigma_2}\Gamma_{\tau_2}) \over x^{2\epsilon-2}_{23}x^{2\epsilon-2}_{14}}
%\\
%&= - {1\over (2\epsilon-2)^2} \left[ - (\partial_1 \cdot \partial_2)^3 \D_{\n_1\n_2} + 2 (\partial_1 \cdot \partial_2)^2 (\partial_1)_{(\n_1} (\partial_2)_{\n_2)} \right.
%\\
%& \hspace{.5in} \left. + (\partial_1 \cdot \partial_2) ( \partial_1^2 \partial_2^2 \D_{\n_1\n_2} - \partial_1^2 (\partial_2)_{\n_1} (\partial_2)_{\n_2} - \partial_2^2 (\partial_1)_{\n_1} (\partial_1)_{\n_2} ) \right] {1 \over x^{2\epsilon-2}_{23}x^{2\epsilon-2}_{14}}
%\\
&= 2^{\lfloor\epsilon\rfloor+1} \times \Bigg\{ 8 \epsilon^2 \D_{\n_1\n_2} \left[ -3 \epsilon {x_{23} \cdot x_{14} \over x_{23}^{2\epsilon+2} x_{14}^{2\epsilon+2} } + 2 (\epsilon+1)^2 {(x_{23} \cdot x_{14})^3 \over x_{23}^{2\epsilon+4} x_{14}^{2\epsilon+4} } - 2 {(x_{23} \cdot x_{14}) \over x_{23}^{2\epsilon+2} x_{14}^{2\epsilon+2} } \right]
\\
& \qquad\qquad\qquad\qquad + 2 (\partial_1)_{(\n_1} (\partial_2)_{\n_2)} \left[ {2(\epsilon-1) \over x_{23}^{2\epsilon} x_{14}^{2\epsilon}} - 4\epsilon^2 {(x_{23} \cdot x_{14})^2 \over x_{23}^{2\epsilon+2} x_{14}^{2\epsilon+2} } \right]
\\
& \qquad\qquad\qquad\qquad - 4\epsilon \left[ (\partial_2)_{\n_1} (\partial_2)_{\n_2} {x_{23} \cdot x_{14} \over x^{2\epsilon}_{23}x^{2\epsilon+2}_{14}} + (\partial_1)_{\n_1} (\partial_1)_{\n_2} {x_{23} \cdot x_{14} \over x^{2\epsilon+2}_{23}x^{2\epsilon}_{14}} \right] \Bigg\}.
\fe
Then we have
\ie
& \vev{ T^{13}_{\m_1\n_1} T^{13}_{\m_2\n_2} } = (\partial_2)_{(\underline{\m_1}} (\partial_1)_{(\m_2} K_{\n_2)\underline{\n_1})}, \quad \vev{ T^{13}_{\m_1\n_1} T^{14}_{\m_2\n_2} } = (\partial_2)_{(\underline{\m_1}} (\partial_2)_{(\m_2} K_{\n_2)\underline{\n_1})},
\\
& \vev{ T^{14}_{\m_1\n_1} T^{13}_{\m_2\n_2} } = (\partial_1)_{(\underline{\m_1}} (\partial_1)_{(\m_2} K_{\n_2)\underline{\n_1})}, \quad \vev{ T^{14}_{\m_1\n_1} T^{14}_{\m_2\n_2} } = (\partial_1)_{(\underline{\m_1}} (\partial_2)_{(\m_2} K_{\n_2)\underline{\n_1})}.
\fe

The two-point functions \eqref{eqn:TiTj} allow us to identity the three-dimensional space of primary operators as the space orthogonal to the descendants.  In unitary theories, a primary operator with scaling dimension saturating the unitarity bound must be conserved, but this is false in non-unitary theories.  Indeed, using the explicit two-point functions \eqref{eqn:TiTj}, we find that there are two conserved spin-two primaries and one non-conserved spin-two primary.  The stress tensor is a particular linear combination of the two conserved spin-two primaries that satisfies the $T_{\m\n}\psi$ OPE \eqref{FermionOPE}.  A consequence of this OPE is that in the large $x_2$ limit,
\ie\label{eqn:TPsiPsi}
& x_1^\m x_2^\n \vev{ T_{\m\n}(x_1) \bar\psi(x_2) {\not\!x_1} \psi(0) } %&= {2^{\epsilon+1} \Delta_\psi (x_1\cdot x_2)^2\over V_{\widehat{\rm S}^{d-1}} x_1^{d} x_2^{2\Delta_\psi+1}}+ {d\over 4V_{\widehat{\rm S}^{d-1}} x_1^{d}x_2^{2\Delta_\psi+1} } \Tr( {\not\!x_2}  {\not\!x_1} x^\rho_1 x^\n_2\Gamma_{\rho\n})+ {\cal O}(x_2^{-2\Delta_\psi})
%\\
% &= {2^{\epsilon+1} \Delta_\psi (x_1\cdot x_2)^2\over V_{\widehat{\rm S}^{d-1}} x_1^{d} x_2^{2\Delta_\psi+1}}+ {2^{\epsilon+1}d (x_1^2 x_2^2-(x_1\cdot x_2)^2)\over 4V_{\widehat{\rm S}^{d-1}} x_1^{d}x_2^{2\Delta_\psi+1} } + {\cal O}(x_2^{-2\Delta_\psi})
% \\
\\
& \qquad = { 2^{\lfloor\epsilon\rfloor+1} \left(\Delta_\psi -{d\over 4}\right)(x_1\cdot x_2)^2\over V_{\widehat{\rm S}^{d-1}} x_1^{d} x_2^{2\Delta_\psi+1}}+ { 2^{\lfloor\epsilon\rfloor+1} d \over 4V_{\widehat{\rm S}^{d-1}} x_1^{d-2}x_2^{2\Delta_\psi-1} } + {\cal O}(x_2^{-2\Delta_\psi}).
\fe
To find the stress tensor, we compute the three-point functions
\ie\label{eqn:TiPsiPsi}
x_1^\m x_2^\n \vev{ T^i_{\m\n}(x_1) \bar\psi(x_2) {\not\!x_1} \psi(0) },
\fe
and identify the correct linear combination of conserved primaries to match with \eqref{eqn:TPsiPsi}.  This computation can be done by taking derivatives on the three-point functions
\ie
&\vev{:\bar\psi(x_4)\Gamma_\m \psi(x_1): \bar\psi(x_2) {\not\!x_3} \psi(0)} 
%\\
%&= {\Tr\left(\Gamma_\m {\not\!x}_{12}{\not\!x}_{3}{\not\!x}_{4}\right)\over x_{12}^{2\epsilon}x_{4}^{2\epsilon}}
%\\
%&= 2^{\epsilon+1}{x_{4,\m}(x_{12}\cdot x_3)-x_{3,\m}(x_{12}\cdot x_4)+x_{12,\m}(x_{3}\cdot x_4)\over x_{12}^{2\epsilon}x_{4}^{2\epsilon}}
%\\
\\
& \qquad = - 2^{\lfloor\epsilon\rfloor+1} {x_{4,\m}( x_2\cdot x_3)-x_{3,\m}(x_2\cdot x_4)+x_{2,\m}(x_3\cdot x_4)\over x_{2}^{2\epsilon}x_{4}^{2\epsilon}}+{\cal O}(x_2^{-2\epsilon}),
%\\
%&\quad+2^{\epsilon+1}{x_{4,\m}(x_{1}\cdot x_3)-x_{3,\m}(x_{1}\cdot x_4)+x_{1,\m}(x_{3}\cdot x_4)\over x_{2}^{2\epsilon}x_{4}^{2\epsilon}}
%\\
%&\quad-2^{\epsilon+1}{x_{4,\m}( x_2\cdot x_3)-x_{3,\m}(x_2\cdot x_4)+x_{2,\m}(x_3\cdot x_4)\over x_{2}^{2\epsilon+2}x_{4}^{2\epsilon}}(2\epsilon)(x_1\cdot x_2)+{\cal O}(x_2^{-2\epsilon-1})
\\
&\vev{:\bar\psi(x_4)\Gamma_{\m\n\rho} \psi(x_1): \bar\psi(x_2) {\not\!x_3} \psi(0)}
%\\
%&= {\Tr\left(\Gamma_{\m\n\rho} {\not\!x}_{12}{\not\!x}_{3}{\not\!x}_{4}\right)\over x_{12}^{2\epsilon}x_{4}^{2\epsilon}}
%\\
= 2^{\lfloor\epsilon\rfloor+1} { 6x_{2,[\underline{\m}}x_{3,\underline{\n}}x_{4,\underline{\rho}]}\over x_{2}^{2\epsilon}x_{4}^{2\epsilon}}+{\cal O}(x_2^{-2\epsilon}),
\fe
followed by the limit $x_3,x_4\to x_1$.  For example, we have
\ie
x_1^\m x_2^\n\vev{T^{13}_{\m\n}(x_1)\bar\psi(x_2) {\not\!x_1} \psi(0)} = 0+{\cal O}(x_2^{-2\epsilon+1}),
\\
x_1^\m x_2^\n\vev{T^{14}_{\m\n}(x_1)\bar\psi(x_2) {\not\!x_1} \psi(0)} = 0+{\cal O}(x_2^{-2\epsilon+1}).
\fe

In four spacetime dimensions, the stress tensor $T_{\m\n}$, the spin-two conserved primary $\widetilde T_{\m\n}$ orthogonal to $T_{\m\n}$, and the spin-two non-conserved primary $\Theta_{\m\n}$ orthogonal to both $T_{\m\n}$ and $\widetilde T_{\m\n}$ are
\ie
&T_{\m\n}={1\over 48\pi^2}\big(-2T^1_{\m\n}-T^2_{\m\n}+7T^3_{\m\n}+3T^4_{\m\n}+9T^5_{\m\n}+9T^6_{\m\n}-9T^7_{\m\n}
\\
&\quad\quad\quad\quad\quad~\,-9T^8_{\m\n}-3T^9_{\m\n}-7T^{10}_{\m\n}+T^{11}_{\m\n}+2T^{12}_{\m\n}+3T^{13}_{\m\n}-3T^{14}_{\m\n}\big),
\\
&\widetilde T_{\m\n}=5T^1_{\m\n}+2T^2_{\m\n}-17T^3_{\m\n}-7T^4_{\m\n}-22T^5_{\m\n}-21T^6_{\m\n}+21T^7_{\m\n}
\\
&\quad\quad\quad\quad\quad~\,+22T^8_{\m\n}+7T^9_{\m\n}+17T^{10}_{\m\n}-2T^{11}_{\m\n}-5T^{12}_{\m\n}-8T^{13}_{\m\n}+8T^{14}_{\m\n},
\\
&\Theta_{\m\n}=2T^1_{\m\n}-3T^2_{\m\n}-5T^3_{\m\n}-2T^4_{\m\n}-9T^5_{\m\n}-7T^6_{\m\n}+7T^7_{\m\n}
\\
&\quad\quad\quad\quad\quad~\,+9T^8_{\m\n}+2T^9_{\m\n}+5T^{10}_{\m\n}+3T^{11}_{\m\n}-2T^{12}_{\m\n}-3T^{13}_{\m\n}+3T^{14}_{\m\n}.
\fe
In six spacetime dimensions, they are
\ie
&T_{\m\n}={1\over 80\pi^3}\big(-2T^1_{\m\n}-4T^2_{\m\n}+8T^3_{\m\n}+5T^4_{\m\n}+10T^5_{\m\n}+10T^6_{\m\n}-10T^7_{\m\n}
\\
&\quad\quad\quad\quad\quad~\,-10T^8_{\m\n}-5T^9_{\m\n}-8T^{10}_{\m\n}+4T^{11}_{\m\n}+2T^{12}_{\m\n}+5T^{13}_{\m\n}-5T^{14}_{\m\n}\big),
\\
&\widetilde T_{\m\n}=2T^1_{\m\n}+2T^2_{\m\n}-6T^3_{\m\n}-3T^4_{\m\n}-8T^5_{\m\n}-6T^6_{\m\n}+6T^7_{\m\n}
\\
&\quad\quad\quad\quad\quad~\,+8T^8_{\m\n}+3T^9_{\m\n}+6T^{10}_{\m\n}-2T^{11}_{\m\n}-2T^{12}_{\m\n}-5T^{13}_{\m\n}+5T^{14}_{\m\n},
\\
&\Theta_{\m\n}=8T^1_{\m\n}-5T^2_{\m\n}-13T^3_{\m\n}-8T^4_{\m\n}-25T^5_{\m\n}-17T^6_{\m\n}+17T^7_{\m\n}
\\
&\quad\quad\quad\quad\quad~\,+25T^8_{\m\n}+8T^9_{\m\n}+13T^{10}_{\m\n}+5T^{11}_{\m\n}-8T^{12}_{\m\n}-15T^{13}_{\m\n}+15T^{14}_{\m\n}.
\fe
We can read off the central charge $C_T$ from the two-point function \eqref{eqn:TTtpf} of the stress tensor $T_{\m\n}$. In four spacetime dimensions, we find
\ie
C_T = -{8\over 3} \quad \text{(4d three-derivative Dirac fermion)}.
\fe
In six spacetime dimensions, we find
\ie
C_T=- {144 \over 5} \quad \text{(6d three-derivative Dirac fermion)}.
\fe
These values are in agreement with \cite{Fradkin:1981jc,Fradkin:1985am,Beccaria:2017dmw}.

\section{Analytic examples of solutions to crossing}

We write down two analytic solutions to the superconformal crossing equation \eqref{crossing13}, using first generalized free fields (mean field theory) and second a free hypermultiplet.  Since these solutions exist in arbitrary spacetime dimensions, we keep $d = 2\epsilon+2$ general.

\subsection{Generalized free fields}
\label{Sec:GFF}

The four-point function of generalized free fields is the sum over factorized two-point functions in the three channels (mean field theory),
\ie
&\vev{{\cal O}(X_1,Y_1){\cal O}(X_2,Y_2){\cal O}(X_3,Y_3){\cal O}(X_4,Y_4)}
%\\
%&=\left({(Y_1\cdot Y_2)(Y_3\cdot Y_4)\over (X_1\cdot X_2)^{\epsilon}(X_3\cdot X_4)^{\epsilon}}\right)^{k}+\left({(Y_1\cdot Y_3)(Y_2\cdot Y_4)\over (X_1\cdot X_3)^{\epsilon}(X_2\cdot X_4)^{\epsilon}}\right)^{k}+\left({(Y_1\cdot Y_4)(Y_2\cdot Y_3)\over (X_1\cdot X_4)^{\epsilon}(X_2\cdot X_3)^{\epsilon}}\right)^{k}
%\\
&=\left({(Y_1\cdot Y_2)(Y_3\cdot Y_4)\over x_{12}^{2\epsilon}x_{34}^{2\epsilon}}\right)^{k}G_{\text{\tiny gff}}(u,v;w),
\fe
where $G_{\text{\tiny gff}}(u,v;w)$ is given by
\ie\label{eqn:Ggff}
G_{\text{\tiny gff}}(u,v;w)=1+\left((1+w)u^{\epsilon}\over w\right)^k+\left(u^\epsilon\over wv^\epsilon \right)^k.
\fe
It satisfies the crossing equations \eqref{crossing13} and \eqref{crossing12} by construction.  Let us focus on $k=2$, and decompose $G_{\text{\tiny gff}}(u,v;w)$ into $\mathfrak{su}(2)_R$ harmonics,
\ie
G_{\text{\tiny gff}}(u,v;w) &= \left[1+{1\over 3}\left(u^{2\epsilon}+{u^{2\epsilon}\over v^{2\epsilon}}\right)\right]+ {1\over 2}\left[u^{2\epsilon}-{u^{2\epsilon}\over v^{2\epsilon}}\right]P_1(1+\tfrac{2}{ w})
\\
& \quad +{1\over 6}\left[u^{2\epsilon}+{u^{2\epsilon}\over v^{2\epsilon}}\right]P_2(1+\tfrac{2}{ w}).
\fe
The factors multiplying $P_{J_R}(1+\tfrac{2}{ w})$ can be decomposed into bosonic conformal blocks,
\ie
&u^{2\epsilon}+{u^{2\epsilon}\over v^{2\epsilon}} = \sum_{\ell\,{\rm even},\,\ell\ge 0}\sum_{n=0}^{\infty}P_{n,\ell}{\cal G}_{\ell+4\epsilon+2n,\ell}(u,v),
\\
&u^{2\epsilon}-{u^{2\epsilon}\over v^{2\epsilon}} = \sum_{\ell\,{\rm odd},\,\ell\ge 1}\sum_{n=0}^{\infty}P_{n,\ell}{\cal G}_{\ell+4\epsilon+2n,\ell}(u,v),
\fe
where the coefficients $P_{n,\ell}$ are given in\cite{Fitzpatrick:2011dm}
\ie
& P_{n,\ell} = {(\ell+\epsilon)_\epsilon\over (\epsilon)_\epsilon}
\\
& \quad \times {2(-1)^\ell(\epsilon)^2_n (2\epsilon)^2_{n+\ell}\over \Gamma(\ell+1) \Gamma(n+1)(\ell+\epsilon+1)_n(2\epsilon+n-1)_n (4\epsilon+2n+\ell-1)_\ell(3\epsilon+n+\ell-1)_n}.
\fe
$G_{\text{\tiny gff}}(u,v;w)$ can also be decomposed into superconformal blocks,
\ie
G_{\text{\tiny gff}}(u,v;w) &= 1+\lambda^2_{{\cal D}[4]} {\cal A}_{{\cal D}[4]} +\sum_{\ell\,{\rm odd},\,\ell\ge 1}\lambda^2_{{\cal B}[2]_{\ell}} {\cal A}_{{\cal B}[2]_{\ell}}
\\
&\hspace{1in} + \sum_{\ell\,{\rm even},\,\ell\ge 0}\sum_{n=0}^\infty\lambda^2_{{\cal L}[0]_{\ell+4\epsilon+2n,\ell}} {\cal A}_{{\cal L}[0]_{\ell+4\epsilon+2n,\ell}},
\fe
where the OPE coefficients are given by
\ie
& \lambda^2_{{\cal D}[4]}={P_{0,0}\over 6}={1\over 3}, \quad \lambda^2_{{\cal B}[2]_{\ell}}={\ell+1\over 8(\ell+2\epsilon)}P_{0,\ell+1},
\\
& \lambda^2_{{\cal L}[0]_{\ell+4\epsilon+2n,\ell}}={(n+1)(n+L+\epsilon+1)\over 16(n+\epsilon)(n+L+2\epsilon)}P_{n+1,\ell}.
\fe
We note that the decomposition of the generalized free field solution is void of any conserved current.

\subsection{Free hypermultiplet}
\label{App:FreeHyper}

A free hypermultiplet consists of a pair of complex scalars transforming in the fundamental representation of $\mathfrak{su}(2)_R$, and a fermion singlet.  The fermion could be Dirac, Majorana, or Weyl depending on the number of spacetime dimensions; in six dimensions, it is a Weyl fermion.  Let us denote the complex scalar doublet by $\phi_A$, and $\bar\phi^A$ its complex conjugate $\bar\phi^A=(\phi_A)^*$.  They are normalized by the two-point function
\ie
\vev{\phi_{A_1}(x_1)\bar\phi^{A_2}(x_2)}={\delta^{A_2}_{A_1}\over x_{12}^{2\epsilon}}.
\fe

The superconformal primaries of a $D[2]$ superconformal multiplet have scaling dimension $2\epsilon$, and can be constructed as scalar bilinears 
\ie
\bar\phi^A (\sigma^a)_A^{~B} \phi_B,
\fe
where $A,B = 1,2$, $a = 1,2,3$, and $(\sigma^a)_A^{~B}$ are the Pauli matrices. To keep track of $\mathfrak{su}(2)_R$, we can contract the scalars with auxiliary variables $Y^A$, and consider the four-point function of ${\cal O} \equiv i\phi_A \bar\phi^B Y^A Y_B$,\footnote{The indices can be raised and lowered by $Y_A = \epsilon_{AB} Y^B$ and $Y^A=Y_B\epsilon^{BA}$.}
\ie
& \langle {\cal O}(x_1, Y_1) {\cal O}(x_2, Y_2) {\cal O}(x_3, Y_3) {\cal O}(x_4, Y_4) \rangle
%\\
%&= 2{(Y_1\cdot Y_2)(Y_2\cdot Y_3)(Y_3\cdot Y_4)(Y_4\cdot Y_1)\over x_{12}^{2\epsilon}x_{23}^{2\epsilon}x_{34}^{2\epsilon}x_{41}^{2\epsilon}}+2{(Y_1\cdot Y_2)(Y_2\cdot Y_4)(Y_4\cdot Y_3)(Y_3\cdot Y_1)\over x_{12}^{2\epsilon}x_{24}^{2\epsilon}x_{43}^{2\epsilon}x_{31}^{2\epsilon}}
%\\
%& \hspace{1in} +2{(Y_1\cdot Y_4)(Y_4\cdot Y_2)(Y_2\cdot Y_3)(Y_3\cdot Y_1)\over x_{14}^{2\epsilon}x_{42}^{2\epsilon}x_{23}^{2\epsilon}x_{31}^{2\epsilon}}
%\\
%& \hspace{.5in} +{(Y_1\cdot Y_2)^2(Y_3\cdot Y_4)^2\over x_{12}^{4\epsilon}x_{34}^{4\epsilon}}+{(Y_1\cdot Y_3)^2(Y_2\cdot Y_4)^2\over x_{13}^{4\epsilon}x_{24}^{4\epsilon}}+{(Y_1\cdot Y_4)^2(Y_2\cdot Y_3)^2\over x_{14}^{4\epsilon}x_{23}^{4\epsilon}}
%\\
&=\left({(Y_1\cdot Y_2)(Y_3\cdot Y_4)\over x_{12}^{2\epsilon}x_{34}^{2\epsilon}}\right)^{2} G_{\text{\tiny hyper}}(u,v,w),
\fe
where $G_{\text{\tiny hyper}}(u,v,w)$ is given by 
\ie
G_{\text{\tiny hyper}}(u,v,w)&=G_{\text{\tiny gff}}(u,v,w)+G_{\text{\tiny extra}}(u,v,w),
\\
G_{\text{\tiny extra}}(u,v,w)&= - {2 \over w} \left( u \over v \right)^{\epsilon} + 2 \left( 1 + {1 \over w} \right) u^{\epsilon} + {2 \over w} \left( 1 + {1 \over w} \right) \left( u^2 \over v \right)^{\epsilon}.
\fe

A single free hypermultiplet has SU(2) flavor symmetry.\footnote{The two complex scalars, regarded as four real scalars, can be rotated by an SO(4) action which is a direct sum of the SU(2)$_R$ R-symmetry and the SU(2) flavor symmetry. The Weyl spinor in six dimensions admits a quaternionic structure, and also transforms as a doublet under the SU(2) flavor symmetry.} We can construct a triplet of $D[2]$ superconformal primaries,
\ie
{\cal O}_1={i\over 2}({\cal O}_+-{\cal O}_-),\quad {\cal O}_3=-{1\over 2}({\cal O}_++{\cal O}_-),\quad {\cal O}_2\equiv {\cal O}=i\phi_A \bar\phi^B Y^A Y_B,
\fe
where ${\cal O}_+=\phi_A\phi_B Y^A Y^B$ and ${\cal O}_-=\bar \phi^A\bar\phi^B Y_A Y_B$.
The SU(2) flavor symmetry can be made manifest by introducing new auxiliary variables $\widetilde Y_A$, and defining
\ie
{\cal O}(x,Y,\widetilde Y)=i(\sigma^a)_{\dot A}{}^{\dot B}\widetilde Y^{\dot A}\widetilde Y_{\dot B}{\cal O}_a(x,Y).
\fe
One can write
\ie
{\cal O}(x,Y,\widetilde Y)=(\phi_{ \dot A A}\widetilde Y^{\dot A}Y^A)^2,
\fe
where $\phi_A$ and $\bar \phi_A$ are expressed in terms of $\phi_{ \dot A A}$ as\footnote{The scalar $\phi_{ \dot A A}$ satisfies the reality condition $(\phi_{ \dot A A})^*=\phi^{ \dot A A}$.}
\ie
\phi_A={1\over\sqrt{2}}(\phi_{1A}+i\phi_{2A}),\quad \bar \phi_A={1\over\sqrt{2}}(-\phi_{2A}-i\phi_{1A}).
\fe
It is now straightforward to compute the four-point function,
\ie
&\langle {\cal O}(x_1, Y_1,\widetilde Y_1) {\cal O}(x_2, Y_2,\widetilde Y_2) {\cal O}(x_3, Y_3,\widetilde Y_3) {\cal O}(x_4, Y_4,\widetilde Y_4) \rangle
\\
&= {(  Y_1\cdot Y_2)^2(Y_3\cdot Y_4)^2(\widetilde Y_1\cdot\widetilde Y_2)^2(\widetilde Y_3\cdot\widetilde Y_4)^2\over x_{12}^{4\epsilon}x_{34}^{4\epsilon}} G_{\text{\tiny hyper}}(u,v,w,\widetilde w),
\fe
where $\widetilde w$ is defined the same way as $w$, and $G_{\text{\tiny hyper}}(u,v,w,\widetilde w)$ is given by 
\ie
G_{\text{\tiny hyper}}(u,v,w,\widetilde w)&=4+4\left((1+w)(1+\widetilde w)u^{\epsilon}\over w\widetilde w\right)^2+4\left(u^\epsilon\over w\widetilde wv^\epsilon \right)^2
\\
&\quad + {16 \over w\widetilde w} \left( u \over v \right)^{\epsilon} + 16 {(1+w)(1+\widetilde w)\over w\widetilde w} u^{\epsilon} + 16{(1+w)(1+\widetilde w)\over w^2\widetilde w^2} \left( u^2 \over v \right)^{\epsilon}.
\fe
The four-point function $G_{\text{\tiny hyper}}(u,v,w,\widetilde w)$ can be further decomposed into Legendre polynomials,
\ie\label{eqn:GhyperSU2}
&G_{\text{\tiny hyper}}(u,v,w,\widetilde w)
\\
&=4\bigg\{P_0(1+\tfrac{2}{\widetilde w})P_0(1+\tfrac{2}{w})\left[1+{1\over 9}\left(u^{2\epsilon}+\left(u\over v\right)^{2\epsilon}\right)+{1\over 9}\left(9u^\epsilon+9\left(u\over v\right)^\epsilon+\left(u^2\over v\right)^\epsilon\right)\right]
\\
&\qquad\qquad +P_1(1+\tfrac{2}{\widetilde w})P_1(1+\tfrac{2}{w})\left[{1\over 4}\left(u^{2\epsilon}+\left(u\over v\right)^{2\epsilon}\right)+\left(u^\epsilon+\left(u\over v\right)^\epsilon\right)\right]
\\
&\qquad\qquad +P_2(1+\tfrac{2}{\widetilde w})P_2(1+\tfrac{2}{w})\left[{1\over 36}\left(u^{2\epsilon}+\left(u\over v\right)^{2\epsilon}\right)+{1\over 9}\left(u^2\over v\right)^\epsilon\right]\bigg\}
\\
&\qquad +4\bigg\{P_0(1+\tfrac{2}{\widetilde w})P_1(1+\tfrac{2}{w})\left[{1\over 6}\left(u^{2\epsilon}-\left(u\over v\right)^{2\epsilon}\right)+\left(u^\epsilon-\left(u\over v\right)^\epsilon\right)\right]
\\
&\qquad\qquad\qquad +P_0(1+\tfrac{2}{\widetilde w})P_2(1+\tfrac{2}{w})\left[{1\over 18}\left(u^{2\epsilon}+\left(u\over v\right)^{2\epsilon}\right)-{1\over 9}\left(u^2\over v\right)^\epsilon\right]
\\
&\qquad\qquad\qquad +P_1(1+\tfrac{2}{\widetilde w})P_2(1+\tfrac{2}{w})\left[{1\over 12}\left(u^{2\epsilon}-\left(u\over v\right)^{2\epsilon}\right)\right]+(w\leftrightarrow\widetilde w)\bigg\}.
\fe
If we define
\ie
G_{\text{\tiny hyper}}(u,v,w,\widetilde w)
=G^{a_1 a_2 a_3 a_4}_{\text{\tiny hyper}}(u,v,w){\bf Y}_{1,a_1}{\bf Y}_{2,a_2}{\bf Y}_{3,a_3}{\bf Y}_{4,a_4},
\fe
where ${\bf Y}^a \equiv i(\sigma^a)_{\dot A}{}^{\dot B}\widetilde Y^{\dot A}\widetilde Y_{\dot B}$, then $G^{a_1 a_2 a_3 a_4}_{\text{\tiny hyper}}(u,v,w)$ admits a superconformal block decomposition,\footnote{The SU(2) projection matrices are
\ie
&P_{\bf 1}^{abcd}={1\over 3}\delta^{ab}\delta^{cd},\quad P_{\bf 3}^{abcd}={1\over 2}(\delta^{ad}\delta^{bc}-\delta^{ac}\delta^{bd}),\quad P_{\bf 5}^{abcd}={1\over 2}(\delta^{ad}\delta^{bc}+\delta^{ac}\delta^{bd})-{1\over 3}\delta^{ab}\delta^{cd},
\fe
which are related to Legendre polynomials by
\ie
&P_{\bf 1}^{a_1 a_2 a_3 a_4}{\bf Y}_{1,a_1}{\bf Y}_{2,a_2}{\bf Y}_{3,a_3}{\bf Y}_{4,a_4}={4\over 3}(\widetilde Y_1\cdot\widetilde Y_2)^2(\widetilde Y_3\cdot\widetilde Y_4)^2P_0(1+\tfrac{2}{\widetilde w}),
\\
&P_{\bf 3}^{a_1 a_2 a_3 a_4}{\bf Y}_{1,a_1}{\bf Y}_{2,a_2}{\bf Y}_{3,a_3}{\bf Y}_{4,a_4}=-2(\widetilde Y_1\cdot\widetilde Y_2)^2(\widetilde Y_3\cdot\widetilde Y_4)^2P_1(1+\tfrac{2}{\widetilde w}),
\\
&P_{\bf 5}^{a_1 a_2 a_3 a_4}{\bf Y}_{1,a_1}{\bf Y}_{2,a_2}{\bf Y}_{3,a_3}{\bf Y}_{4,a_4}={2\over 3}(\widetilde Y_1\cdot\widetilde Y_2)^2(\widetilde Y_3\cdot\widetilde Y_4)^2P_2(1+\tfrac{2}{\widetilde w}).
\fe
}
\ie
G^{a_1 a_2 a_3 a_4}_{\text{\tiny hyper}}(u,v,w)=\sum_{i\in\{{\bf 1},{\bf 3},{\bf 5}\}}P_{i}^{a_1 a_2 a_3 a_4}\lambda_{{\cal X},i}^2{\cal A}_{\cal X}(u,v,w).
\fe
In six dimensions, the low-lying nonzero OPE coefficients are
\ie
\label{HyperOPE}
{\bf 1}: \quad &\lambda_{{\cal D}[0],{\bf 1}}^2=3,\quad \lambda_{{\cal B}[0]_{0},{\bf 1}}^2=6,\quad\lambda_{{\cal B}[0]_{2},{\bf 1}}^2={60\over 7},\quad\lambda_{{\cal B}[0]_{4},{\bf 1}}^2={35\over 11},
\\
&\lambda_{{\cal B}[2]_{1},{\bf 1}}^2={8\over 3},\quad \lambda_{{\cal B}[2]_{3},{\bf 1}}^2={62\over 13},\quad\lambda_{{\cal B}[2]_{5},{\bf 1}}^2={256\over 85},
\\
&\lambda_{{\cal L}[0]_{8,0},{\bf 1}}^2=1,\quad \lambda_{{\cal L}[0]_{10,0},{\bf 1}}^2={4\over 15},\quad\lambda_{{\cal L}[0]_{10,2},{\bf 1}}^2={265\over 66},\quad\cdots,
\\
{\bf 3}: \quad &\lambda_{{\cal D}[2],{\bf 3}}^2=4,\quad \lambda_{{\cal B}[0]_{1},{\bf 3}}^2={32\over 5},\quad\lambda_{{\cal B}[0]_{3},{\bf 3}}^2={80\over 21},\quad\lambda_{{\cal B}[0]_{5},{\bf 3}}^2={448\over 429},
\\
&\lambda_{{\cal B}[2]_{0},{\bf 3}}^2=1,\quad \lambda_{{\cal B}[2]_{2},{\bf 3}}^2={50\over 11},\quad\lambda_{{\cal B}[2]_{4},{\bf 3}}^2={490\over 117},
\\
 &\lambda_{{\cal L}[0]_{9,1},{\bf 3}}^2={16\over 7},\quad \lambda_{{\cal L}[0]_{11,1},{\bf 3}}^2={10\over 9},\quad\lambda_{{\cal L}[0]_{11,3},{\bf 3}}^2={500\over 117},\quad\cdots,
\\
{\bf 5}: \quad &\lambda_{{\cal D}[4],{\bf 5}}^2=1,\quad\lambda_{{\cal B}[2]_{1},{\bf 5}}^2={32\over 9},\quad\lambda_{{\cal L}[0]_{10,0},{\bf 5}}^2={64\over 105},\quad \lambda_{{\cal L}[0]_{10,2},{\bf 5}}^2={10\over 3},\quad\cdots.
\fe

An observation on the above decomposition is the absence of ${\cal B}[0]_{\ell,{\bf 5}}$, which are allowed in general free theories.  This property of the hypermultiplet can be explained as follows.  In order to satisfy the relation between $\Delta$ and $\ell$, the superconformal primary of a ${\cal B}[0]_{\ell}$ multiplet must take the form of a scalar bilinear.  Since a scalar transforms in {\bf 2}, scalar bilinears can only transform in {\bf 1} and {\bf 3}.

\section{Details on numerics}
\label{App:Numerics}

We comment on the parameter settings in the practical implementation of the linear functional method.  The most relevant parameters include the derivative order $\Lambda$, the truncation on spins $\ell_{max}$, and the order $n_r$ to which the $r$ expansion of the superconformal blocks (see Appendix~\ref{App:Recur}) are truncated.  For fixed $\Lambda$, we in principle need to extrapolate to infinite $n_r$ and $\ell_{max}$ to obtain rigorous bounds.  However, in practice, we find that if we set $n_r \geq 2\Lambda$, then the bounds are stable to within numerical precision against further increases in $n_r$.  The numerical bounds in this paper are obtained using $\ell_{max} = 64$, $n_r = 80$ for $\Lambda \leq 40$ and $n_r = 96$ for $40 < \Lambda \leq 48$.  The relevant parameter settings for the SDPB package are 
\ie
& \texttt{precision = 1024},
\\
& \texttt{initialMatrixScalePrimal = initialMatrixScaleDual = 1e20},
\\
& \texttt{dualityGapThreshold = 1e-10}.
\fe

\begin{figure}[h]
\centering
\subfloat{
\includegraphics[width=.47\textwidth]{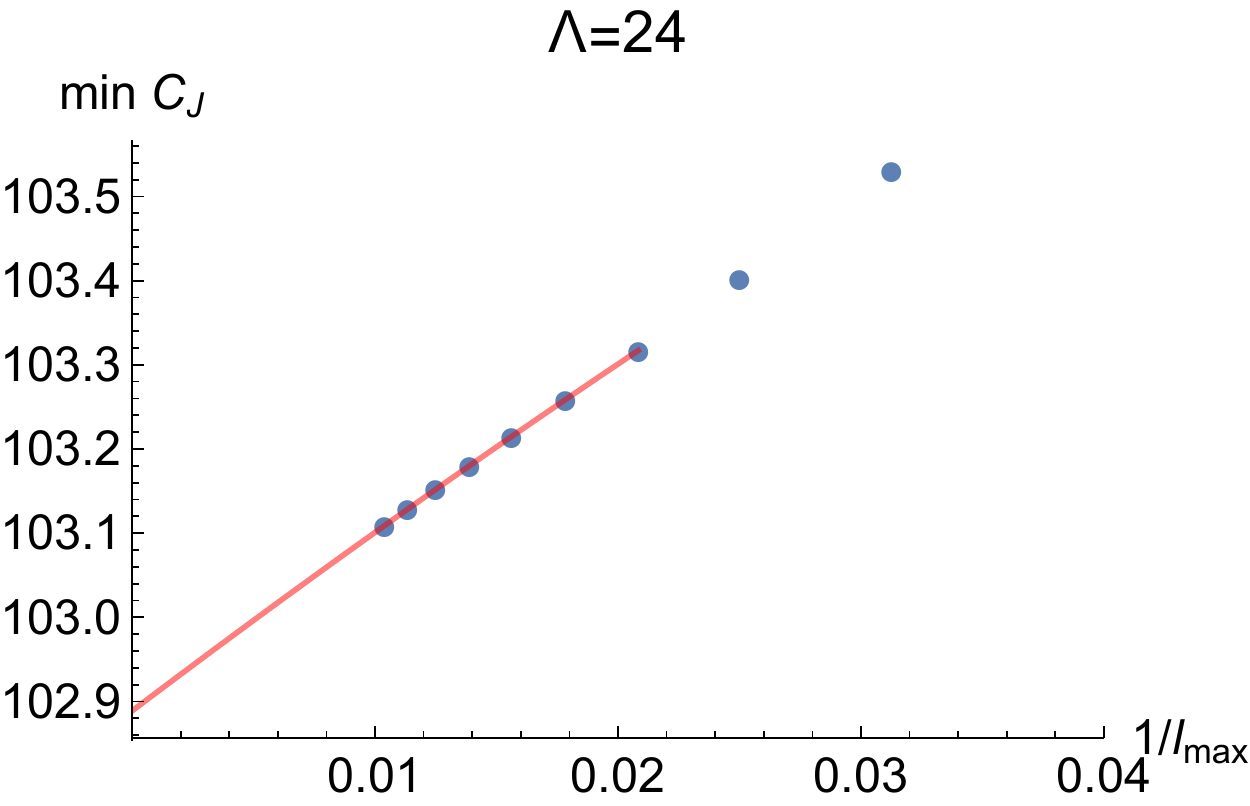}
}
\quad
\subfloat{
\includegraphics[width=.47\textwidth]{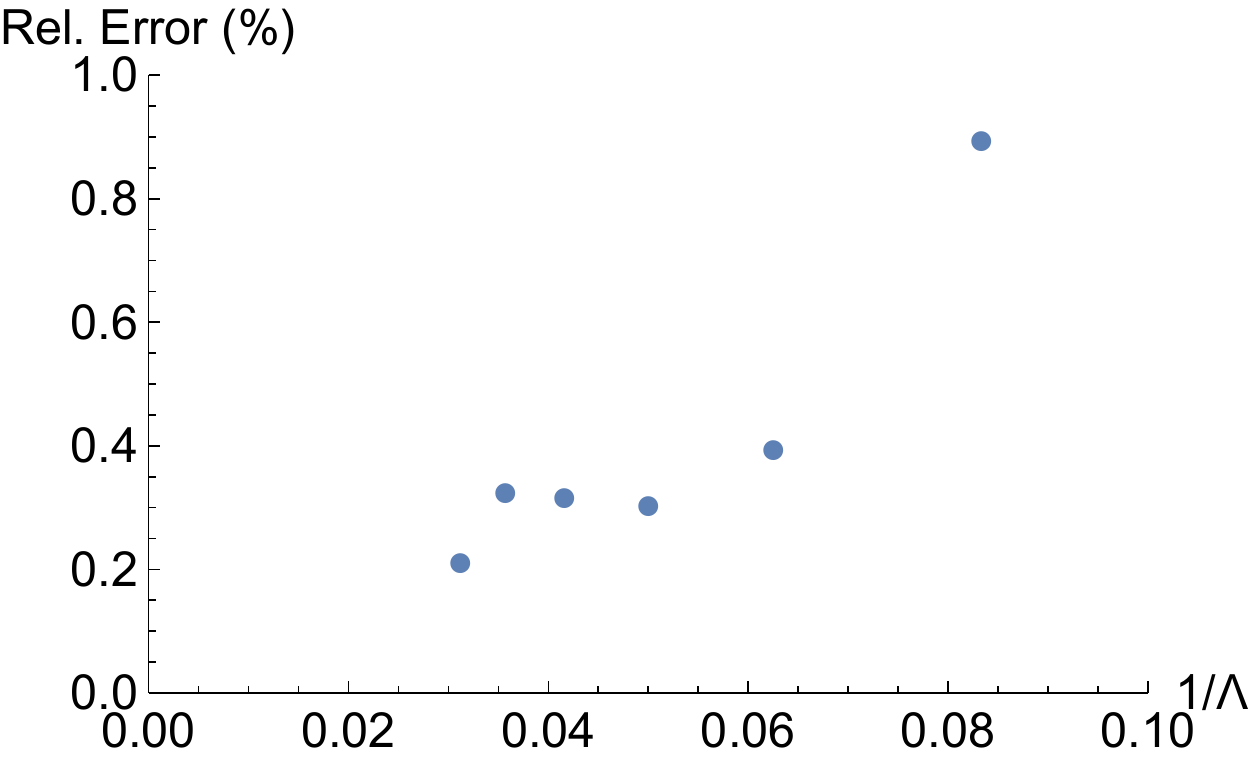}
}
\caption{{\bf Left:} The lower bounds on $C_J$ for interacting theories with $E_8$ flavor group, at derivative order $\Lambda=24$ and across a range of spin truncations $\ell_{max}$.  Also shown is an extrapolation to $\ell_{max} \to \infty$ using the quadratic ansatz \eqref{SpinAnsatz}.  {\bf Right:} The relative errors between $\ell_{max} = 64$ and the extrapolations to $\ell_{max} \to \infty$, at different $\Lambda$.}
\label{Fig:RelError}
\end{figure}

In the past, the weakening of the bounds with increasing $\ell_{max}$ has been handled by imposing non-negativity conditions on functionals acted on a few blocks of very high spin (such as $\ell = 1000, 1001$ in \cite{Poland:2011ey}), in addition to blocks below some $\ell_{max}$.\footnote{In the case of bootstrapping conformal field theories in two spacetime dimensions with an infinite dimensional chiral algebra, it was found that the bounds are stable against increasing $\ell_{max}$ to within numerical precision, as soon as $\ell_{max}$ exceeds some threshold, say $2\Lambda$ \cite{Lin:2015wcg,Collier:2016cls,Lin:2016gcl,Collier:2017shs}.  Some comments on the issue of spin stabilization in higher spacetime dimensions can be found in \cite{Caracciolo:2014cxa}.
}  We find that this approach does not make our bounds stable against increasing $\ell_{max}$.  But numerical extrapolations to infinite $\ell_{max}$ require data with a large range of $\ell_{max}$ for each derivative order, which is computationally intensive and impractical.\footnote{When the flavor group has a large crossing matrix, to run jobs at very high derivative order $\Lambda$ and very high spin truncation $\ell_{max}$, the required RAM for running SDPB exceeds the limitations of the machines we have at hand.
}  Our strategy is to use $\ell_{max} = 64$, and estimate the errors by performing the extrapolations to infinite $\ell_{max}$ in simpler cases.  We shall consider $E_8$ flavor in the absence of higher spin conserved currents.  The left side of Figure~\ref{Fig:RelError} shows the extrapolations for the lower bound on $C_J$ at derivative order $\Lambda = 24$, and the right side shows the relative error between $\ell_{max} = 64$ and extrapolations to $\ell_{max} \to \infty$ using the quadratic ansatz
\ie
\label{SpinAnsatz}
{\rm min} \, C_T = a + {b\over\ell_{max}} + {c\over\ell_{max}^2}, \quad 48 \leq \ell_{max} \leq 96,
\fe
obtained at various derivative orders.  We see that the relative error decreases to below $0.5\%$ as we go to high enough derivative orders.

\begin{figure}[h]
\centering
\subfloat{
\includegraphics[width=.47\textwidth]{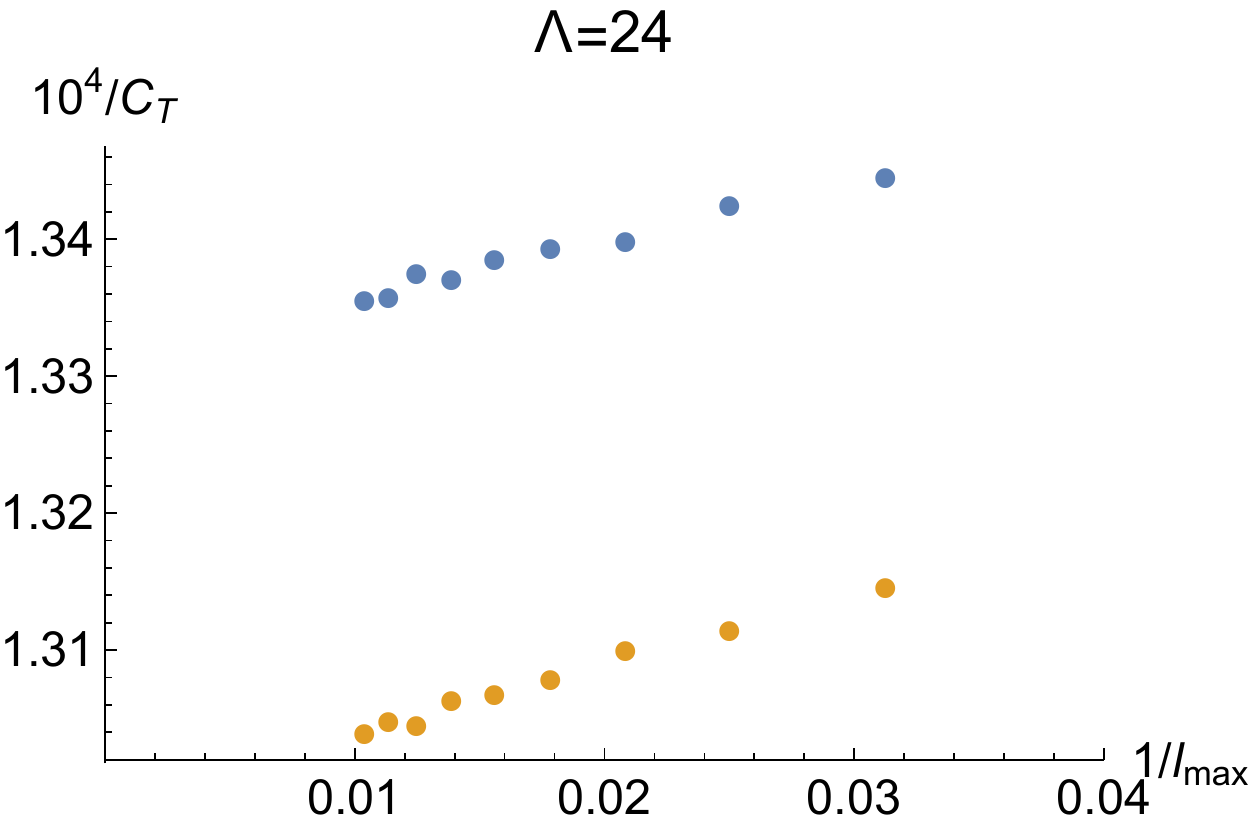}
\quad
\includegraphics[width=.47\textwidth]{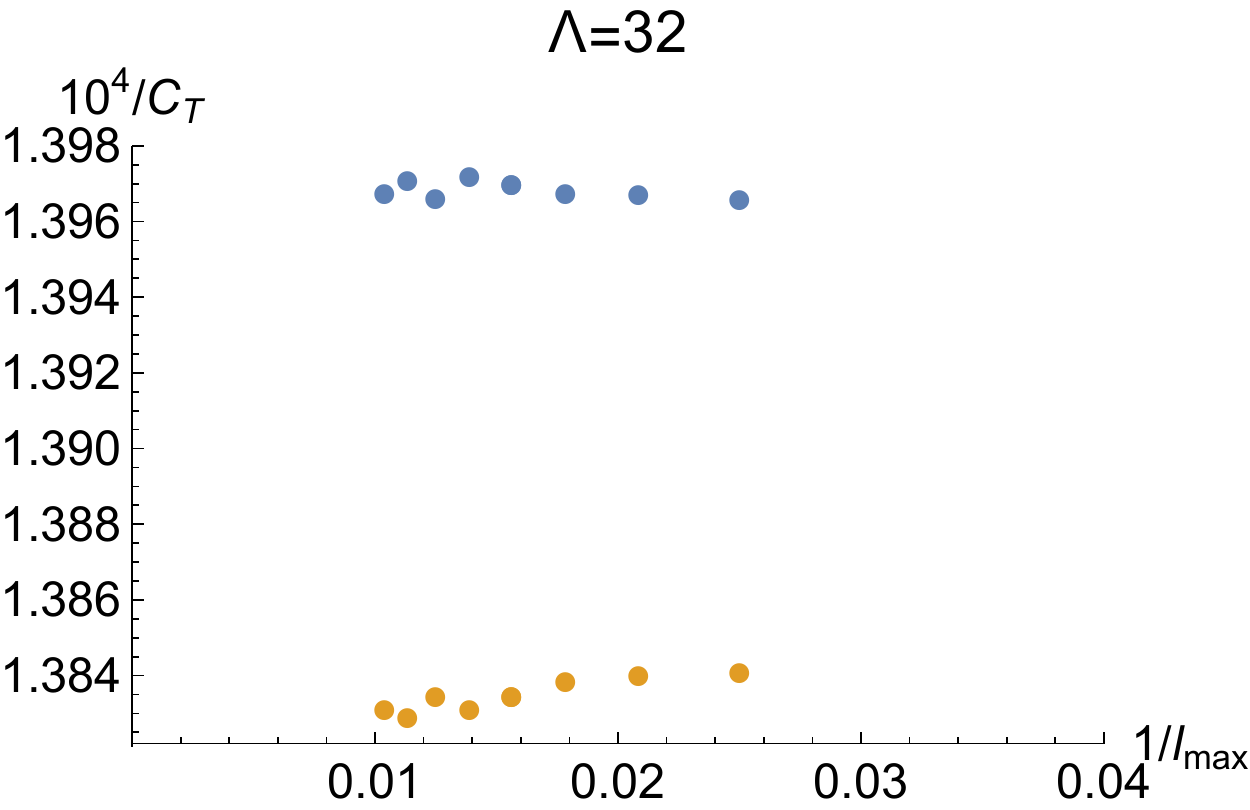}
}
\caption{The upper and lower bounds on the inverse central charge $C_T^{-1}$ when the value of the flavor central charge $C_J$ is set close to saturating the lower bound, $C_J = (1+10^{-4}) \, {\rm min}\,C_J$, for interacting theories with $E_8$ flavor group, at derivative orders $\Lambda=24,32$ and across a range of spin truncations $\ell_{max}$.}
\label{Fig:SpinTruncation}
\end{figure}

In light of the slight discrepancy between the value of $C_T$ at min $C_J$ and the rank-one E-string, as shown in Figure~\ref{Fig:CTMinCJ}, we estimate its error due to spin truncation.  Figure~\ref{Fig:SpinTruncation} shows the upper and lower bounds on $C_T$, when the value of the flavor central charge $C_J$ is set close to saturating the lower bound, $C_J = (1+10^{-4}) \, {\rm min}\,C_J$, at derivative order $\Lambda = 24,32$ and across a range of spin truncations $\ell_{max}$.  The data appears less regular than that for min $C_J$, and extrapolations using the ansatz \eqref{SpinAnsatz} do not look reliable, but we estimate that the error due to truncating spins to $\ell_{max} = 64$ is less than 2\% for $\Lambda\ge 24$.  Similar to min $C_J$, this error decreases with increasing derivative order.

\bibliography{refs} 

\providecommand{\href}[2]{#2}\begingroup\raggedright\begin{thebibliography}{100}

\bibitem{Gaiotto:2009hg}
D.~Gaiotto, G.~W. Moore and A.~Neitzke, \emph{{Wall-crossing, Hitchin Systems,
  and the WKB Approximation}},  \href{http://arxiv.org/abs/0907.3987}{{\tt
  0907.3987}}.

\bibitem{Gaiotto:2009we}
D.~Gaiotto, \emph{{N=2 dualities}},
  \href{http://dx.doi.org/10.1007/JHEP08(2012)034}{\emph{JHEP} {\bf 08} (2012)
  034}, [\href{http://arxiv.org/abs/0904.2715}{{\tt 0904.2715}}].

\bibitem{Apruzzi:2016rny}
F.~Apruzzi, G.~Dibitetto and L.~Tizzano, \emph{{A new 6d fixed point from
  holography}}, \href{http://dx.doi.org/10.1007/JHEP11(2016)126}{\emph{JHEP}
  {\bf 11} (2016) 126}, [\href{http://arxiv.org/abs/1603.06576}{{\tt
  1603.06576}}].

\bibitem{Ooguri:2016pdq}
H.~Ooguri and C.~Vafa, \emph{{Non-supersymmetric AdS and the Swampland}},
  \href{http://arxiv.org/abs/1610.01533}{{\tt 1610.01533}}.

\bibitem{Minwalla:1997ka}
S.~Minwalla, \emph{{Restrictions imposed by superconformal invariance on
  quantum field theories}}, {\emph{Adv. Theor. Math. Phys.} {\bf 2} (1998)
  781--846}, [\href{http://arxiv.org/abs/hep-th/9712074}{{\tt
  hep-th/9712074}}].

\bibitem{Bhattacharya:2008zy}
J.~Bhattacharya, S.~Bhattacharyya, S.~Minwalla and S.~Raju, \emph{{Indices for
  Superconformal Field Theories in 3,5 and 6 Dimensions}},
  \href{http://dx.doi.org/10.1088/1126-6708/2008/02/064}{\emph{JHEP} {\bf 02}
  (2008) 064}, [\href{http://arxiv.org/abs/0801.1435}{{\tt 0801.1435}}].

\bibitem{Louis:2015mka}
J.~Louis and S.~L{\"u}st, \emph{{Supersymmetric AdS$_{7}$ backgrounds in
  half-maximal supergravity and marginal operators of (1, 0) SCFTs}},
  \href{http://dx.doi.org/10.1007/JHEP10(2015)120}{\emph{JHEP} {\bf 10} (2015)
  120}, [\href{http://arxiv.org/abs/1506.08040}{{\tt 1506.08040}}].

\bibitem{Buican:2016hpb}
M.~Buican, J.~Hayling and C.~Papageorgakis, \emph{{Aspects of Superconformal
  Multiplets in $D>4$}},
  \href{http://dx.doi.org/10.1007/JHEP11(2016)091}{\emph{JHEP} {\bf 11} (2016)
  091}, [\href{http://arxiv.org/abs/1606.00810}{{\tt 1606.00810}}].

\bibitem{Cordova:2016emh}
C.~Cordova, T.~T. Dumitrescu and K.~Intriligator, \emph{{Multiplets of
  Superconformal Symmetry in Diverse Dimensions}},
  \href{http://arxiv.org/abs/1612.00809}{{\tt 1612.00809}}.

\bibitem{Cordova:2016xhm}
C.~Cordova, T.~T. Dumitrescu and K.~Intriligator, \emph{{Deformations of
  Superconformal Theories}},
  \href{http://dx.doi.org/10.1007/JHEP11(2016)135}{\emph{JHEP} {\bf 11} (2016)
  135}, [\href{http://arxiv.org/abs/1602.01217}{{\tt 1602.01217}}].

\bibitem{Polyakov:1974gs}
A.~M. Polyakov, \emph{{Nonhamiltonian approach to conformal quantum field
  theory}}, {\emph{Zh. Eksp. Teor. Fiz.} {\bf 66} (1974) 23--42}.

\bibitem{Ferrara:1973yt}
S.~Ferrara, A.~F. Grillo and R.~Gatto, \emph{{Tensor representations of
  conformal algebra and conformally covariant operator product expansion}},
  \href{http://dx.doi.org/10.1016/0003-4916(73)90446-6}{\emph{Annals Phys.}
  {\bf 76} (1973) 161--188}.

\bibitem{Mack:1975jr}
G.~Mack, \emph{{Duality in quantum field theory}},
  \href{http://dx.doi.org/10.1016/0550-3213(77)90238-3}{\emph{Nucl. Phys.} {\bf
  B118} (1977) 445--457}.

\bibitem{Belavin:1984vu}
A.~A. Belavin, A.~M. Polyakov and A.~B. Zamolodchikov, \emph{{Infinite
  Conformal Symmetry in Two-Dimensional Quantum Field Theory}},
  \href{http://dx.doi.org/10.1016/0550-3213(84)90052-X}{\emph{Nucl. Phys.} {\bf
  B241} (1984) 333--380}.

\bibitem{Rattazzi:2008pe}
R.~Rattazzi, V.~S. Rychkov, E.~Tonni and A.~Vichi, \emph{{Bounding scalar
  operator dimensions in 4D CFT}},
  \href{http://dx.doi.org/10.1088/1126-6708/2008/12/031}{\emph{JHEP} {\bf 12}
  (2008) 031}, [\href{http://arxiv.org/abs/0807.0004}{{\tt 0807.0004}}].

\bibitem{Rychkov:2009ij}
V.~S. Rychkov and A.~Vichi, \emph{{Universal Constraints on Conformal Operator
  Dimensions}}, \href{http://dx.doi.org/10.1103/PhysRevD.80.045006}{\emph{Phys.
  Rev.} {\bf D80} (2009) 045006}, [\href{http://arxiv.org/abs/0905.2211}{{\tt
  0905.2211}}].

\bibitem{Poland:2010wg}
D.~Poland and D.~Simmons-Duffin, \emph{{Bounds on 4D Conformal and
  Superconformal Field Theories}},
  \href{http://dx.doi.org/10.1007/JHEP05(2011)017}{\emph{JHEP} {\bf 05} (2011)
  017}, [\href{http://arxiv.org/abs/1009.2087}{{\tt 1009.2087}}].

\bibitem{Poland:2011ey}
D.~Poland, D.~Simmons-Duffin and A.~Vichi, \emph{{Carving Out the Space of 4D
  CFTs}}, \href{http://dx.doi.org/10.1007/JHEP05(2012)110}{\emph{JHEP} {\bf 05}
  (2012) 110}, [\href{http://arxiv.org/abs/1109.5176}{{\tt 1109.5176}}].

\bibitem{ElShowk:2012ht}
S.~El-Showk, M.~F. Paulos, D.~Poland, S.~Rychkov, D.~Simmons-Duffin and
  A.~Vichi, \emph{{Solving the 3D Ising Model with the Conformal Bootstrap}},
  \href{http://dx.doi.org/10.1103/PhysRevD.86.025022}{\emph{Phys. Rev.} {\bf
  D86} (2012) 025022}, [\href{http://arxiv.org/abs/1203.6064}{{\tt
  1203.6064}}].

\bibitem{ElShowk:2012hu}
S.~El-Showk and M.~F. Paulos, \emph{{Bootstrapping Conformal Field Theories
  with the Extremal Functional Method}},
  \href{http://dx.doi.org/10.1103/PhysRevLett.111.241601}{\emph{Phys. Rev.
  Lett.} {\bf 111} (2013) 241601}, [\href{http://arxiv.org/abs/1211.2810}{{\tt
  1211.2810}}].

\bibitem{Beem:2013qxa}
C.~Beem, L.~Rastelli and B.~C. van Rees, \emph{{The $\mathcal N=4$
  Superconformal Bootstrap}},
  \href{http://dx.doi.org/10.1103/PhysRevLett.111.071601}{\emph{Phys. Rev.
  Lett.} {\bf 111} (2013) 071601}, [\href{http://arxiv.org/abs/1304.1803}{{\tt
  1304.1803}}].

\bibitem{Kos:2013tga}
F.~Kos, D.~Poland and D.~Simmons-Duffin, \emph{{Bootstrapping the $O(N)$ vector
  models}}, \href{http://dx.doi.org/10.1007/JHEP06(2014)091}{\emph{JHEP} {\bf
  06} (2014) 091}, [\href{http://arxiv.org/abs/1307.6856}{{\tt 1307.6856}}].

\bibitem{El-Showk:2014dwa}
S.~El-Showk, M.~F. Paulos, D.~Poland, S.~Rychkov, D.~Simmons-Duffin and
  A.~Vichi, \emph{{Solving the 3d Ising Model with the Conformal Bootstrap II.
  c-Minimization and Precise Critical Exponents}},
  \href{http://dx.doi.org/10.1007/s10955-014-1042-7}{\emph{J. Stat. Phys.} {\bf
  157} (2014) 869}, [\href{http://arxiv.org/abs/1403.4545}{{\tt 1403.4545}}].

\bibitem{Chester:2014fya}
S.~M. Chester, J.~Lee, S.~S. Pufu and R.~Yacoby, \emph{{The $ \mathcal{N}=8 $
  superconformal bootstrap in three dimensions}},
  \href{http://dx.doi.org/10.1007/JHEP09(2014)143}{\emph{JHEP} {\bf 09} (2014)
  143}, [\href{http://arxiv.org/abs/1406.4814}{{\tt 1406.4814}}].

\bibitem{Kos:2014bka}
F.~Kos, D.~Poland and D.~Simmons-Duffin, \emph{{Bootstrapping Mixed Correlators
  in the 3D Ising Model}},
  \href{http://dx.doi.org/10.1007/JHEP11(2014)109}{\emph{JHEP} {\bf 11} (2014)
  109}, [\href{http://arxiv.org/abs/1406.4858}{{\tt 1406.4858}}].

\bibitem{Caracciolo:2014cxa}
F.~Caracciolo, A.~Castedo~Echeverri, B.~von Harling and M.~Serone,
  \emph{{Bounds on OPE Coefficients in 4D Conformal Field Theories}},
  \href{http://dx.doi.org/10.1007/JHEP10(2014)020}{\emph{JHEP} {\bf 10} (2014)
  020}, [\href{http://arxiv.org/abs/1406.7845}{{\tt 1406.7845}}].

\bibitem{Chester:2014mea}
S.~M. Chester, J.~Lee, S.~S. Pufu and R.~Yacoby, \emph{{Exact Correlators of
  BPS Operators from the 3d Superconformal Bootstrap}},
  \href{http://dx.doi.org/10.1007/JHEP03(2015)130}{\emph{JHEP} {\bf 03} (2015)
  130}, [\href{http://arxiv.org/abs/1412.0334}{{\tt 1412.0334}}].

\bibitem{Bae:2014hia}
J.-B. Bae and S.-J. Rey, \emph{{Conformal Bootstrap Approach to O(N) Fixed
  Points in Five Dimensions}},  \href{http://arxiv.org/abs/1412.6549}{{\tt
  1412.6549}}.

\bibitem{Beem:2014zpa}
C.~Beem, M.~Lemos, P.~Liendo, L.~Rastelli and B.~C. van Rees, \emph{{The $
  \mathcal{N}=2 $ superconformal bootstrap}},
  \href{http://dx.doi.org/10.1007/JHEP03(2016)183}{\emph{JHEP} {\bf 03} (2016)
  183}, [\href{http://arxiv.org/abs/1412.7541}{{\tt 1412.7541}}].

\bibitem{Chester:2014gqa}
S.~M. Chester, S.~S. Pufu and R.~Yacoby, \emph{{Bootstrapping $O(N)$ vector
  models in 4 $< d <$ 6}},
  \href{http://dx.doi.org/10.1103/PhysRevD.91.086014}{\emph{Phys. Rev.} {\bf
  D91} (2015) 086014}, [\href{http://arxiv.org/abs/1412.7746}{{\tt
  1412.7746}}].

\bibitem{Simmons-Duffin:2015qma}
D.~Simmons-Duffin, \emph{{A Semidefinite Program Solver for the Conformal
  Bootstrap}}, \href{http://dx.doi.org/10.1007/JHEP06(2015)174}{\emph{JHEP}
  {\bf 06} (2015) 174}, [\href{http://arxiv.org/abs/1502.02033}{{\tt
  1502.02033}}].

\bibitem{Kos:2015mba}
F.~Kos, D.~Poland, D.~Simmons-Duffin and A.~Vichi, \emph{{Bootstrapping the
  O(N) Archipelago}},
  \href{http://dx.doi.org/10.1007/JHEP11(2015)106}{\emph{JHEP} {\bf 11} (2015)
  106}, [\href{http://arxiv.org/abs/1504.07997}{{\tt 1504.07997}}].

\bibitem{Chester:2015qca}
S.~M. Chester, S.~Giombi, L.~V. Iliesiu, I.~R. Klebanov, S.~S. Pufu and
  R.~Yacoby, \emph{{Accidental Symmetries and the Conformal Bootstrap}},
  \href{http://dx.doi.org/10.1007/JHEP01(2016)110}{\emph{JHEP} {\bf 01} (2016)
  110}, [\href{http://arxiv.org/abs/1507.04424}{{\tt 1507.04424}}].

\bibitem{Beem:2015aoa}
C.~Beem, M.~Lemos, L.~Rastelli and B.~C. van Rees, \emph{{The (2, 0)
  superconformal bootstrap}},
  \href{http://dx.doi.org/10.1103/PhysRevD.93.025016}{\emph{Phys. Rev.} {\bf
  D93} (2016) 025016}, [\href{http://arxiv.org/abs/1507.05637}{{\tt
  1507.05637}}].

\bibitem{Iliesiu:2015qra}
L.~Iliesiu, F.~Kos, D.~Poland, S.~S. Pufu, D.~Simmons-Duffin and R.~Yacoby,
  \emph{{Bootstrapping 3D Fermions}},
  \href{http://dx.doi.org/10.1007/JHEP03(2016)120}{\emph{JHEP} {\bf 03} (2016)
  120}, [\href{http://arxiv.org/abs/1508.00012}{{\tt 1508.00012}}].

\bibitem{Lemos:2015awa}
M.~Lemos and P.~Liendo, \emph{{Bootstrapping $ \mathcal{N}=2 $ chiral
  correlators}}, \href{http://dx.doi.org/10.1007/JHEP01(2016)025}{\emph{JHEP}
  {\bf 01} (2016) 025}, [\href{http://arxiv.org/abs/1510.03866}{{\tt
  1510.03866}}].

\bibitem{Lin:2015wcg}
Y.-H. Lin, S.-H. Shao, D.~Simmons-Duffin, Y.~Wang and X.~Yin, \emph{{N=4
  Superconformal Bootstrap of the K3 CFT}},
  \href{http://arxiv.org/abs/1511.04065}{{\tt 1511.04065}}.

\bibitem{Kos:2016ysd}
F.~Kos, D.~Poland, D.~Simmons-Duffin and A.~Vichi, \emph{{Precision islands in
  the Ising and O(N ) models}},
  \href{http://dx.doi.org/10.1007/JHEP08(2016)036}{\emph{JHEP} {\bf 08} (2016)
  036}, [\href{http://arxiv.org/abs/1603.04436}{{\tt 1603.04436}}].

\bibitem{Li:2016wdp}
Z.~Li and N.~Su, \emph{{Bootstrapping Mixed Correlators in the Five Dimensional
  Critical O(N) Models}},  \href{http://arxiv.org/abs/1607.07077}{{\tt
  1607.07077}}.

\bibitem{Collier:2016cls}
S.~Collier, Y.-H. Lin and X.~Yin, \emph{{Modular Bootstrap Revisited}},
  \href{http://arxiv.org/abs/1608.06241}{{\tt 1608.06241}}.

\bibitem{Lin:2016gcl}
Y.-H. Lin, S.-H. Shao, Y.~Wang and X.~Yin, \emph{{(2,2) Superconformal
  Bootstrap in Two Dimensions}},  \href{http://arxiv.org/abs/1610.05371}{{\tt
  1610.05371}}.

\bibitem{Lemos:2016xke}
M.~Lemos, P.~Liendo, C.~Meneghelli and V.~Mitev, \emph{{Bootstrapping
  $\mathcal{N}=3$ superconformal theories}},
  \href{http://dx.doi.org/10.1007/JHEP04(2017)032}{\emph{JHEP} {\bf 04} (2017)
  032}, [\href{http://arxiv.org/abs/1612.01536}{{\tt 1612.01536}}].

\bibitem{Li:2017ddj}
D.~Li, D.~Meltzer and A.~Stergiou, \emph{{Bootstrapping Mixed Correlators in 4D
  $\mathcal{N}=1$ SCFTs}},  \href{http://arxiv.org/abs/1702.00404}{{\tt
  1702.00404}}.

\bibitem{Collier:2017shs}
S.~Collier, P.~Kravchuk, Y.-H. Lin and X.~Yin, \emph{{Bootstrapping the
  Spectral Function: On the Uniqueness of Liouville and the Universality of
  BTZ}},  \href{http://arxiv.org/abs/1702.00423}{{\tt 1702.00423}}.

\bibitem{Ganor:1996mu}
O.~J. Ganor and A.~Hanany, \emph{{Small E(8) instantons and tensionless
  noncritical strings}},
  \href{http://dx.doi.org/10.1016/0550-3213(96)00243-X}{\emph{Nucl. Phys.} {\bf
  B474} (1996) 122--140}, [\href{http://arxiv.org/abs/hep-th/9602120}{{\tt
  hep-th/9602120}}].

\bibitem{Seiberg:1996vs}
N.~Seiberg and E.~Witten, \emph{{Comments on string dynamics in
  six-dimensions}},
  \href{http://dx.doi.org/10.1016/0550-3213(96)00189-7}{\emph{Nucl. Phys.} {\bf
  B471} (1996) 121--134}, [\href{http://arxiv.org/abs/hep-th/9603003}{{\tt
  hep-th/9603003}}].

\bibitem{Seiberg:1996bd}
N.~Seiberg, \emph{{Five-dimensional SUSY field theories, nontrivial fixed
  points and string dynamics}},
  \href{http://dx.doi.org/10.1016/S0370-2693(96)01215-4}{\emph{Phys. Lett.}
  {\bf B388} (1996) 753--760}, [\href{http://arxiv.org/abs/hep-th/9608111}{{\tt
  hep-th/9608111}}].

\bibitem{Morrison:1996xf}
D.~R. Morrison and N.~Seiberg, \emph{{Extremal transitions and five-dimensional
  supersymmetric field theories}},
  \href{http://dx.doi.org/10.1016/S0550-3213(96)00592-5}{\emph{Nucl. Phys.}
  {\bf B483} (1997) 229--247}, [\href{http://arxiv.org/abs/hep-th/9609070}{{\tt
  hep-th/9609070}}].

\bibitem{Ganor:1996pc}
O.~J. Ganor, D.~R. Morrison and N.~Seiberg, \emph{{Branes, Calabi-Yau spaces,
  and toroidal compactification of the N=1 six-dimensional E(8) theory}},
  \href{http://dx.doi.org/10.1016/S0550-3213(96)00690-6}{\emph{Nucl. Phys.}
  {\bf B487} (1997) 93--127}, [\href{http://arxiv.org/abs/hep-th/9610251}{{\tt
  hep-th/9610251}}].

\bibitem{Heckman:2016xdl}
J.~J. Heckman, P.~Jefferson, T.~Rudelius and C.~Vafa, \emph{{Punctures for
  theories of class $ {\mathcal{S}}_{\varGamma } $}},
  \href{http://dx.doi.org/10.1007/JHEP03(2017)171}{\emph{JHEP} {\bf 03} (2017)
  171}, [\href{http://arxiv.org/abs/1609.01281}{{\tt 1609.01281}}].

\bibitem{Razamat:2016dpl}
S.~S. Razamat, C.~Vafa and G.~Zafrir, \emph{{4d N=1 from 6d (1,0)}},
  \href{http://arxiv.org/abs/1610.09178}{{\tt 1610.09178}}.

\bibitem{DelZotto:2014hpa}
M.~Del~Zotto, J.~J. Heckman, A.~Tomasiello and C.~Vafa, \emph{{6d Conformal
  Matter}}, \href{http://dx.doi.org/10.1007/JHEP02(2015)054}{\emph{JHEP} {\bf
  02} (2015) 054}, [\href{http://arxiv.org/abs/1407.6359}{{\tt 1407.6359}}].

\bibitem{Heckman:2013pva}
J.~J. Heckman, D.~R. Morrison and C.~Vafa, \emph{{On the Classification of 6D
  SCFTs and Generalized ADE Orbifolds}},
  \href{http://dx.doi.org/10.1007/JHEP06(2015)017,
  10.1007/JHEP05(2014)028}{\emph{JHEP} {\bf 05} (2014) 028},
  [\href{http://arxiv.org/abs/1312.5746}{{\tt 1312.5746}}].

\bibitem{Heckman:2014qba}
J.~J. Heckman, \emph{{More on the Matter of 6D SCFTs}},
  \href{http://dx.doi.org/10.1016/j.physletb.2015.05.046}{\emph{Phys. Lett.}
  {\bf B747} (2015) 73--75}, [\href{http://arxiv.org/abs/1408.0006}{{\tt
  1408.0006}}].

\bibitem{Heckman:2015bfa}
J.~J. Heckman, D.~R. Morrison, T.~Rudelius and C.~Vafa, \emph{{Atomic
  Classification of 6D SCFTs}},
  \href{http://dx.doi.org/10.1002/prop.201500024}{\emph{Fortsch. Phys.} {\bf
  63} (2015) 468--530}, [\href{http://arxiv.org/abs/1502.05405}{{\tt
  1502.05405}}].

\bibitem{Harvey:1998bx}
J.~A. Harvey, R.~Minasian and G.~W. Moore, \emph{{NonAbelian tensor multiplet
  anomalies}},
  \href{http://dx.doi.org/10.1088/1126-6708/1998/09/004}{\emph{JHEP} {\bf 09}
  (1998) 004}, [\href{http://arxiv.org/abs/hep-th/9808060}{{\tt
  hep-th/9808060}}].

\bibitem{Intriligator:2000eq}
K.~A. Intriligator, \emph{{Anomaly matching and a Hopf-Wess-Zumino term in 6d,
  N=(2,0) field theories}},
  \href{http://dx.doi.org/10.1016/S0550-3213(00)00148-6}{\emph{Nucl. Phys.}
  {\bf B581} (2000) 257--273}, [\href{http://arxiv.org/abs/hep-th/0001205}{{\tt
  hep-th/0001205}}].

\bibitem{Ohmori:2014pca}
K.~Ohmori, H.~Shimizu and Y.~Tachikawa, \emph{{Anomaly polynomial of E-string
  theories}}, \href{http://dx.doi.org/10.1007/JHEP08(2014)002}{\emph{JHEP} {\bf
  08} (2014) 002}, [\href{http://arxiv.org/abs/1404.3887}{{\tt 1404.3887}}].

\bibitem{Ohmori:2014kda}
K.~Ohmori, H.~Shimizu, Y.~Tachikawa and K.~Yonekura, \emph{{Anomaly polynomial
  of general 6d SCFTs}},
  \href{http://dx.doi.org/10.1093/ptep/ptu140}{\emph{PTEP} {\bf 2014} (2014)
  103B07}, [\href{http://arxiv.org/abs/1408.5572}{{\tt 1408.5572}}].

\bibitem{Intriligator:2014eaa}
K.~Intriligator, \emph{{6d, $ \mathcal{N}=\left(1,\;0\right) $ Coulomb branch
  anomaly matching}},
  \href{http://dx.doi.org/10.1007/JHEP10(2014)162}{\emph{JHEP} {\bf 10} (2014)
  162}, [\href{http://arxiv.org/abs/1408.6745}{{\tt 1408.6745}}].

\bibitem{Mekareeya:2016yal}
N.~Mekareeya, T.~Rudelius and A.~Tomasiello, \emph{{T-branes, Anomalies and
  Moduli Spaces in 6D SCFTs}},  \href{http://arxiv.org/abs/1612.06399}{{\tt
  1612.06399}}.

\bibitem{Shimizu:2017kzs}
H.~Shimizu, Y.~Tachikawa and G.~Zafrir, \emph{{Anomaly matching on the Higgs
  branch}},  \href{http://arxiv.org/abs/1703.01013}{{\tt 1703.01013}}.

\bibitem{Cordova:2015fha}
C.~Cordova, T.~T. Dumitrescu and K.~Intriligator, \emph{{Anomalies,
  renormalization group flows, and the a-theorem in six-dimensional (1, 0)
  theories}}, \href{http://dx.doi.org/10.1007/JHEP10(2016)080}{\emph{JHEP} {\bf
  10} (2016) 080}, [\href{http://arxiv.org/abs/1506.03807}{{\tt 1506.03807}}].

\bibitem{Beccaria:2015uta}
M.~Beccaria and A.~A. Tseytlin, \emph{{Conformal a-anomaly of some non-unitary
  6d superconformal theories}},
  \href{http://dx.doi.org/10.1007/JHEP09(2015)017}{\emph{JHEP} {\bf 09} (2015)
  017}, [\href{http://arxiv.org/abs/1506.08727}{{\tt 1506.08727}}].

\bibitem{Beccaria:2015ypa}
M.~Beccaria and A.~A. Tseytlin, \emph{{Conformal anomaly c-coefficients of
  superconformal 6d theories}},
  \href{http://dx.doi.org/10.1007/JHEP01(2016)001}{\emph{JHEP} {\bf 01} (2016)
  001}, [\href{http://arxiv.org/abs/1510.02685}{{\tt 1510.02685}}].

\bibitem{Dumitrescu}
C.~Cordova, T.~T. Dumitrescu and K.~Intriligator, ``To appear.''.

\bibitem{Yankielowicz:2017xkf}
S.~Yankielowicz and Y.~Zhou, \emph{{Supersymmetric Renyi Entropy and Anomalies
  in Six-Dimensional (1,0) Superconformal Theories}},
  \href{http://arxiv.org/abs/1702.03518}{{\tt 1702.03518}}.

\bibitem{Beccaria:2017dmw}
M.~Beccaria and A.~A. Tseytlin, \emph{{$C_T$ for higher derivative conformal
  fields and anomalies of (1,0) superconformal 6d theories}},
  \href{http://arxiv.org/abs/1705.00305}{{\tt 1705.00305}}.

\bibitem{Alba:2015upa}
V.~Alba and K.~Diab, \emph{{Constraining conformal field theories with a higher
  spin symmetry in $d > 3$ dimensions}},
  \href{http://dx.doi.org/10.1007/JHEP03(2016)044}{\emph{JHEP} {\bf 03} (2016)
  044}, [\href{http://arxiv.org/abs/1510.02535}{{\tt 1510.02535}}].

\bibitem{Maldacena:2011jn}
J.~Maldacena and A.~Zhiboedov, \emph{{Constraining Conformal Field Theories
  with A Higher Spin Symmetry}},
  \href{http://dx.doi.org/10.1088/1751-8113/46/21/214011}{\emph{J. Phys.} {\bf
  A46} (2013) 214011}, [\href{http://arxiv.org/abs/1112.1016}{{\tt
  1112.1016}}].

\bibitem{Alba:2013yda}
V.~Alba and K.~Diab, \emph{{Constraining conformal field theories with a higher
  spin symmetry in d=4}},  \href{http://arxiv.org/abs/1307.8092}{{\tt
  1307.8092}}.

\bibitem{Dolan:2004mu}
F.~A. Dolan, L.~Gallot and E.~Sokatchev, \emph{{On four-point functions of
  1/2-BPS operators in general dimensions}},
  \href{http://dx.doi.org/10.1088/1126-6708/2004/09/056}{\emph{JHEP} {\bf 09}
  (2004) 056}, [\href{http://arxiv.org/abs/hep-th/0405180}{{\tt
  hep-th/0405180}}].

\bibitem{Ferrara:2001uj}
S.~Ferrara and E.~Sokatchev, \emph{{Universal properties of superconformal OPEs
  for 1/2 BPS operators in 3 $<$= D $<$= 6}},
  \href{http://dx.doi.org/10.1088/1367-2630/4/1/302}{\emph{New J. Phys.} {\bf
  4} (2002) 2}, [\href{http://arxiv.org/abs/hep-th/0110174}{{\tt
  hep-th/0110174}}].

\bibitem{Bobev:2017jhk}
N.~Bobev, E.~Lauria and D.~Mazac, \emph{{Superconformal Blocks for SCFTs with
  Eight Supercharges}},  \href{http://arxiv.org/abs/1705.08594}{{\tt
  1705.08594}}.

\bibitem{cvitanovic2008group}
P.~Cvitanovi{\'c}, \emph{Group Theory: Birdtracks, Lies, and Exceptional
  Groups}.
\newblock Princeton University Press, 41 William St, Princeton, NJ 08540 USA,
  2008.

\bibitem{Osborn:1993cr}
H.~Osborn and A.~C. Petkou, \emph{{Implications of conformal invariance in
  field theories for general dimensions}},
  \href{http://dx.doi.org/10.1006/aphy.1994.1045}{\emph{Annals Phys.} {\bf 231}
  (1994) 311--362}, [\href{http://arxiv.org/abs/hep-th/9307010}{{\tt
  hep-th/9307010}}].

\bibitem{Bonora:1985cq}
L.~Bonora, P.~Pasti and M.~Bregola, \emph{{Weyl Cocycles}},
  \href{http://dx.doi.org/10.1088/0264-9381/3/4/018}{\emph{Class. Quant. Grav.}
  {\bf 3} (1986) 635}.

\bibitem{Deser:1993yx}
S.~Deser and A.~Schwimmer, \emph{{Geometric classification of conformal
  anomalies in arbitrary dimensions}},
  \href{http://dx.doi.org/10.1016/0370-2693(93)90934-A}{\emph{Phys. Lett.} {\bf
  B309} (1993) 279--284}, [\href{http://arxiv.org/abs/hep-th/9302047}{{\tt
  hep-th/9302047}}].

\bibitem{Bastianelli:2000hi}
F.~Bastianelli, S.~Frolov and A.~A. Tseytlin, \emph{{Conformal anomaly of (2,0)
  tensor multiplet in six-dimensions and AdS / CFT correspondence}},
  \href{http://dx.doi.org/10.1088/1126-6708/2000/02/013}{\emph{JHEP} {\bf 02}
  (2000) 013}, [\href{http://arxiv.org/abs/hep-th/0001041}{{\tt
  hep-th/0001041}}].

\bibitem{Giombi:2016fct}
S.~Giombi, G.~Tarnopolsky and I.~R. Klebanov, \emph{{On $C_{J}$ and $C_{T}$ in
  Conformal QED}}, \href{http://dx.doi.org/10.1007/JHEP08(2016)156}{\emph{JHEP}
  {\bf 08} (2016) 156}, [\href{http://arxiv.org/abs/1602.01076}{{\tt
  1602.01076}}].

\bibitem{Osborn:2016bev}
H.~Osborn and A.~Stergiou, \emph{{C$_{T}$ for non-unitary CFTs in higher
  dimensions}}, \href{http://dx.doi.org/10.1007/JHEP06(2016)079}{\emph{JHEP}
  {\bf 06} (2016) 079}, [\href{http://arxiv.org/abs/1603.07307}{{\tt
  1603.07307}}].

\bibitem{Penedones:2015aga}
J.~Penedones, E.~Trevisani and M.~Yamazaki, \emph{{Recursion Relations for
  Conformal Blocks}},
  \href{http://dx.doi.org/10.1007/JHEP09(2016)070}{\emph{JHEP} {\bf 09} (2016)
  070}, [\href{http://arxiv.org/abs/1509.00428}{{\tt 1509.00428}}].

\bibitem{Hogervorst:2013kva}
M.~Hogervorst, H.~Osborn and S.~Rychkov, \emph{{Diagonal Limit for Conformal
  Blocks in $d$ Dimensions}},
  \href{http://dx.doi.org/10.1007/JHEP08(2013)014}{\emph{JHEP} {\bf 08} (2013)
  014}, [\href{http://arxiv.org/abs/1305.1321}{{\tt 1305.1321}}].

\bibitem{Gaiotto:2008nz}
D.~Gaiotto, A.~Neitzke and Y.~Tachikawa, \emph{{Argyres-Seiberg duality and the
  Higgs branch}},
  \href{http://dx.doi.org/10.1007/s00220-009-0938-6}{\emph{Commun. Math. Phys.}
  {\bf 294} (2010) 389--410}, [\href{http://arxiv.org/abs/0810.4541}{{\tt
  0810.4541}}].

\bibitem{Beem:2013sza}
C.~Beem, M.~Lemos, P.~Liendo, W.~Peelaers, L.~Rastelli and B.~C. van Rees,
  \emph{{Infinite Chiral Symmetry in Four Dimensions}},
  \href{http://dx.doi.org/10.1007/s00220-014-2272-x}{\emph{Commun. Math. Phys.}
  {\bf 336} (2015) 1359--1433}, [\href{http://arxiv.org/abs/1312.5344}{{\tt
  1312.5344}}].

\bibitem{Seiberg:1994aj}
N.~Seiberg and E.~Witten, \emph{{Monopoles, duality and chiral symmetry
  breaking in N=2 supersymmetric QCD}},
  \href{http://dx.doi.org/10.1016/0550-3213(94)90214-3}{\emph{Nucl. Phys.} {\bf
  B431} (1994) 484--550}, [\href{http://arxiv.org/abs/hep-th/9408099}{{\tt
  hep-th/9408099}}].

\bibitem{Argyres:1996eh}
P.~C. Argyres, M.~R. Plesser and N.~Seiberg, \emph{{The Moduli space of vacua
  of N=2 SUSY QCD and duality in N=1 SUSY QCD}},
  \href{http://dx.doi.org/10.1016/0550-3213(96)00210-6}{\emph{Nucl. Phys.} {\bf
  B471} (1996) 159--194}, [\href{http://arxiv.org/abs/hep-th/9603042}{{\tt
  hep-th/9603042}}].

\bibitem{Argyres:2012fu}
P.~C. Argyres, K.~Maruyoshi and Y.~Tachikawa, \emph{{Quantum Higgs branches of
  isolated N=2 superconformal field theories}},
  \href{http://dx.doi.org/10.1007/JHEP10(2012)054}{\emph{JHEP} {\bf 10} (2012)
  054}, [\href{http://arxiv.org/abs/1206.4700}{{\tt 1206.4700}}].

\bibitem{Gubser:2002tv}
S.~S. Gubser, I.~R. Klebanov and A.~M. Polyakov, \emph{{A Semiclassical limit
  of the gauge / string correspondence}},
  \href{http://dx.doi.org/10.1016/S0550-3213(02)00373-5}{\emph{Nucl. Phys.}
  {\bf B636} (2002) 99--114}, [\href{http://arxiv.org/abs/hep-th/0204051}{{\tt
  hep-th/0204051}}].

\bibitem{Gromov:2009zb}
N.~Gromov, V.~Kazakov and P.~Vieira, \emph{{Exact Spectrum of Planar ${\cal
  N}=4$ Supersymmetric Yang-Mills Theory: Konishi Dimension at Any Coupling}},
  \href{http://dx.doi.org/10.1103/PhysRevLett.104.211601}{\emph{Phys. Rev.
  Lett.} {\bf 104} (2010) 211601}, [\href{http://arxiv.org/abs/0906.4240}{{\tt
  0906.4240}}].

\bibitem{Anselmi:1996mq}
D.~Anselmi, M.~T. Grisaru and A.~Johansen, \emph{{A Critical behavior of
  anomalous currents, electric - magnetic universality and CFT in
  four-dimensions}},
  \href{http://dx.doi.org/10.1016/S0550-3213(97)00108-9}{\emph{Nucl. Phys.}
  {\bf B491} (1997) 221--248}, [\href{http://arxiv.org/abs/hep-th/9601023}{{\tt
  hep-th/9601023}}].

\bibitem{Chang}
C.-M. Chang, M.~Fluder, Y.-H. Lin and Y.~Wang, ``Work in progress.''.

\bibitem{Hama:2011ea}
N.~Hama, K.~Hosomichi and S.~Lee, \emph{{SUSY Gauge Theories on Squashed
  Three-Spheres}}, \href{http://dx.doi.org/10.1007/JHEP05(2011)014}{\emph{JHEP}
  {\bf 05} (2011) 014}, [\href{http://arxiv.org/abs/1102.4716}{{\tt
  1102.4716}}].

\bibitem{Imamura:2011wg}
Y.~Imamura and D.~Yokoyama, \emph{{N=2 supersymmetric theories on squashed
  three-sphere}},
  \href{http://dx.doi.org/10.1103/PhysRevD.85.025015}{\emph{Phys. Rev.} {\bf
  D85} (2012) 025015}, [\href{http://arxiv.org/abs/1109.4734}{{\tt
  1109.4734}}].

\bibitem{Closset:2012ru}
C.~Closset, T.~T. Dumitrescu, G.~Festuccia and Z.~Komargodski,
  \emph{{Supersymmetric Field Theories on Three-Manifolds}},
  \href{http://dx.doi.org/10.1007/JHEP05(2013)017}{\emph{JHEP} {\bf 05} (2013)
  017}, [\href{http://arxiv.org/abs/1212.3388}{{\tt 1212.3388}}].

\bibitem{Nishioka:2013gza}
T.~Nishioka and K.~Yonekura, \emph{{On RG Flow of $tau_{RR}$ for Supersymmetric
  Field Theories in Three-Dimensions}},
  \href{http://dx.doi.org/10.1007/JHEP05(2013)165}{\emph{JHEP} {\bf 05} (2013)
  165}, [\href{http://arxiv.org/abs/1303.1522}{{\tt 1303.1522}}].

\bibitem{Imamura:2012xg}
Y.~Imamura, \emph{{Supersymmetric theories on squashed five-sphere}},
  \href{http://dx.doi.org/10.1093/ptep/pts052}{\emph{PTEP} {\bf 2013} (2013)
  013B04}, [\href{http://arxiv.org/abs/1209.0561}{{\tt 1209.0561}}].

\bibitem{Lockhart:2012vp}
G.~Lockhart and C.~Vafa, \emph{{Superconformal Partition Functions and
  Non-perturbative Topological Strings}},
  \href{http://arxiv.org/abs/1210.5909}{{\tt 1210.5909}}.

\bibitem{Imamura:2012bm}
Y.~Imamura, \emph{{Perturbative partition function for squashed $S^5$}},
  \href{http://dx.doi.org/10.1093/ptep/ptt044}{\emph{PTEP} {\bf 2013} (2013)
  073B01}, [\href{http://arxiv.org/abs/1210.6308}{{\tt 1210.6308}}].

\bibitem{Kim:2012qf}
H.-C. Kim, J.~Kim and S.~Kim, \emph{{Instantons on the 5-sphere and
  M5-branes}},  \href{http://arxiv.org/abs/1211.0144}{{\tt 1211.0144}}.

\bibitem{Spiridonov:2012de}
V.~P. Spiridonov, \emph{{Modified elliptic gamma functions and 6d
  superconformal indices}},
  \href{http://dx.doi.org/10.1007/s11005-013-0678-6}{\emph{Lett. Math. Phys.}
  {\bf 104} (2014) 397--414}, [\href{http://arxiv.org/abs/1211.2703}{{\tt
  1211.2703}}].

\bibitem{Alday:2014rxa}
L.~F. Alday, M.~Fluder, P.~Richmond and J.~Sparks, \emph{{Gravity Dual of
  Supersymmetric Gauge Theories on a Squashed Five-Sphere}},
  \href{http://dx.doi.org/10.1103/PhysRevLett.113.141601}{\emph{Phys. Rev.
  Lett.} {\bf 113} (2014) 141601}, [\href{http://arxiv.org/abs/1404.1925}{{\tt
  1404.1925}}].

\bibitem{Alday:2014bta}
L.~F. Alday, M.~Fluder, C.~M. Gregory, P.~Richmond and J.~Sparks,
  \emph{{Supersymmetric gauge theories on squashed five-spheres and their
  gravity duals}}, \href{http://dx.doi.org/10.1007/JHEP09(2014)067}{\emph{JHEP}
  {\bf 09} (2014) 067}, [\href{http://arxiv.org/abs/1405.7194}{{\tt
  1405.7194}}].

\bibitem{Bobev:2017asb}
N.~Bobev, P.~Bueno and Y.~Vreys, \emph{{Comments on Squashed-sphere Partition
  Functions}},  \href{http://arxiv.org/abs/1705.00292}{{\tt 1705.00292}}.

\bibitem{Brandhuber:1999np}
A.~Brandhuber and Y.~Oz, \emph{{The D-4 - D-8 brane system and five-dimensional
  fixed points}},
  \href{http://dx.doi.org/10.1016/S0370-2693(99)00763-7}{\emph{Phys. Lett.}
  {\bf B460} (1999) 307--312}, [\href{http://arxiv.org/abs/hep-th/9905148}{{\tt
  hep-th/9905148}}].

\bibitem{Kao:1995gf}
H.-C. Kao, K.-M. Lee and T.~Lee, \emph{{The Chern-Simons coefficient in
  supersymmetric Yang-Mills Chern-Simons theories}},
  \href{http://dx.doi.org/10.1016/0370-2693(96)00119-0}{\emph{Phys. Lett.} {\bf
  B373} (1996) 94--99}, [\href{http://arxiv.org/abs/hep-th/9506170}{{\tt
  hep-th/9506170}}].

\bibitem{Kapustin:1999ha}
A.~Kapustin and M.~J. Strassler, \emph{{On mirror symmetry in three-dimensional
  Abelian gauge theories}},
  \href{http://dx.doi.org/10.1088/1126-6708/1999/04/021}{\emph{JHEP} {\bf 04}
  (1999) 021}, [\href{http://arxiv.org/abs/hep-th/9902033}{{\tt
  hep-th/9902033}}].

\bibitem{Gaiotto:2007qi}
D.~Gaiotto and X.~Yin, \emph{{Notes on superconformal Chern-Simons-Matter
  theories}},
  \href{http://dx.doi.org/10.1088/1126-6708/2007/08/056}{\emph{JHEP} {\bf 08}
  (2007) 056}, [\href{http://arxiv.org/abs/0704.3740}{{\tt 0704.3740}}].

\bibitem{Hogervorst:2013sma}
M.~Hogervorst and S.~Rychkov, \emph{{Radial Coordinates for Conformal Blocks}},
  \href{http://dx.doi.org/10.1103/PhysRevD.87.106004}{\emph{Phys. Rev.} {\bf
  D87} (2013) 106004}, [\href{http://arxiv.org/abs/1303.1111}{{\tt
  1303.1111}}].

\bibitem{Dolan:2003hv}
F.~A. Dolan and H.~Osborn, \emph{{Conformal partial waves and the operator
  product expansion}},
  \href{http://dx.doi.org/10.1016/j.nuclphysb.2003.11.016}{\emph{Nucl. Phys.}
  {\bf B678} (2004) 491--507}, [\href{http://arxiv.org/abs/hep-th/0309180}{{\tt
  hep-th/0309180}}].

\bibitem{Zamolodchikov:1985ie}
A.~B. Zamolodchikov, \emph{{CONFORMAL SYMMETRY IN TWO-DIMENSIONS: AN EXPLICIT
  RECURRENCE FORMULA FOR THE CONFORMAL PARTIAL WAVE AMPLITUDE}},
  \href{http://dx.doi.org/10.1007/BF01214585}{\emph{Commun. Math. Phys.} {\bf
  96} (1984) 419--422}.

\bibitem{Zamolodchikov:1995aa}
A.~B. Zamolodchikov and A.~B. Zamolodchikov, \emph{{Structure constants and
  conformal bootstrap in Liouville field theory}},
  \href{http://dx.doi.org/10.1016/0550-3213(96)00351-3}{\emph{Nucl. Phys.} {\bf
  B477} (1996) 577--605}, [\href{http://arxiv.org/abs/hep-th/9506136}{{\tt
  hep-th/9506136}}].

\bibitem{Penedones:2016voo}
J.~Penedones, \emph{{TASI lectures on AdS/CFT}},  in \emph{{Proceedings,
  Theoretical Advanced Study Institute in Elementary Particle Physics: New
  Frontiers in Fields and Strings (TASI 2015): Boulder, CO, USA, June 1-26,
  2015}}, pp.~75--136, 2017.
\newblock \href{http://arxiv.org/abs/1608.04948}{{\tt 1608.04948}}.
\newblock \href{http://dx.doi.org/10.1142/9789813149441_0002}{DOI}.

\bibitem{Fradkin:1981jc}
E.~S. Fradkin and A.~A. Tseytlin, \emph{{One Loop Beta Function in Conformal
  Supergravities}},
  \href{http://dx.doi.org/10.1016/0550-3213(82)90481-3}{\emph{Nucl. Phys.} {\bf
  B203} (1982) 157--178}.

\bibitem{Fradkin:1985am}
E.~S. Fradkin and A.~A. Tseytlin, \emph{{CONFORMAL SUPERGRAVITY}},
  \href{http://dx.doi.org/10.1016/0370-1573(85)90138-3}{\emph{Phys. Rept.} {\bf
  119} (1985) 233--362}.

\bibitem{Fitzpatrick:2011dm}
A.~L. Fitzpatrick and J.~Kaplan, \emph{{Unitarity and the Holographic
  S-Matrix}}, \href{http://dx.doi.org/10.1007/JHEP10(2012)032}{\emph{JHEP} {\bf
  10} (2012) 032}, [\href{http://arxiv.org/abs/1112.4845}{{\tt 1112.4845}}].

\end{thebibliography}\endgroup
\bibliographystyle{JHEP}

\end{document}